\def\Xint#1{\mathchoice
   {\XXint\displaystyle\textstyle{#1}}%
   {\XXint\textstyle\scriptstyle{#1}}%
   {\XXint\scriptstyle\scriptscriptstyle{#1}}%
   {\XXint\scriptscriptstyle\scriptscriptstyle{#1}}%
   \!\int}
\def\XXint#1#2#3{{\setbox0=\hbox{$#1{#2#3}{\int}$}
     \vcenter{\hbox{$#2#3$}}\kern-.5\wd0}}

\def\dashint{\Xint-}

\documentclass[aps,prb,twocolumn,showpacs,floatfix,eqsecnum]{revtex4}

\pdfoutput=1

\usepackage{times}
\usepackage{amsfonts}
\usepackage{amssymb}
\usepackage{amsmath}
\usepackage{graphicx}
\usepackage{bm}
\usepackage{verbatim}

\usepackage{hyperref}

\usepackage{bm} 
\usepackage{color}
\usepackage{stackrel}
\usepackage{accents}

\usepackage{latexsym}

\newcommand{\bsub}{\begin{subequations}}
\newcommand{\esub}{\end{subequations}}

\newcommand{\sgn}{\operatorname{sgn}}

\newcommand \bea {\begin{eqnarray} }
\newcommand \eea {\end{eqnarray}}
 
\newcommand{\beg}{\begin{equation}}
\newcommand{\en}{\end{equation}}
\newcommand{\bp}{\mathbf p}
\newcommand{\bq}{\mathbf q}

\newcommand \bel  {\begin{align}}
\newcommand \enl  {\end{align}}

\newcommand{\eps}{\varepsilon}
\newcommand{\lam}{\lambda}

\newcommand{\re}[1]{(\ref{#1})}

\newcommand{\up}{\uparrow}
\newcommand{\dn}{\downarrow}
\newcommand{\dg}{^\dagger}

\newcommand{\eref}[1]{Eq.~(\ref{#1})}

\newcommand{\esref}[1]{Eqs.~(\ref{#1})}

\renewcommand{\Re}{\mathrm{Re}}
\renewcommand{\Im}{\mathrm{Im}}

\newcommand{\pmat}{\begin{pmatrix}}
\newcommand{\epmat}{\end{pmatrix}}

\def\q{{\bf q}}
\def\p{{\bf p}}
\def\k{{\bf k}}

\def\8{\infty}

\def\undertext#1{\vtop{\hbox{#1}\kern 1pt \hrule}}

\def\EV#1{\langle#1\rangle}

\def\be{\begin{equation}}
\def\ee{\end{equation}}
\def\bea{\begin{eqnarray} & &}
\def\eea{\end{eqnarray}}

\def\a{{\hat a}}
\def\ad{\hat a^\dagger}
\def\b{{\hat b}}
\def\bd{\hat b^\dagger}
\def\p {{\bf p}}
\def\q {{\bf q}}

\newcommand{\dis}{\displaystyle}


\begin{document}

\title{Quantum quench phase diagrams of    an $s$-wave BCS-BEC condensate}

\author{E. A. Yuzbashyan$^1$, M. Dzero$^{2}$, V. Gurarie$^3$, and M. S. Foster$^{1,4}$}
\affiliation{$^1$Center for Materials Theory, Rutgers University, Piscataway, NJ 08854, USA\\
$^2$ Department of Physics, Kent State University, Kent, OH 44240 USA\\ 
$^3$ Department of Physics, University of Colorado, Boulder, CO 80309, USA\\
$^4$ Department of Physics and Astronomy,  Rice University,  Houston, Texas 7700, USA}

\begin{abstract} 
We study the dynamic response of  an $s$-wave BCS-BEC (atomic-molecular) condensate to detuning quenches within the two channel model beyond the weak coupling BCS limit. At long times after the quench, the condensate ends up in one of three main asymptotic states (nonequilibrium phases), which are qualitatively similar to those in other fermionic condensates defined by a global complex order parameter.  In phase I the amplitude of the order parameter vanishes as a power law, in phase II it goes to a nonzero constant, and in phase III it oscillates
persistently. We construct exact quench phase diagrams that predict the asymptotic state (including the many-body wavefunction) depending on the initial and final detunings and on the Feshbach resonance width. Outside of the weak coupling regime, both the mechanism and the time dependence of the relaxation of the amplitude of the order parameter in phases I and II are modified. Also, quenches from arbitrarily weak initial to sufficiently strong final coupling do not produce persistent oscillations in contrast to the behavior in the BCS regime.  The most remarkable feature  of coherent  condensate dynamics in various fermion superfluids is an effective  reduction in the number of dynamic  degrees of freedom as the evolution time goes to infinity. As a result, the long time dynamics can be fully  described in terms of just a few new collective dynamical variables governed by the same Hamiltonian only with ``renormalized" parameters. Combining this feature with the integrability of the  underlying (e.g. the two channel) model, we develop and consistently   present a general method that explicitly obtains the exact asymptotic state of the system.

 \end{abstract}

\pacs{67.85.-d, 67.85.De, 34.90.+q, 74.40.Gh, 02.30.Ik, 05.45.Xt, 05.30.Fk, 05.45.Yv}

\maketitle

 \tableofcontents

\section{Introduction}

The problem of a superconductor driven out of  equilibrium by a sudden perturbation goes back  many decades. Early studies\cite{Anderson1958,Galaiko1972,Kogan1973,Galperin1981,Varma1982,Shumeiko} addressed small deviations from equilibrium using linearized equations of motion. An important result was obtained by Volkov and Kogan \cite{Kogan1973}, who discovered a power law oscillatory attenuation of the Bardeen-Cooper-Schriffer (BCS) order parameter for non-equilibrium initial conditions close to the superconducting ground state.

In the last decade it was realized that even large  deviations from equilibrium are within the reach of appropriate theoretical methods.  Recent studies, motivated by experiments in cold atomic fermions,  focused on quantum quenches -- nonequilibrium conditions created by a sudden change in the superconducting coupling strength.
Barankov et. al. \cite{Spivak2004} in a paper that set off  a surge of modern research in this long-standing problem
\cite{Amin2004,Andreev2004,Levitov2004,Burnett2005,Leggett2005,Enolski2005,Enolski2005a,Altshuler2005,Yuzbashyan2006,Barankov2006,Dzero2006,Coleman2007,Levitov2007,Fazio2008,Bettelheim2008,Gurarie2009,Sebastien2009} in the context of quantum gases, found that for initial conditions close to the unstable normal state, the order parameter exhibits large unharmonic periodic oscillations. 

Subsequently, Yuzbashyan et. al.\cite{Yuzbashyan2006} developed  an analytical method to predict the state of the system at large times based on the integrability of the underlying BCS model. This work extended Volkov and Kogan's result to large deviations from equilibrium and showed that the oscillation frequency is twice the \textit{nonequilibrium} asymptotic value of the order parameter, a conclusion confirmed by recent terahertz pump pulse experiments in ${\mathrm{Nb}}_{1\mathrm{\text{-}}x}{\mathrm{Ti}}_{x}\mathbf{N}$ films \cite{Shimano2013,Matsunaga2014}.  
Later studies \cite{Barankov2006,Dzero2006} mapped out the full quantum quench ``phase diagram'' for weakly coupled s-wave BCS superconductors    finding that three distinct regimes occur depending on the strength of the quench: Volkov and Kogan like behavior, persistent oscillations, and exponential vanishing of the order parameter.  Most recent research\cite{Kuhn2007,Axt2009,Schnyder2013,Aoki2014} fueled by experimental breakthroughs\cite{Shimano2012,Demsar2013,Shimano2013} investigates  non-adiabatic dynamics of $s$-wave BCS superconductors in response to fast electromagnetic perturbations. Closely related subjects developing in parallel are exciton dynamics\cite{2011Eastham}, collective neutrino  oscillations\cite{Takashi2011,Georg2013}, quenched p-wave superfluids\cite{Yuzbashyan2013,2014PwavePRL} etc.

Most existing work  addressed the dynamics in the BCS regime  and, in particular,  quenches such that the interaction strength is weak both before and after the quench. This was so that the system  always remains in the BCS regime, since the  physics of the condensate beyond this regime was not sufficiently well understood. However,   a superfluid made up of cold atoms can be as well quenched from the BCS to the Bose-Einstein Condensation (BEC) regime, or within the BEC regime. With few exceptions~\cite{Gurarie2009,Yuzbashyan2013,2014PwavePRL},  these types of quenches are not adequately studied  in the existing literature.  
 
 Our paper aims to close this gap and analyze all possible interaction quenches throughout the  BCS-BEC crossover in a paired superfluid, including BCS to BEC, BEC to BCS and BEC to BEC quenches. We fully determine the steady state of the system at large times after the quench -- the asymptote of the order parameter as well as the approach to the asymptote, the many-body wave function,   and certain observables, such as the radio-frequency absorption spectrum and the momentum distribution. In the BCS limit, we recover previous results. Beyond this limit the dynamics is quantitatively  and sometimes qualitatively different. For example, the power law in the  Volkov and Kogan like attenuation changes in the BEC regime, exponential vanishing is replaced with a power law, and  persistent oscillations first change their form and then disappear altogether after a certain threshold for quenches from any initial   (e.g. arbitrarily weak) to sufficiently strong final coupling. We believe  an experimental verification of the
predictions of this work is within a reach of current experiments in cold atomic systems.    

The long time dynamics can be determined explicitly due to a remarkable reduction mechanism at work, so that at large times the system is governed by an effective interacting  Hamiltonian with just a few   classical collective spin or oscillator degrees of freedom. In a sense, the system ``flows in time'' to a much simpler Hamiltonian. This observation combined with the integrability of the original Hamiltonian (see below) lead to a method originally proposed in Ref.~\onlinecite{Yuzbashyan2006} for obtaining the long time asymptote (steady state) of integrable  Hamiltonian dynamics in the continuum (thermodynamic) limit.
Here we  improve this method  as well as provide its comprehensive  and self-contained review including many  previously  unpublished results and  steps. We do so in the context of the 
$s$-wave BCS (one channel) and inhomogeneous Dicke (two channel) models, but  with some modifications the same method also applies  to all known integrable pairing models 
\cite{Gaudin_book,Richardson2002,Zhao2009,Zhao2010,Ortiz2010,Sierra2004,Rombouts2005}  , such as   $p+ip$ superfluids\cite{Yuzbashyan2013,2014PwavePRL}, integrable fermion or boson pairing models with nonuniform interactions\cite{Schuck2001,Osterloh2001},  Gaudin  magnets (central spin models), and potentially can be extended to a much broader class of integrable nonlinear equations.

 The purpose of  this paper is therefore twofold. First, it serves as an encyclopedia of quantitatively exact predictions, new and old, for the quench dynamics  of real $s$-wave BCS-BEC condensates in two and three spatial dimensions. Readers primarily interested in this aspect of our work will find most of the relevant information in the Introduction, Sect.~\ref{rf}, and Conclusion. In particular,  Sect.~\ref{mainres}  concisely summarizes our main results and provides a guide to other  sections  that contain further results and details.  Our second goal is to develop and thoroughly review a method for determining the far from equilibrium dynamics in a certain class of integrable models. We refer readers interested in learning about the method to Sect.~\ref{method}.  Also, from this viewpoint, Sects.~\ref{qpsd} and \ref{1channel} should be considered   as   applications of our approach and Sect.~\ref{linear} as a  related development.

A major experimental breakthrough with ultracold atoms was achieved in 2004, when they were used to emulate $s$-wave superconductors with an interaction strength that can
be varied at will \cite{Jin2004,Ketterle2004}. Experimental control parameter is the detuning $\omega$ -- the binding energy of a two fermion bound state (molecule). This parameter determines the strength of the effective interaction between fermions and can be varied both slowly and abruptly with the help of a Feshbach resonance. Moreover, it is straightforward to make time resolved measurements of the subsequent evolution of the system. Thus cold atoms provide a natural platform to study quenches in superfluids and in a variety of other setups\cite{Cardy2006,Kollath2007}.  

At large $\omega$ we have fermionic atoms with weak effective attraction that form a paired superfluid, an analog of the superconducting state of electrons in a metal.  As $\omega$ is decreased, the atoms   pair up into bosonic molecules which then Bose condense. It was argued for a long time
that both the paired superfluid and the Bose-condensed molecules are in the same phase of the fermionic gas, named the BCS-BEC  condensate
\cite{Leggett1980,Nozieres1985}. As $\omega$ decreases, the  strength of the effective interaction (coupling) between  fermions increases from weak to strong and the system  undergoes a BCS-BEC crossover. At $\omega\gg 2\eps_F$, where $\eps_F$ is the Fermi energy, the system is deep in the BCS regime, while at large negative $\omega$ it is deep in the BEC regime. It is not known how to recreate such a crossover in a conventional solid-state superconductor since the interaction strength cannot be easily adjusted.

In a quantum quench setup the system is prepared in the ground state at a detuning $\omega_i$. At $t=0$ the detuning is suddenly changed, $\omega_i\to \omega_f$. At $t>0$ the system  evolves with a new Hamiltonian $H(\omega_f)$.  The main goal is to determine the state of the system  at large times, $t\to\infty$.

\subsection{Models and  approximations}
\label{modelsmf}

We consider two closely related models in this paper  in both two and three dimensions.
The first one is the well-known two channel model that describes two species of fermionic atoms interacting via an $s$-wave Feshbach resonance 
\begin{eqnarray} \label{eq:twochannel} 
\hat H_{\mathrm{2ch}} &=&\!\! \sum_{\bf p, \sigma=\uparrow, \downarrow}\!\!\!\!\! \epsilon_\p \ad_{\p\sigma} \a_{\p\sigma} + \sum_\q \left( \omega+ \frac{q^2}{4m} \right)\bd_\q\b_\q + \\ && g \sum_{\p \q}\!\! \left( \hat b^\dagger_\q \a_{\frac \q 2 + \p, \uparrow} \a_{\frac \q 2 - \p,\downarrow} + \b_\q \ad_{\frac \q 2 - \p,\downarrow} \ad_{\frac \q 2 + \p,\uparrow} \right). \nonumber
\end{eqnarray}
It is convenient to think of the two types of fermions of mass $m$ and energy $\epsilon_\p=p^2/2m$ as spin up and down, created and annihilated by  operators $\ad_{\p \sigma}$ and $\a_{\p \sigma}$. The interaction term converts two fermions  into a bosonic molecule and vise versa at a rate controlled by the parameter $g$.
 Molecules are created and annihilated by $\bd_\q$ and $\b_\q$ and have  a binding energy $\omega$. The parameter  $g$ is set by the type of atoms and 
the specifics of a particular Feshbach resonance and cannot be changed in a single experiment; $\omega$ can be varied at will by varying the  magnitude of the magnetic field applied during the experiment. 
This model describes atoms in the BCS regime when $\omega$ is large, which undergo a crossover to the BEC regime as $\omega$ is decreased. 

A parameter with dimensions of energy important for our analysis of this model is $ g^2 \nu_F$, where $\nu_F$ is the bulk density of states (proportional to the total volume) at the Fermi energy $\epsilon_F$.  A well known   parameter 
\be 
\gamma= \frac{g^2 \nu_F}{\epsilon_F}
\label{width}
\ee 
controls whether the resonance is narrow $\gamma \ll 1$ or broad $\gamma \gg 1$. This parameter is the dimensionless atom-molecule interaction strength or, equivalently, the resonance width.

A very convenient feature of the narrow resonance is that regardless of the regime of the system, controlled by $\omega$, the system is adequately described with  mean field theory  \cite{Gurarie2007}. This is already  clear from the form of the Hamiltonian: small $\gamma$ implies that interaction $g$ is small. 

Broad resonances on the other hand correspond to large $g$. Under those conditions it is possible to integrate out the molecules $\b_\q$ to arrive at a simpler Hamiltonian  \cite{Gurarie2007} describing fermions interacting via a short range attractive  interaction with variable strength
\begin{eqnarray}  \label{eq:onechannel}  && \hat H_\mathrm{1ch} = \sum_{\bf p, \sigma=\uparrow, \downarrow}  \epsilon_\p\ad_{\p\sigma} \a_{\p\sigma} - \cr && \frac{ \lambda}{\nu_F}\sum_{\p \p' \q}
 \ad_{\frac \q 2 - \p,\downarrow} \ad_{\frac \q 2 + \p,\uparrow}  \a_{\frac \q 2 + \p', \uparrow} \a_{\frac \q 2 - \p',\downarrow},
\end{eqnarray}
where 
\be 
\label{eq:parmap} 
\lambda = \frac{g^2\nu_F}{\omega}=\frac{\gamma\eps_F}{\omega}.
\ee
 This is the single (one) channel or BCS model, which is the second model we analyze in this paper. It also describes the BCS-BEC crossover as $\omega$ is decreased ($\lambda$ is increased). However, while in the BCS and (to some extent) in the BEC
regimes corresponding to large and small $\lambda$, respectively, mean field theory holds in equilibrium, for the intermediate values of $\lambda$ (neither large nor small) the mean field theory
is known to break down. A special value of $\lambda$ in the middle of the regime unaccessible to the mean field theory already in equilibrium is called the unitary point.  It corresponds to the interaction strength where molecules are about to be formed. Non-condensed molecules play an important role in the description of the unitary point and its special properties  are a subject of many studies in the literature\cite{Nozieres1985,Troyer2006}. 
 
Just as  earlier work on the far from equilibrium superconductivity, we analyze the quench dynamics in the  mean field approximation where no molecules are transferred
into or out of  the BCS-BEC condensate after the quench, i.e. the dynamics of the condensate is decoupled from the non-condensed modes. We analyze the validity of this approximation for nonequilibrium steady states produced by  quenches in two channel model in Appendix~\ref{pairbreak}. We find that the situation is similar to that in equilibrium \cite{Gurarie2007}.
In the case of  a broad Feshbach resonance, mean field   is  expected to hold for  quenches  where both initial and final detunings are far from the unitary point.     A quench into the unitary point is a very interesting problem addressed by some publications before \cite{Bulgac2009}, but the method we employ here is not applicable to this case. 

Nevertheless, a variety of  quenches are still accessible to our description even when the resonance is broad, including BCS$\rightarrow$ BCS, BCS $\rightarrow$ BEC, BEC $\rightarrow$ BCS, and BEC $\rightarrow$ BEC, where BCS and BEC stand for the value of the interaction strength far weaker or far stronger than that at the unitary point. 
In the case of BCS-BEC superfluids formed with interactions generated by narrow Feshbach resonances, the mean field theory treatment is valid even at the 
 threshold
of the formation of the bound state and throughout the BCS-BEC crossover. Here we consider quenches of the detuning $\omega$ for both narrow and  broad resonances within mean field. 
 Note that in the case of the one channel model we expect the mean field on the BEC side to be valid only in the far BEC limit where the ground state essentially consists of non-interacting Bose condensed molecules\cite{Levinsen2006}.

In the mean field treatment the condensate is described by the ${\bf q}=0$ part of the Hamiltonian \re{eq:twochannel}, which is decoupled from  ${\bf q}\ne 0$ terms in this approximation. The Hamiltonian therefore becomes 
\beg\label{Model}
\hat H_{\mathrm{2ch}}=\sum\limits_{\bp}2 \epsilon_\bp \hat{s}_\bp^z +\omega\hat{b}\dg \hat{b}  
	+
	g\sum\limits_{\bp} \left( \hat{b}\dg \hat{s}_\bp^- +\hat{s}_\bp^+ \hat{b} \right),
\en
where
\beg
\hat{s}_\bp^-= \hat{a}_{\bp\uparrow}\hat{a}_{-\bp\downarrow},\,\, \hat{s}_\bp^z=\frac{1}{2}\bigl(\hat{a}_{\bp\uparrow}^\dagger\hat{a}_{\bp\uparrow}+\hat{a}_{-\bp\downarrow}^\dagger\hat{a}_{-\bp\downarrow}-1\bigr)
\label{pseudodef}
\en
are Anderson pseudospin-1/2 operators \cite{Anderson1958} and 
$$
\hat b=\hat b_{\q=0}.
$$
Hamiltonian \re{Model} is also known as  inhomogeneous Dicke or Tavis-Cummings model. In a quantum quench problem we need to solve  Heisenberg equations of motion for this Hamiltonian for given initial conditions
\beg\label{Bloch1}
\begin{split}
\frac{d\hat{\vec s}_\bp}{dt}=\hat{\vec{B}}_\bp\times\hat{\vec s}_\bp, \quad \frac{d\hat{b}}{dt}=-i\omega \hat b-ig \hat J_-,\\
\hat{\vec J}=\sum_\p \hat{\vec s}_\p,\quad  \hat{{\vec B}}_\bp=2g\hat{\vec{b}}+2\epsilon_\bp\hat{\bf z},\\
\end{split}
\en
where   $\hat{\vec{b}}=\hat b_x \hat{\bf x}+\hat b_y\hat{\bf y}$, $\hat b_x$ and $-\hat b_y$ are  Hermitian and anti-Hermitian parts of the operator $\hat{b}=\hat b_x-i\hat b_y$, and $\hat{\bf x}, \hat{\bf y}, \hat{\bf z}$ are coordinate unit vectors.

The second step in the mean field treatment of the two-channel model is to replace  Heisenberg operator $\hat b(t)$ in the first equation of motion in \re{Bloch1} with its time-dependent quantum mechanical average, $\hat b(t)\to \langle \hat b(t)\rangle\equiv b(t)$, which is expected to be exact in thermodynamic limit as long as the ${\bf q}=0$ state is macroscopically occupied at all times.   This replacement can be shown to be exact in  equilibrium using the exact solution  for the spectrum of  inhomogeneous Dicke  model \cite{Richardson1977,Gaudin_book}  and numerically for the time dependent problem 
\cite{Faribault2012}. Upon this replacement equations of motion become linear in operators and taking their quantum mechanical average, we obtain 
\beg\label{Bloch}
\begin{split}
\dot{\vec s}_\bp={\vec{B}}_\bp\times{\vec s}_\bp, \quad \dot{b}=-i\omega  b-ig J_-,\\
 {\vec J}=\sum_\p {\vec s}_\p,\quad  {{\vec B}}_\bp=2g{\vec{b}}+2\epsilon_\bp\hat{\bf z},\\
\end{split}
\en
where $\vec s_\p=\langle \hat{\vec s}_\p\rangle$. These are Hamiltonian equations of motion for a classical Hamiltonian
\beg\label{Model1}
	H_{\mathrm{2ch}}=\sum\limits_{\bp}2 \epsilon_\bp {s}_\bp^z +\omega{\bar b} {b}  	
	+
	g\sum\limits_{\bp} \left({\bar b} {s}_\bp^- +b{s}_\bp^+\right) ,
\en
which describes a set of angular momenta (classical spins or vectors) coupled to a  harmonic oscillator. 
Here, $\bar{b}$ denotes the complex conjugate of $b$.
These dynamical variables obey the following Poisson brackets:
\beg
\left\{ s_\p^a, s_\k^b\right\}=-\varepsilon_{abc}\delta_{\p\k} s_\p^c,\quad \{b,\bar b\}=i,
\label{poisson}
\en
where $a$, $b$, and $c$ stand for  spatial indecies $x$, $y$, and $z$.

Similar steps in the case of the single channel model \re{eq:onechannel} lead to a classical spin Hamiltonian
\beg\label{Model2}
 H_{\mathrm{1ch}}=\sum\limits_{\bp}2 \epsilon_\bp {s}_\bp^z  -\frac{\lambda}{\nu_F}\sum\limits_{\bp,\p'} {s}_\bp^-{s}_{\bp'}^+ 
\en
together with the corresponding equations of motion. 
 
An important characteristic of the system both in and out of equilibrium is the superfluid order parameter or the gap function defined in the two channel model as
\be 
\Delta(t) = -g  \EV{\hat b(t)}=-gb(t)\equiv \Delta_x(t)-i\Delta_y(t). 
\label{gap}
\ee 
In the one channel limit this expression turns into
\be 
\Delta_{\mathrm{1ch}}(t) = \frac{\lam}{\nu_F} \sum_\bp \EV{\hat{a}_{\bp\uparrow}(t)\hat{a}_{-\bp\downarrow}(t)}=\frac{\lam}{\nu_F}\sum_\bp s_\bp^-. 
\label{gap1ch}
\ee 
 The magnitude $|\Delta(t)|$ of the order parameter   is known as the Higgs or amplitude mode for its similarity with the Higgs boson\cite{Varma2014,Levitov2007} and its time-dependent phase represents a Goldstone mode.
Note however that out of equilibrium the gap function does not entirely determine the state of the system. It specifies the effective magnetic field acting on each spin according to \eref{Bloch}, but there is still a certain freedom in how the spin moves in this field. For example, even for a constant field the spin can precess around it making an arbitrary constant angle with its direction.

  In above models we took a free single particle spectrum, $\eps_\p=p^2/2m$ and labeled states with momenta $\p$. 
  This choice is not essential for our analysis. We can as well consider an arbitrary   spectrum $\eps_i$. The pairing is then between pairs
  of time reversed states\cite{Anderson1959,Altshuler2000}, see also the first two pages in Ref.~\onlinecite{Enolski2005} for more details. For example, in Hamiltonian \re{Model}  this results in relabeling  $\hat{\vec s}_\p\to \hat{\vec s}_i$, $\hat{a}_{\p\uparrow}\hat{a}_{-\p\downarrow}\to\hat{a}_{i\uparrow}\hat{a}_{i\downarrow}$, $\hat{a}_{\p\uparrow}^\dagger\hat{a}_{\p\uparrow}\to\hat{a}_{i\uparrow}^\dagger\hat{a}_{i\uparrow}$
  etc., where the state $|i\downarrow\rangle$ is the time reversed counterpart of $|i\uparrow\rangle$. Our results below depend only on the density of the single particle states $\nu(\eps)$ in the continuum limit regardless of whether these states are characterized by momenta $\p$ or any other set of quantum numbers $i$.

 \begin{figure}[h!]
\includegraphics[scale=0.26,angle=0]{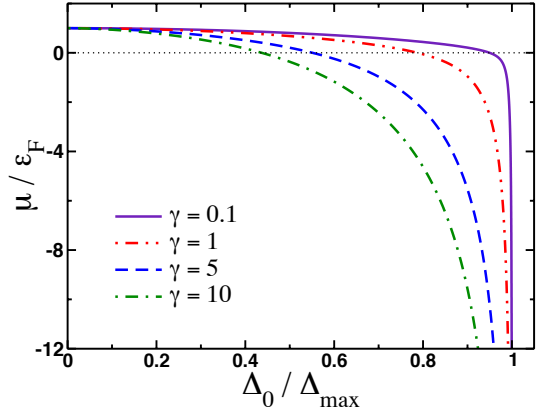}
\caption{(color online) Ground state chemical potential $\mu$ for two channel model   in 3d in units of the Fermi energy $\eps_F$ as a function of the ground state gap $\Delta_0$ for various resonance width
$\gamma$. $\mu(\Delta_0)$ is calculated from \esref{mucont} and \re{gapeqcont}. Note that in the two channel model   $\Delta_0$ is bounded from above by $\Delta_{\max}$.}
\label{mudelta}
\end{figure}

\subsection{Ground state}

In the ground state
\beg
\Delta(t)=\Delta_0e^{-2i\mu t}, 
\label{grdeltat}
\en
where the magnitude $\Delta_0$ is time-independent. 
Apart from an overall rotation about the $z$-axis with frequency $2\mu$, the ground state is a static solution of the equations of motion that minimizes $H_{2\mathrm{ch}}$.  
The minimum is achieved when each spin is directed against its effective magnetic field, i.e.
\beg
s_\p^-=\frac{\Delta_0 e^{-2i\mu t}}{2E(\eps_\p;\Delta_0, \mu)} ,\quad s_\p^z=-\frac{\eps_\p-\mu}{2E(\eps_\p;\Delta_0, \mu)},  
\label{gspin}
\en
where
\beg
E(\eps;\Delta, \mu)\equiv\sqrt{(\eps-\mu)^2+\Delta^2}.
\label{Ep}
\en
Note that the length of the spin $s_\p=1/2$. This is because the ground state is a tensor product of single spin-1/2 wave functions and $\vec s_\p=\langle \hat{\vec s}_\p\rangle$.

The equation of motion \re{Bloch} for $b$ yields
\beg
|J_-|=\frac{(\omega-2\mu)\Delta_0}{g^2},
\en
which implies a self-consistency equation for $\Delta_0$
\beg
\frac{(\omega-2\mu)}{g^2}=\sum_\p\frac{1}{2E(\eps_\p;\Delta_0, \mu)}.
\label{gapeq}
\en

Further, the Hamiltonian \re{Model1} conserves 
\beg
n=\overline{b}b+\sum\limits_{\bp}\left(s_\bp^z+\frac{1}{2}\right),
\label{muargument}
\en
which is the average total number of bosons and fermion pairs. This number is related to $\Delta_0$ and the chemical potential $\mu$ as
\beg
2n=\frac{2\Delta_0^2}{g^2}+\sum_\p\left(1-\frac{\eps_\p-\mu}{E(\eps_\p;\Delta_0, \mu)}\right).
\label{mu}
\en

The Fermi energy $\eps_F$ is the chemical potential of the fermionic atoms at zero temperature in the absence of any interaction, when only fermions are present. It provides an overall energy scale and it is convenient to measure all energies in units of the Fermi energy. Thus, from now on, we set everywhere below
\beg
\eps_F=1.
\en

Below we   often  switch from discrete to continuum (thermodynamic limit) formulations. In the former version, there are $N$  discrete single-particle energy levels $\eps_\p$ with certain degeneracy 
each. Any quantity $A_\p$ we consider in this paper depends on $\p$ only through $\eps_\p$, $A_\p=A(\eps_\p)$. For example, all spins $\vec s_\p$ on a degenerate  level   $\eps_\p$ are parallel at all times and effectively merge into a single vector.  There are $N$ such vectors, so we count $N$ distinct classical spins. 

In thermodynamic limit,   energies $\eps_\p$  form a continuum on the positive real axis, i.e. are described by a continuous variable $\eps$ with a  density of states $\nu(\eps)$ that depends on the dimensionality of the problem
\beg
\nu(\eps)=\nu_F f(\eps),
\label{dos}
\en
where $\nu_F$ is the \textit{bulk} density of states (proportional to the system volume) at the Fermi energy, $f(\eps)=1$ in 2d, and $f(\eps)=\sqrt{\eps}$ in 3d. Summations over $\p$
turn into integrations,  
\beg
\sum_\p A_\p\to\nu_F\int_0^\infty A(\eps) f(\eps) d\eps.
\en

With only fermions present, the total particle number    is
\beg
2n=\int_0^{1} 2\nu(\eps)d\eps=\frac{4}{d}\nu_F,
\label{redistribute}
\en
where $d=2,3$ is the number of spatial dimensions. Interaction  redistributes this number between fermions and bosons as in \eref{mu}. Combining \esref{mu} and \re{redistribute} and taking the continuum limit, we obtain
\beg
\frac{4}{ d}=\frac{2\Delta_0^2}{\gamma }+ \int_0^{\infty} \biggl(1-\frac{\eps-\mu}{ \sqrt{(\eps-\mu)^2+\Delta_0^2}}\biggr) f(\eps)d\eps,
\label{mucont}
\en
where $\gamma$ is the dimensionless resonance width defined in \eref{width}.  

Similarly, \eref{gapeq} becomes in the thermodynamic limit
\beg
\frac{2\omega-4\mu}{\gamma}= \int_0^{\eps_\Lambda}\frac{f(\eps)d\eps}{ \sqrt{(\eps-\mu)^2+\Delta_0^2}},
\label{gapeqcont}
\en
where $\eps_\Lambda$ is the high energy cutoff. In 3d it can be eliminated by an additive renormalization of the detuning $\omega$, see e.g. Ref.~\onlinecite{Gurarie2007}. This however does not affect our results for the quench dynamics as they depend on the difference between the initial and final values of the detuning. 

  \esref{mucont} and \re{gapeqcont}  contain two independent parameters not counting the cutoff. For example, we can choose $\gamma$ and $\omega$ and determine $\mu$ and $\Delta_0$ from these equations, or choose $\gamma$ and $\Delta_0$ and determine $\mu$ and $\omega$ etc., see Fig.~\ref{mudelta} for a plot of $\mu(\Delta_0)$ for various $\gamma$ in 3d.  Note also that $\Delta_0^2=g^2 \bar b b$ is proportional to the number of bosons and is therefore limited by the total number of particles.    \eref{mucont}   implies
\beg
 \Delta_0 \le \sqrt{\frac{2\gamma}{d}}=\Delta_{\max}.
 \label{delta0max}
\en

\subsection{Quench setup and initial conditions}
\label{setup}

  In a quantum quench setup we prepare the system in a ground state at a certain detuning $\omega_i$, i.e. the initial state  is 
\beg
\begin{split}
s_\p^-(t=0)=\frac{\Delta_{0i} }{2E(\eps_\p;\Delta_{0i}, \mu_i)} ,\\
s_\p^z(t=0)=-\frac{\eps_\p-\mu_i}{2E(\eps_\p;\Delta_{0i}, \mu_i)},  
\end{split}
\label{ini}
\en
where $\Delta_{0i}, \mu_i$ are the ground state values determined by  \esref{mucont} and \re{gapeqcont} with $\omega=\omega_i$. We then quench the detuning $\omega_i\to\omega_f$ and evolve the system with the two-channel Hamiltonian \re{Model1} starting from the initial state \re{ini} at $t=0$.

The state of the system is fully determined by the many-body wavefunction, which in the mean field treatment is at all times a product state of the form
\beg
|\Psi(t)\rangle=|\psi(t)\rangle\otimes\bigl(\hat{b}^\dagger\bigr)^{n(t)}|0\rangle, 
\label{fullwav}
\en
where $n(t)=|b(t)|^2$ and $|\psi(t)\rangle$ is the fermionic part of the wave function:
\beg
|\psi(t)\rangle=\prod\limits_{\bp}\left[\overline{u}_\bp(t)+\overline{v}_\bp(t)\hat{a}_{\bp\uparrow}^\dagger\hat{a}_{-\bp\downarrow}^\dagger\right]|0\rangle.
\label{BCSWaveFunction}
\en
Bogoliubov amplitudes $u_\bp(t),v_\bp(t)$ obey the Bogoliubov de Gennes (BdG) equations
\beg\label{BdG}
i\frac{\partial}{\partial t}\left(\begin{matrix} u_\bp(t) \\ v_\bp(t) \end{matrix}\right)=
\left(\begin{matrix} \epsilon_\bp & \Delta(t) \\ \bar\Delta(t) & -\epsilon_\bp \end{matrix}\right)
\left(\begin{matrix} u_\bp(t) \\ v_\bp(t) \end{matrix}\right),
\en
with the normalization condition $|u_\p|^2+|v_\p|^2=1$. Apart from an overall time-dependent phase (which is important for certain observables), these equations are equivalent to the classical spin equations of motion \re{Bloch} and spins are related to the amplitudes as
\beg\label{lupvp}
\frac{s_\bp^-}{s_\bp}=2u_\bp\overline{v}_\bp, \quad \frac{s_\bp^z}{s_\bp}=|v_\bp|^2-|u_\bp|^2,
\en
where $s_\p$ is the length of the spin. For quench initial conditions $s_\p=1/2$, as explained below \eref{Ep}.

 Each quench is uniquely characterized by three parameters -- the resonance width $\gamma=g^2\nu_F$, the initial $\omega_i$ and final $\omega_f$ values of the detuning in units of the Fermi energy. Indeed, $\omega_i$ and $\gamma$ determine  $\Delta_{0i}$ and  $\mu_i$ and thus the initial condition, while the equations of motion \re{Bloch1} in the thermodynamic limit depend only on $\omega_f$ and $\gamma$. To see the latter, note that model parameters enter the equation of motion for spin $\vec s_\p\equiv \vec s(\eps_\p)$ only through $\Delta=-gb$, while the equation of motion for the bosonic field $b$ can be equivalently written as 
 \beg
 \dot\Delta=-i\omega_f\Delta +i\gamma\int_0^\infty s^-(\eps)f(\eps)d\eps.
 \label{123}
 \en

Instead of $\omega_i, \omega_f$ we find it more convenient to characterize the quench by $\Delta_{0i}, \Delta_{0f}$ -- the ground state gaps corresponding to these values of the detuning. As discussed below \eref{gapeqcont}, for a given $\gamma$, the detuning $\omega$ uniquely determines $\Delta_0$ and vice versa. Note that $\Delta_{0f}$ has nothing to do with the time-dependent gap function $\Delta(t)$ and in particular with the large time asymptote $\Delta(t\to\infty)$. Whenever $\Delta(t)$ goes to a constant at large times, we denote this constant  $\Delta_\infty$.

\subsection{Main results}
\label{mainres}

Our main result is a complete description of the long time dynamics of two and one   channel models \re{Model1} and \re{Model2} in two and three spatial dimensions following a  quench of the detuning $\omega_i\to\omega_f$ (coupling $\lam_i\to\lam_f$ in the one channel model) in the thermodynamic limit. A key effect that makes such a description possible is a drastic reduction in the number of effective degrees of freedom as $t\to\infty$. It turns out  that the large time dynamics can be expressed in terms of
just a few new collective spins plus the oscillator in the two channel case that are governed by the same Hamiltonians  \re{Model1} and \re{Model2} only with new effective parameters
replacing $\eps_\p$ and $\omega$. The number of collective spins is $m=0, 1$ or 2  and $m=-1, 0$ or 2  for one and two channel models, respectively, depending on the quench. The difference is due to the presence of the oscillator degree of freedom in the latter case. For example, $m=-1$ means that the effective large time Hamiltonian $H_\mathrm{red}$ not only has no spins, but also the oscillator $b$ is absent, i.e. $H_\mathrm{red}=0$. This reduction effect combined with integrability of classical Hamiltonians  \re{Model1} and \re{Model2}
allows us to  determine the state of the system (its many-body wave function) at $t\to\infty$. We explain this method in detail in Sect.~\ref{method}. This subsection provides a summary of  main results obtained with the help of this method.

\begin{figure}[h!]
\includegraphics[scale=0.26,angle=0]{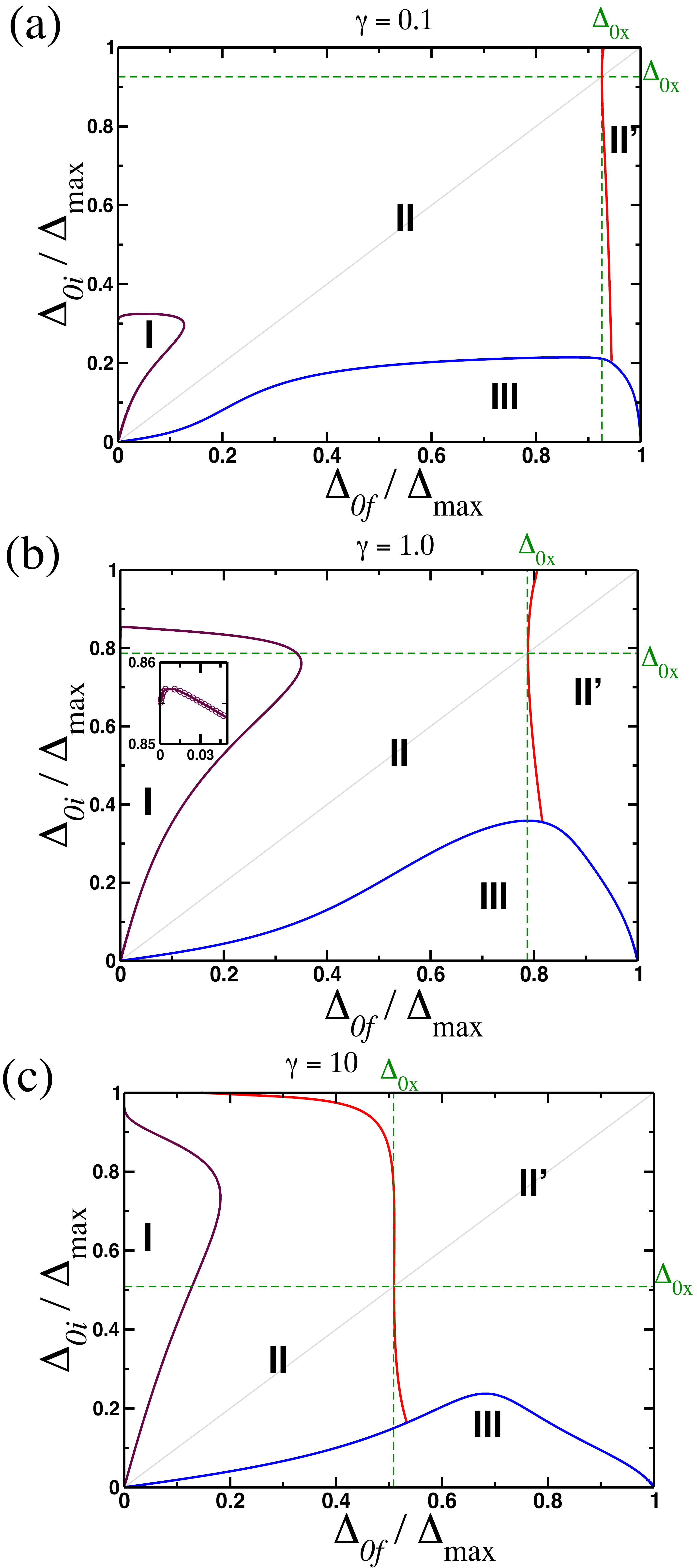}
\caption{(color online) Detuning  quench phase diagrams for two-channel model \re{eq:twochannel} in  2d for an assortment of resonance widths $\gamma$. Each point represents a single quench labeled by $\Delta_{0i}$ (vertical axis) and $\Delta_{0f}$ (horizonal axis) -- pairing gaps the system would have in the ground state for initial and final detunings. At large times the system ends up in one of three steady states shown as regions I, II (including II'), and III.   For quenches in region I the order parameter vanishes, $\Delta(t)\to0$. In II
  $\Delta(t)\to\Delta_\infty e^{-2i\mu_\infty t-2i\varphi}$ and in III $|\Delta(t)|$ oscillates persistently. Subregions II and II' differ in the sign of  $\mu_\infty$ (out of equilibrium analog of the chemical potential): $\mu_\infty>0$ in II
 and $\mu_\infty<0$ in II'. The diagonal, $\Delta_{0i}=\Delta_{0f}$, is the no quench line. To the left of it are strong to weak coupling quenches, to the right  -- weak to strong. $\Delta_{\max}=\eps_F\sqrt{\gamma}$ in 2d is the maximum possible ground state gap and $\Delta_{0\times}$ is the ground state gap corresponding to zero chemical potential, i.e. $\Delta_{0\times}$ is given by \eref{mucont} for $\mu=0$.}
\label{psd2ch2d}
\end{figure}

\begin{figure}[h]
\includegraphics[scale=0.26,angle=0]{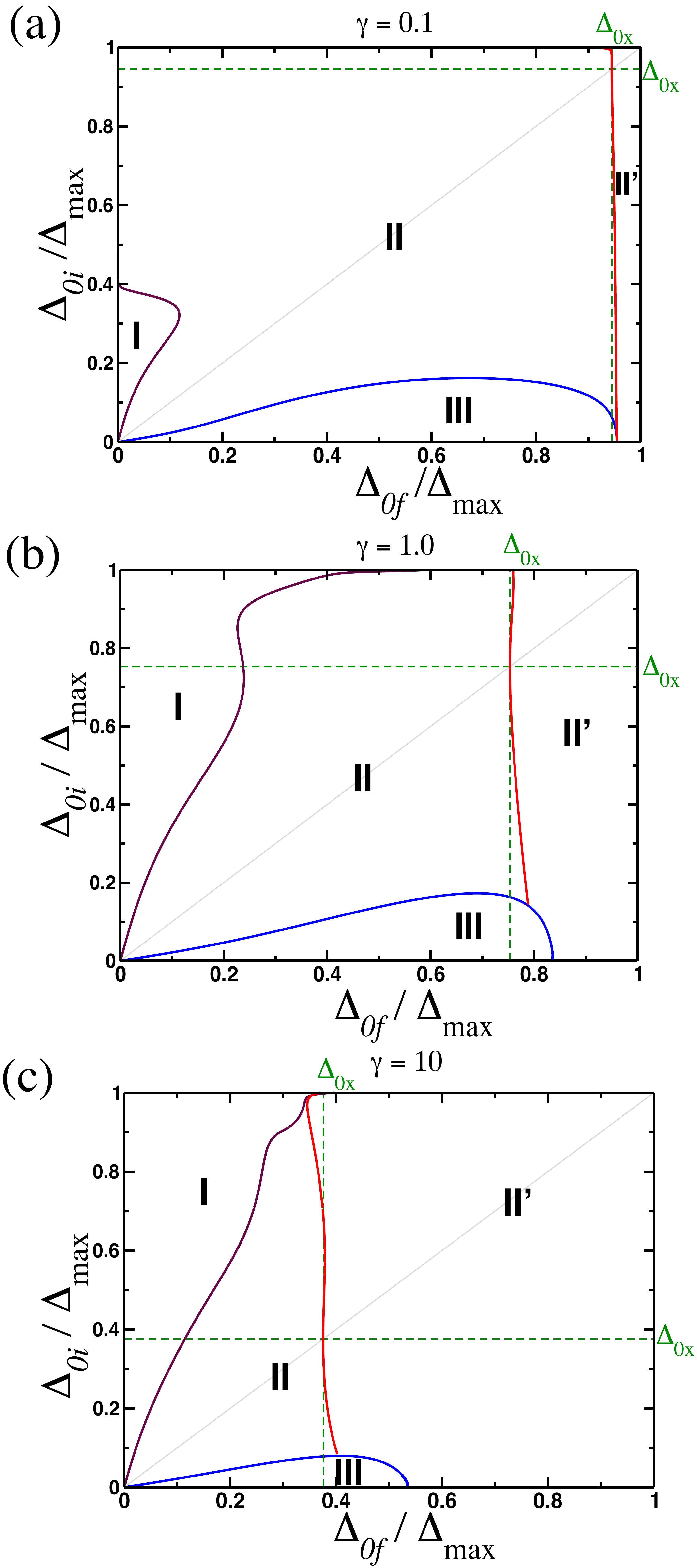}
\caption{(color online)  Same as Fig.~\ref{psd2ch2d} but in three spatial dimensions.}
\label{psd2ch3d}
\end{figure}

\begin{figure}[h]
\includegraphics[scale=0.26,angle=0]{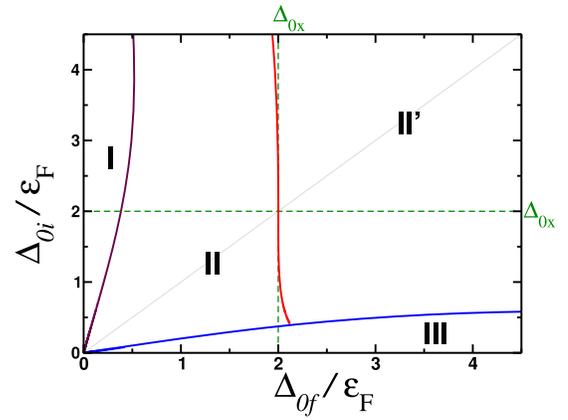}
\caption{(color online) Interaction ($\lambda$) quench phase diagram for one-channel model \re{Model2} in two dimensions. Otherwise same as Fig.~\ref{psd2ch2d}}.
\label{psd1ch2d}
\end{figure}

\begin{figure}[h]
\includegraphics[scale=0.26,angle=0]{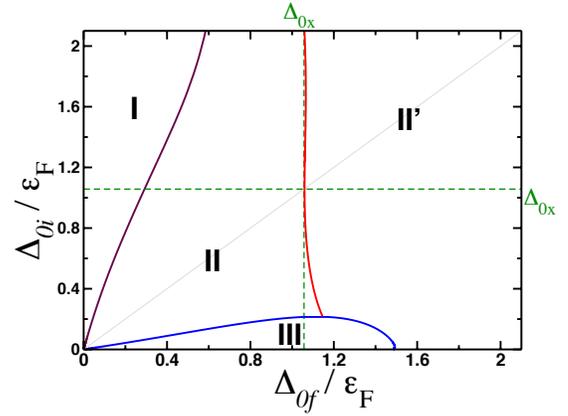}
\caption{(color online) Interaction   quench phase diagram for one-channel model \re{Model2} in 3d. Otherwise same as Fig.~\ref{psd2ch2d}.  Consider e.g. quenches from fixed infinitesimal coupling $\lam_i$ (small $\Delta_{0i}$) to various final couplings $\lam_f$.   Increasing $\lam_f$ ($\Delta_{0f}$)  we move through gapless (I), gapped (II), then oscillating (III) steady states. As $\lam_f$ increases further oscillations disappear and we again end up in a steady state  characterized by constant asymptotic $|\Delta(t)|$ (II').  }
\label{psd1ch3d}
\end{figure}

In Sects.~\ref{qpsd} and \ref{1channel}, we construct exact quench phase diagrams shown in Figs.~\ref{psd2ch2d} - \ref{psd1ch3d}. 
Depending on the values of  $\omega_i$ and $\omega_f$ either system reaches one of  three distinct   steady states labeled by I, II (including subregion II') and III that can be thought about as nonequilibrium phases with second order phase transition lines between them ($t\to\infty$ limit of the order parameter $\Delta(t)$ is continuous along lines separating different regions). These steady states correspond to $m=0, 1$ or 2 collective spins, respectively, for one channel model and to $m=-1, 0$ or 2 in the case of two channels.

Each point in the quench phase diagrams represents a particular quench  specified by a pair of  values $(\Delta_{0i}, \Delta_{0f})$. Here  $\Delta_0$  is the   gap that the system would have in  the ground state at detuning $\omega$, which is a known function of   $\omega$.    Values   $\Delta_{0i}$ and $\Delta_{0f}$ --   ground state  gaps for $\omega=\omega_i$ and $\omega_ f$, respectively -- uniquely determine $\omega_i$ and $\omega_f$ at fixed resonance width $\gamma$.   Note that $\Delta_{0f}$ is \textit{not} the magnitude of the actual steady state gap function $|\Delta(t)|$.
Each quench $\omega_i\to\omega_f$ (or $\lam_i\to\lam_f$) therefore maps to a single point $(\Delta_{0i}, \Delta_{0f})$ and vise versa. 

\begin{figure}[h]
\includegraphics[scale=0.26,angle=0]{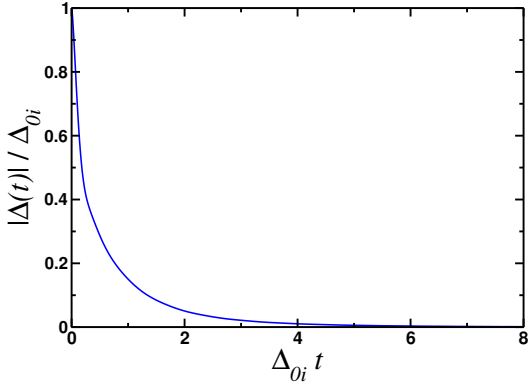}
\caption{(color online) $|\Delta(t)|$  in region I for 3d two-channel model, $\gamma=1$, obtained from numerical  evolution of   $N=5024$ spins following a detuning quench $\omega_i\to\omega_f$. Here
$\Delta_{0i}=0.27\Delta_{\max}$,  $\Delta_{0f}=4.30\times 10^{-2}\Delta_{\max}$ [cf. Fig.~\ref{psd2ch3d}(b)]. From these two values all other parameters obtain, e.g. $\mu_i=0.90\eps_F$, and $\omega_f-\omega_i=1.97\eps_F$. }
\label{regIdeltaplot}
\end{figure}

Steady states I, II, and III reached by the system at $t\to\infty$ can be described in terms of the superfluid order parameter $\Delta(t)$. In region I of phase diagrams 
in Figs.~\ref{psd2ch2d} - \ref{psd1ch3d} the gap function vanishes at large times, $\Delta(t)\to 0$, see  Fig.~\ref{regIdeltaplot}. 

\begin{figure}[h]
\includegraphics[scale=0.26,angle=0]{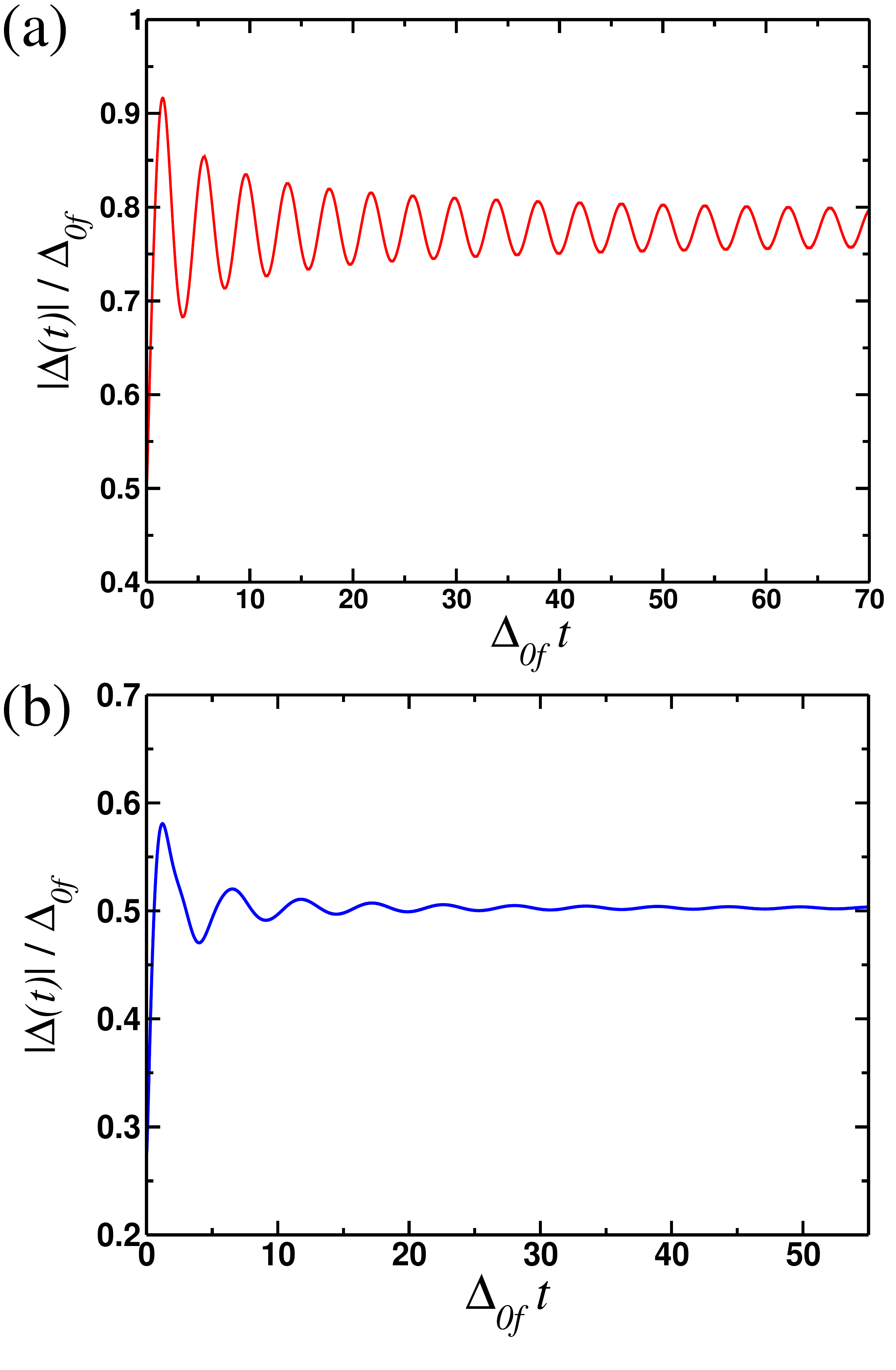}
\caption{(color online)  $|\Delta(t)|$  in regions II (top) and II' (bottom) for 3d two-channel model, $\gamma=1$, obtained from numerical  evolution of $N=5024$ spins after quenching the detuning $\omega$.   $\Delta_{0i}=0.27\Delta_{\max}, \mu_i=0.90\eps_F$ in both panels, same as in Fig.~\ref{regIdeltaplot}. The final detuning corresponds to: (a)
$\Delta_{0f}=0.56\Delta_{\max}=2.07\Delta_{0i}$ and (b) $\Delta_{0f}=0.97\Delta_{\max}=3.59\Delta_{0i}$. See also Fig.~\ref{psd2ch3d}(b).}
\label{regIIdeltaplot}
\end{figure}

In region II (including subregion II') the magnitude of the order parameter asymptotes to a nonzero constant $\Delta_\infty$ as illustrated in Fig.~\ref{regIIdeltaplot},
\be 
\label{eq:reg2} 
\Delta(t) \rightarrow \Delta_{\infty} e^{-2 i \mu_{\infty} t-2i\varphi},
\ee
where $\Delta_\infty, \mu_\infty$ are  functions of $\omega_i, \omega_f$ (or, equivalently, of $\Delta_{0i}$ and $\Delta_{0f}$), and $\gamma$ to be determined below, and
$\varphi$ is a contstant phase.  Plots of $\Delta_\infty$ and $\mu_\infty$  as  functions of $\Delta_{0f}$ for fixed $\Delta_{0i}$  
 are shown in Figs.~\ref{dinfplot}, \ref{delinfpl}, and \ref{muinfpl}.  The quantity $\mu_{\infty}$  plays the role of the out-of-equilibrium chemical potential. Subregions II and II' of region II correspond to $\mu_\infty>0$ and $\mu_\infty<0$, respectively.

\begin{figure}[h]
\includegraphics[scale=0.26,angle=0]{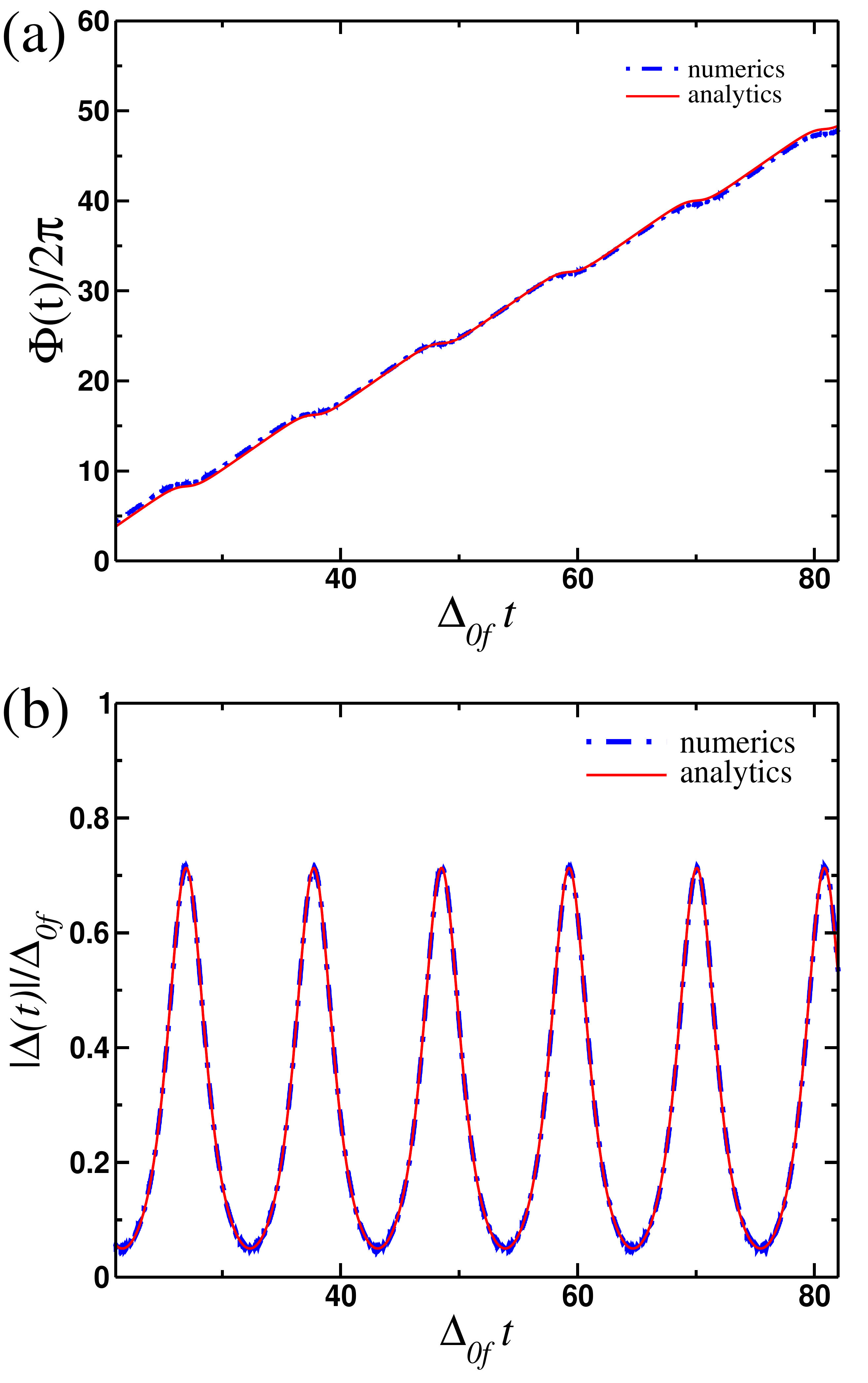}
\caption{(color online) Amplitude (Higgs mode)  and phase $\Phi(t)$ of the order parameter $\Delta(t)$ in region III of Fig.~\ref{psd2ch3d}(c) after detuning quench from deep BCS to BEC in 3d two-channel model for $\gamma=10$.  Numerical evolution with 5024 spins vs. \esref{dnintro} and \re{phiintro}.    $\Delta_{0i}=3.20\times 10^{-3}\Delta_{\max}, \Delta_{0f}=0.45\Delta_{\max}$, and $\delta\omega=-5.86\gamma$.  }
\label{regIIIdeltaplot}
\end{figure}

\begin{figure}[h]
\includegraphics[scale=0.26,angle=0]{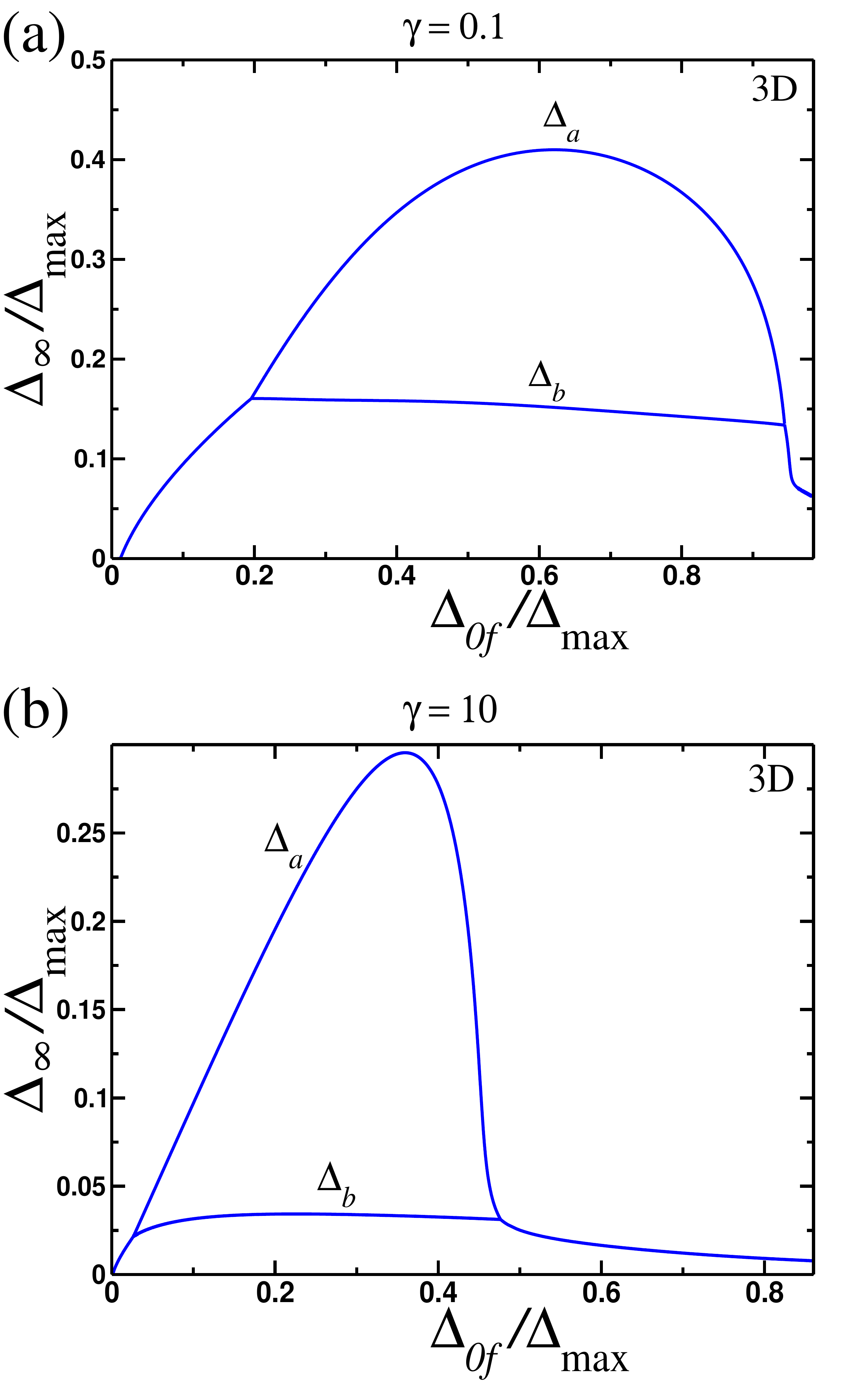}
\caption{(color online) Limiting values of $|\Delta(t)|$ for 3d two-channel model at large times  after a detuning  quench as functions of $\Delta_{0f}$ (or, equivalently, of final detuning $\omega_f$) at fixed small
 $\Delta_{0i}=0.05\Delta_{\max}$ (fixed initial detuning deep in the BCS regime). This corresponds to moving along a horizontal   line (not shown)    in Fig.~\ref{psd2ch3d}(a) and (c) going through regions I where $|\Delta(t)|\to0$, II where $|\Delta(t)|\to\Delta_\infty>0$, III where $|\Delta(t)|$ oscillates periodically between $\Delta_a$ and $\Delta_b$, and into region II' where again  $|\Delta(t)|\to\Delta_\infty>0$. Note that persistent oscillations appear and then disappear again as we decrease $\omega_f-\omega_i$ (i.e. increase $\Delta_{0f}$ at fixed $\Delta_{0i}$). The same behavior is observed in the 3d one-channel model, see Fig.~\ref{psd1ch3d}.  }
\label{dinfplot}
\end{figure}

\begin{figure}[h]
\includegraphics[scale=0.26,angle=0]{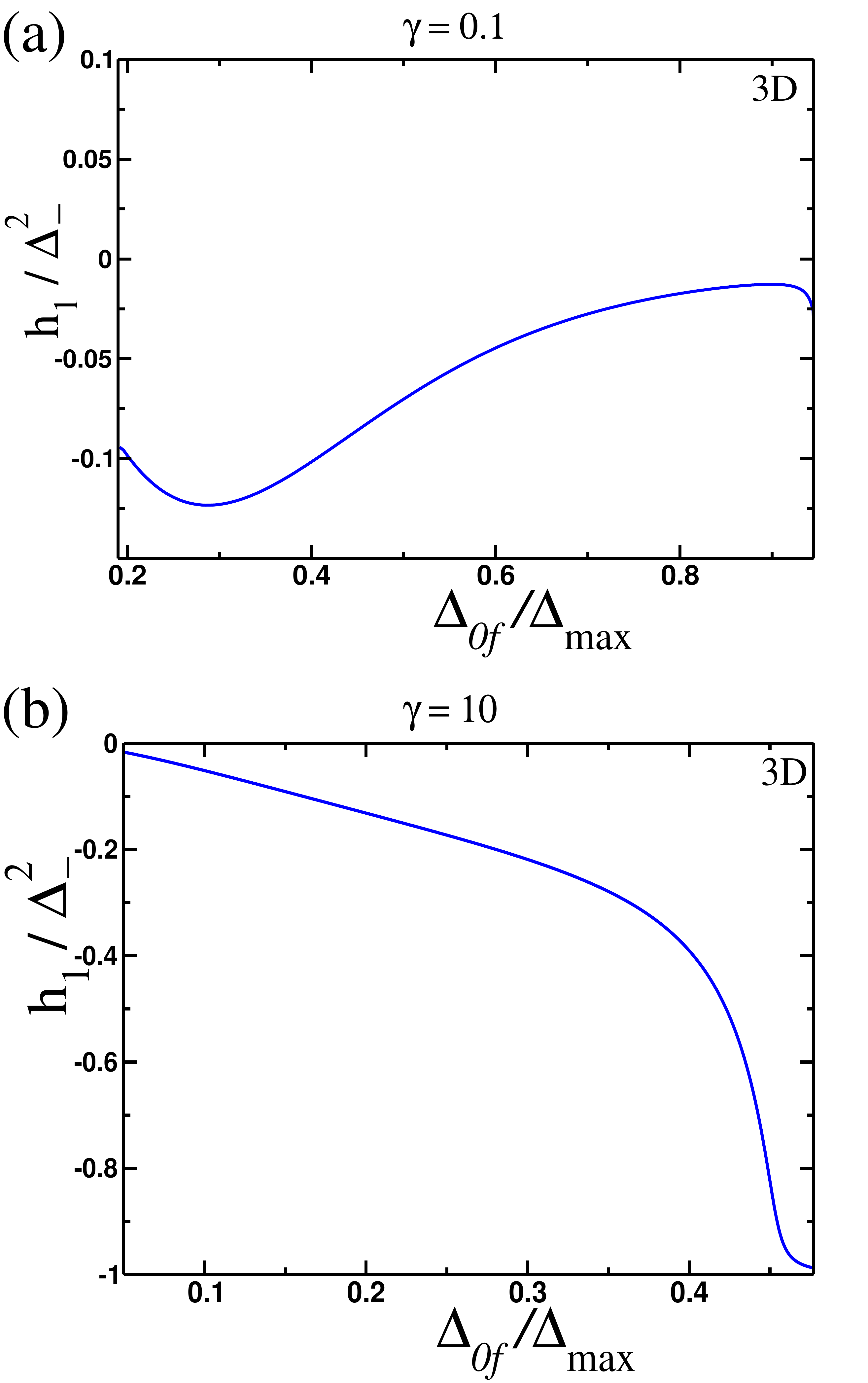}
\caption{(color online) Parameter $h_1$ in  \eref{dnintro} for asymptotic $|\Delta(t)|$ in phase III as a function of $\Delta_{0f}$ at fixed small
 $\Delta_{0i}=0.05\Delta_{\max}$, same as in Fig.~\ref{dinfplot}.  For quenches  within weak coupling limit  $h_1=0$, so nonzero $h_1$ quantifies deviations from this limit.  Note that one must have $h_1\ge -\Delta_-^2$, so that the expression under the square root in  \eref{dnintro} is nonnegative.}
\label{h1111plot}
\end{figure}

In region III of quench phase diagrams the amplitude of the order parameter oscillates persistently at large times as shown in Fig.~\ref{regIIIdeltaplot},
\beg
\Delta(t)\to \sqrt{\Lambda^2(t)+h_1}e^{-i\Phi(t)},
\label{dnintro}
\en
where
\beg
\Lambda(t)=\Delta_+ \mbox{dn}\left[ \Delta_+(t-t_0), k'\right],\quad k'=\frac{\Delta_-}{\Delta_+},
\en
  dn is the  Jacobi elliptic function and $t_0$ is an   integration constant. The magnitude of the order parameter oscillates periodically between $\Delta_b=(\Delta_-^2+h_1)^{1/2}$ 
    and $\Delta_a=(\Delta_+^2+h_1)^{1/2}$. The phase  contains  linear and periodic parts\cite{notelin}
 \beg
 \Phi(t)=2\mu t-\int \frac{\kappa dt}{\Lambda^2(t)+h_1}. 
 \label{phiintro}
 \en
Constants $h_1, \Delta_+, \Delta_-, \mu,$ and $\kappa$ are known functions of $\Delta_{0i}, \Delta_{0f}$ (or $\omega_i,\omega_f$), and $\gamma$ to be specified below, see also Figs.~\ref{dinfplot} and \ref{h1111plot} and refer to Sect.~\ref{asym1} for more information about the periodic solution.

 \begin{figure}[h]
\includegraphics[scale=0.26,angle=0]{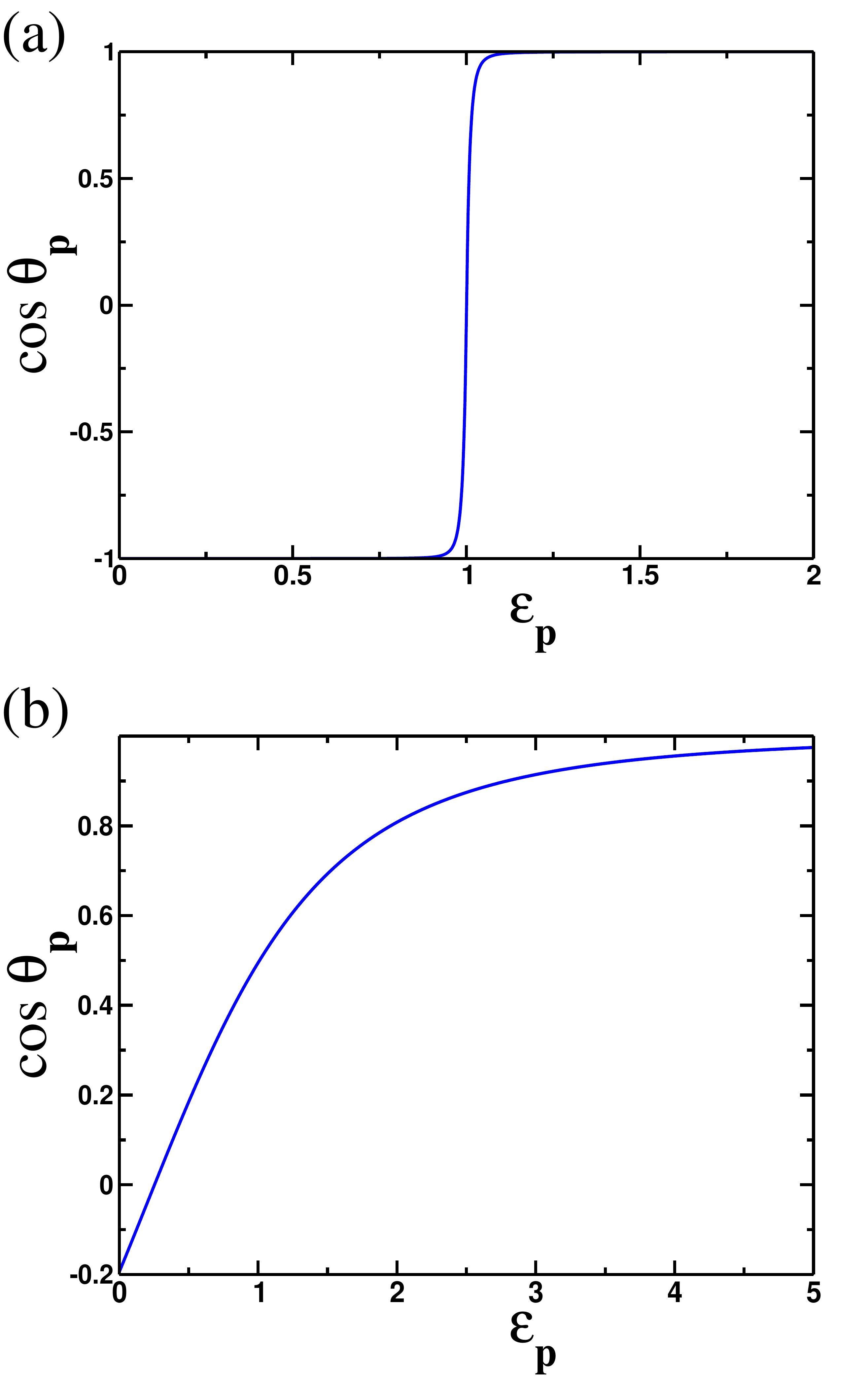}
\caption{(color online)    Spin distribution 
 $\cos\theta_\p$ as a function of $\eps_\p$ (in units of Fermi energy) at large times after the quench in 3d two channel model. In phases I and II,   $-\cos\theta_\p/2$ is the projection of the spin $\vec s_\p$  onto its effective magnetic field ($z$-axis in phase I) around which it precesses. In equilibrium $\cos\theta_\p=\pm 1$ (1 in the ground state) for all momenta and in phase I $\cos\theta_\p=-1$ and 1 correspond to doubly occupied and unoccupied states, respectively. Quench parameters are $\gamma=1$ and: (a) $\Delta_{0i} = 0.05\Delta_{\max}, \Delta_{0f} = 0.002\Delta_{\max}$ (BCS to  deep BCS quench in phase I); (b) $\Delta_{0i} = 0.78\Delta_{\max}, \Delta_{0f} = 0.001\Delta_{\max}$ (BEC to  deep BCS quench in phase I). In both cases $\mu_\infty\approx \eps_F$. Note the Fermi-like shape of the distribution function in (a). Note that $\cos\theta_\p\to 1$ as $\eps_\p\to\infty$, as it should, indicating that states at very high energies are empty. }
\label{dist_plot}
\end{figure}

Previous studies of the BCS dynamics\cite{Kogan1973,Spivak2004,Yuzbashyan2006,Barankov2006,Dzero2006,Coleman2007}
were performed in the weak coupling regime when both $\Delta_{0i}$ and $\Delta_{0f}$ are 
much smaller than a characteristic high energy scale (Fermi energy for cold gases and Debye energy for conventional superconductors). This limit corresponds to an infinitesimal vicinity of the origin $\Delta_{0i}=\Delta_{0f}=0$ in our quench phase diagrams in Figs.~\ref{psd2ch2d} - \ref{psd1ch3d} . The weak coupling limit is universal in that it is independent of the resonance width and dimensionality and thus is the same in all diagrams. Critical lines separating regions I and II, and II and III are straight lines in this case coming out of the origin with slopes
\beg
\frac{\Delta_{0i}}{\Delta_{0f}}=e^{\pm \pi/2}.
\en
Further, $h_1=0$ in \eref{dnintro} and $\Delta_\infty$, $\Delta_\pm$ take a simpler form given by \esref{coth1} -- \re{qwrty}, and \re{III}.

There are several qualitatively new effects beyond the weak coupling regime.  At smaller resonance width $\gamma<\gamma_c=16/\pi^2$, gapless region I   terminates below $\Delta_{\mathrm{max}}$ at $\Delta_{0i}=\gamma\pi/4$ along the vertical axis in 2d.  This means that as  initial coupling  gets stronger ($\Delta_{0i}$ increases), even quenches to arbitrarily weak final coupling (small
$\Delta_{0f}$) do not result in vanishing $\Delta(t)$ at large times in contrast to the weak coupling regime where quenches with sufficiently large $\Delta_{0f}/\Delta_{0i}$   always do. The I-II critical line also displays an interesting backwards bending behavior for $\gamma<\gamma_c=16/\pi^2$, see the inset in Fig.~\ref{psd2ch2d}(b) and \esref{shapebelow} and \re{shapeabove}. 

Region III of persistent oscillations terminates at a   threshold value of $\Delta_{0f}<\Delta_{\max}$ in 3d, 
see Figs.~\ref{psd2ch3d} and \ref{psd1ch3d}. This means that even quenches from an infinitesimally weak initial coupling ($\lam_i=0+$ in  the one channel model, which corresponds to a vicinity of the normal state) to final couplings stronger then a certain threshold value produce no oscillations and $|\Delta(t)|$ instead goes to a constant. At finite but small initial gap $\Delta_{0i}$ (e.g. along the dashed line in Fig.~\ref{psd2ch3d}) there is a reentrant behavior in both 2d and 3d as the final coupling ($\Delta_{0f}$) increases when first there are no oscillations, then they appear, and then disappear again. The threshold value of $\Delta_{0f}$ where the
critical line separating regions II and III terminates is given by \eref{termpt3d} (plotted as a function of the resonance width in Fig.~\ref{232ch3D}) and \eref{termpt3d1ch} for one and two channel models, respectively. For more details about quench diagrams, such as the shape of the critical lines, various thresholds and termination points, values of  parameters (e. g. $\Delta_\infty, \mu_\infty, \Delta_+$, and $\Delta_-$)  characterizing asymptotic $\Delta(t)$,  see Sects.~\ref{qpsd} and \ref{1channel}. 

The large time asymptote of $\Delta(t)$ does not fully specify the steady state. One also needs to know the Bogoliubov amplitudes $u_\p(t\to\infty), v_\p(t\to\infty)$. We calculate them in Sect.~\ref{asym-1} in all three steady states. In terms of spin vectors, this translates into  steady state spin distribution. Even in regions I and II where $|\Delta(t)|$ goes to a constant, the steady state of the system is far from any equilibrium state. Time-independent $|\Delta(t)|$ means that in a frame that rotates around $z$-axis with frequency $2\mu_\infty$ the magnetic field $\vec B_\p$
that acts on spin $\vec s_\p$ in \eref{Bloch1} is constant. In equilibrium $\vec s_\p$ aligns  with $\vec B_\p$ or $-\vec B_\p$ (ground state). In steady states I and II it instead rotates around 
 $\vec B_\p$ making a constant angle with it. Let   $\theta_\p$  be the angle between $\vec s_\p$ and $-\vec B_\p$ (negative $z$-axis in steady state I), so that  in the ground state $\theta_\p=0$. Out of equilibrium $\theta_\p$ determines the steady state spin distribution function and is given by \eref{distAA}. This expression for 
 $\cos\theta(\eps_\p)$ applies in all three steady states, but its interpretation in region III is slightly different and will be explained below. A  plot of the distribution function
 $\cos\theta_\p$ is shown in Fig.~\ref{dist_plot}. We explore the asymptotic states produced by  detuning or interaction quenches in detail in Sect.~\ref{asym-1}. In Sect.~\ref{rf} we provide further insight into their physical nature   and discuss their experimental signatures.

 We perform detailed   analysis of linearized equations of motion   that goes much beyond previous work even in the weak coupling regime
 and yields a range of new results. Small quenches of the detuning correspond to a small neighborhood of the diagonal in quench diagrams in Figs.~\ref{psd2ch2d} - \ref{psd1ch3d},
 i.e. they fall within region II where $|\Delta(t)|\to\Delta_\infty$ and \eref{eq:reg2}  applies. We show that within linear approximation $\Delta_\infty=\Delta_{0f}$ and $\mu_\infty=\mu_f$, i.e. there are no  corrections to these equations linear in the change of detuning or, equivalently, in $\delta\Delta_0=\Delta_{0f}-\Delta_{0i}$. This is in fact a general result that
 has been overlooked by previous work -- to first order in deviations from the ground state $\Delta(t)$ always asymptotes to its ground state form for the Hamiltonian with which the system evolves at $t>0$. Note however that when quadratic correction is taken into account one gets $\Delta_\infty < \Delta_{0f}$. For example, in the weak coupling regime we find
 \beg
\Delta_\infty=\Delta_{0f}-\frac{(\delta\Delta_0)^2}{6\Delta_{0f}}.
\label{old1intro}
\en
 
We obtain an exact expression for $\Delta(t)$, \esref{deltalincont}, \re{x1}, and \re{x2}, valid at all times and arbitrary coupling strength for both one and two channel models. In weak coupling regime this expression simplifies so that
\beg
|\Delta(t)|=\Delta_{0f} -2\delta\Delta_0 \int\limits_0^{\infty}
\frac{dx}{\pi} \frac{\cos\left[ 2\Delta_0 t \cosh(\pi x/2)\right]}{1+x^2}, 
\label{dweak1intro}
\en 
 From here short and long times asymptotes follow. At short times
the order parameter amplitude rises or falls sharply as
\begin{equation}
|\Delta(t)|=\Delta_{0i}+\frac{\delta\Delta_0}{|\ln (\Delta_0 t) |}.
\label{gapshortintro}
\end{equation}
And the long time behavior in the weak coupling limit is
\beg
|\Delta(t)|= \Delta_{0f}-\frac{2\delta\Delta_{0}}{\pi^{3/2}}\frac{\cos(2\Delta_0 t+\pi/4)}{\sqrt{\Delta_0 t}}.
\label{vk11intro}
\en

At stronger coupling in region II (but not II') the long time asymptote is still given by \eref{vk11intro}, only the coefficient in front of the second term on the right hand side is more involved. 

Regions II and II' differ in the sign of the phase frequency  $\mu_\infty$,  
$\mu_\infty>0$ in II and $\mu_\infty<0$ in II'. We will see below that frequency (Fourier) spectrum of quench dynamics in regions II and II' is $E_\infty(\eps_\p) = \sqrt{(\eps_\p-\mu_\infty)^2+\Delta_\infty^2}$, so that the Fourier transform of a dynamical quantity reads
$\int_0^\infty A(\eps) e^{-2i E_\infty(\eps) t} d\eps$. For $\mu_\infty>0$ the phase has a stationary point on
the integration path at $\eps=\mu_\infty$, while for $\mu_\infty <0$ it is absent. As a result,
  the long time behavior in three dimensions  in region II' changes
\beg
|\Delta(t)|=\Delta_{0f}\left( 1-c\frac{\delta\omega}{\gamma}\frac{\cos(2E_\mathrm{min}t+\pi/4)}{(2|\mu|t)^{3/2}}\right),
\label{ccoeffinintro}
\en
where $E_\mathrm{min}=\sqrt{\mu^2+\Delta_0^2}$, $c$ is of order one, and $\delta\omega=\omega_f-\omega_i$. The same expression holds for the one channel model after a replacement $\delta\omega/\gamma\to 1/\lam_f-1/\lam_i$. Oscillation frequency $E_\mathrm{min}$ and $1/t^{3/2}$ decay are in agreement with Ref.~\cite{Gurarie2009} and reflect the fact that in the absence of a stationary point, the long time asymptote is dominated by the end point of integration at $\eps=0$, $E(0)=E_\mathrm{min}$, and the density of states
in 3d vanishes as $\sqrt{\eps}$ at small $\eps$. 

In two dimensions  linear analysis yields a different approach to the asymptote in region II'
\beg
|\Delta(t)|=\Delta_{0f}\left(1-\frac{\delta\omega}{\gamma}\frac{\sin (2E_\mathrm{min} t)}{|\mu| t\ln^2 t}\right),
\label{2dexintro}
\en
because of a constant density of states and $\ln\eps$ divergence of the Fourier amplitude of $|\Delta(t)|$ at small $\eps$ (see below). We also determine the time-dependent phase of the order parameter $\Phi(t)$ in all cases corresponding to \esref{dweak1intro} -- 
\re{2dexintro}, asymptotes of individual spins $\vec s_\p(t)$
as $t\to\infty$,  and many other  new results for the linearized dynamics in Sect.~\ref{linear}.

Finally, we extend  some of the above results for the long time behavior of $|\Delta(t)|$ to the nonlinear 
regime, though unlike the linear analysis these results are not rigorous. In region II
\beg
|\Delta(t)|=\Delta_{\infty}+c'\frac{\cos(2\Delta_\infty t+\pi/4)}{\sqrt{\Delta_\infty t}},
\label{statptexnonlinintro}
\en
where $c'$ is a dimensionless coefficient. This answer holds for both one and two channel models in either dimension.

For region II' we argue that the  answer depends on dimensionality similarly to the linear analysis and
\beg
|\Delta(t)|=\Delta_{\infty}\left(1- c_1\frac{\sin (2E_\infty^\mathrm{min} t)}{ t\ln^2 t}\right)\quad\mbox{in 2d,}
\label{2dexnlinintro}
\en
\beg
|\Delta(t)|=\Delta_{\infty}\left( 1-c_2 \frac{\cos(2E_\infty^\mathrm{min}t+\pi/4)}{t^{3/2}}\right)
\quad\mbox{in 3d,}
\label{ccoeffinnlinintro}
\en
where $E_\infty^\mathrm{min}=\sqrt{\mu_\infty^2+\Delta_\infty^2}$.

The approach to the gapless steady state (region I)   is expected to be
\beg
|\Delta(t)|=\frac{c_4}{t\ln^r t}\quad\mbox{in 2d,}
\label{cccc2intro}
\en
where  $r=1$ or $r=2$, and
\beg
|\Delta(t)|= \frac{c_3}{t^{3/2}}\quad\mbox{in 3d.}
\label{ccoeffinnlinglessintro}
\en
We discuss these nonlinear large time asymptotes in more detail in Sect.~\ref{nonlinear}.

\section{Method}
\label{method}

 Here we describe a method that allows one to determine the  asymptotic state of the system at long times.
Both the quantum \re{Model} and classical \re{Model1} two-channel models are integrable meaning that there are as many nontrivial conservation laws as there are degrees of freedom. There is an exact Bethe Ansatz type solution for the quantum spectrum \cite{Gaudin_book}. In the classical case integrability implies a formal inexplicit solution of the equations of motion  in terms of certain multivariable special (hyperelliptic) functions\cite{Altshuler2005} that can be  helpful for understanding certain general features of the dynamics.  Evaluating specific dynamical quantities of interest for realistic initial conditions with this solution is however roughly equivalent to just  solving the equations of motion numerically. But the latter could be as well done directly without the formal exact solution. This is a typical situation in the standard theory of nonlinear integrable systems.

Fortunately, it was realized that at least for the BCS type models the large time dynamics dramatically simplifies in the thermodynamic limit, so that the number of evolving degrees of freedom effectively drops to just a few spins. Building on this insight, Yuzbashyan et. al. \cite{Yuzbashyan2006} were able to   develop a method that goes beyond the standard theory   and explicitly predicts the long time dynamics in  the thermodynamic limit. 

The main idea of this method is as follows. First, we construct a special class of \textit{reduced} solutions of the classical equations of motion for the two-channel model such that the dynamics reduces to that of just few effective spins. Then, we choose a suitable reduced solution and fix its parameters  so that  its integrals of motion match those for a given quench   in the thermodynamic limit.  Reduced solutions have only few additional arbitrary constants and cannot generally satisfy all of the quench initial conditions \re{ini}. There are $2N+2$ initial conditions (two angles per spin
plus two initial conditions for the oscillator mode $b$) and only $N+1$ correspond to the integrals of motion. 

 Next, exploiting the fact that for fixed $\Delta(t)$ BdG equations \re{BdG} are linear in the amplitudes $u_\p$ and $v_\p$, we  derive  the most general $t\to\infty$ asymptotic solution that has the same $\Delta(t)$ as the reduced one.  It has the same integrals
 as the quench dynamics by construction and, in addition, $N+1$ arbitrary independent constants to match the remaining initial conditions. We    \textit{conjecture} that so constructed asymptotic solution  is
 the true large time asymptote of the actual quench dynamics. To verify this \textit{few spin conjecture} it is sufficient to show that the large time asymptote of the actual $\Delta(t)$ matches that of the reduced (and therefore general asymptotic) solution. We do so numerically
 in the nonlinear case and  analytically for infinitesimal quenches when the dynamics can be linearized.

 We consider the two channel model in this and the following sections  and then obtain similar results for the one channel (BCS) model in Sect.~\ref{1channel}  by taking the broad resonance, $\gamma\to\infty$, limit.  

\subsection{Integrability and  Lax vector construction}

An object called  Lax vector plays a key role in our approach. It encodes all the information about the integrals of motion and turns out to be especially useful in analyzing the quench dynamics in the thermodynamic limit. The Lax vector is defined as
\beg\label{Lax}
\begin{split}
{\vec L}(u)= \sum\limits_\bp\frac{{\vec s}_\bp}{u-\eps_\bp}-\frac{(\omega-2\mu)}{g^2}\hat{\bf z}+\\
\frac{2}{g^2}\left[(u-\mu)\hat{\bf z}-\vec{\Delta}\right],
\end{split}
\en
where $u$ is an auxiliary  complex variable and $\vec\Delta\equiv \Delta_x\hat{\bf x}+\Delta_y\hat{\bf y}$. Poisson brackets of components of $\vec L(u)$ satisfy the following Gaudin algebra:
\beg
\left\{ L^a(u), L^b(v)\right\}=\varepsilon_{abc}\frac{L^c(u)-L^c(v)}{u-v}.
\label{fundament}
\en
This implies an important equality 
\beg
\left\{\vec L^2(u), \vec L^2(v)\right\}=0.
\label{l2com}
\en
Explicit evaluation of $\vec L^2(u)$ yields
\beg\label{zL2}
\begin{split}
\vec L^2(u)&=\frac{(2u-\omega)}{g^4}+\frac{4{\cal H}_b}{\omega g^2}+\\
&+\sum\limits_{\bp}\left(\frac{2{\cal H}_\bp}{g^2(u-\eps_\bp)}+\frac{s_\bp^2}{(u-\eps_\bp)^2}\right),
\end{split}
\en
where  
\beg\label{Hp}
\begin{split}
{\cal H}_\bp &=
g^2\sum\limits_{\bq\not=\bp}\frac{{\vec s}_\bp\cdot{\vec s}_\bq}{(\eps_\bp-\eps_\bq)}+(2\epsilon_\bp-\omega){s_\bp^z}+
g\left(\overline{b}s_\bp^-+bs_\bp^+\right), \\
{\cal H}_b &= \overline{b}b+\sum\limits_{\bp}s_\bp^z.
\end{split}
\en
It follows from \eref{l2com} that these spin Hamiltonians mutually Poisson commute, i.e.
\beg
\{ {\cal H}_\bp, {\cal H}_{\bp'} \}=\{ {\cal H}_\bp, {\cal H}_b\}=0.
\label{allcommute}
\en
Moreover, the two-channel Hamiltonian \re{Model1} is
\beg
H_{2\mathrm{ch}}=\omega {\cal H}_b+\sum_\p {\cal H}_\bp.
\en
This implies that ${\cal H}_\bp$ and ${\cal H}_b$ are conserved by $H_{2\mathrm{ch}}$ and establishes the integrability of the two-channel Hamiltonian. Note that $\vec L^2(u)$ is also conserved for any value of $u$ and serves as a generator of the integrals of motion for the two-channel model. The same construction works in the quantum case as well; one only needs to promote  classical dynamical variables to  corresponding quantum operators and replace Poisson brackets with commutators.

Equations of motion can be conveniently and compactly written in terms of the Lax vector as
\beg
\dot{\vec L}=\left(-2\vec\Delta+2u\hat{\bf z}\right)\times \vec L.
\label{Lmotion}
\en
Comparing the residues at the poles at both sides of this equation, we see that it is equivalent the equations of motion for spins \re{Bloch}.

The square of the Lax vector is of the form
\beg
\vec L^2(u)=\frac{Q_{2N+2}(u)}{g^4 \prod_{\eps_\p} (u-\eps_\p)^2},
\label{qdef}
\en
where $N$ is the total number of  distinct single particle energies $\eps_\p$, the product is similarly over nondegenerate values of $\eps_\p$, and $Q_{2N+2}(u)$ is a polynomial in $u$ of degree $2N+2$. The roots of this \textit{spectral} polynomial (or equivalently of $\vec L^2(u)$) play an important role in further analysis of the asymptotic behavior.
Note that since $\vec L^2(u)$ is conserved, so are its roots. They thus constitute a set of integrals of motion alternative to \eref{Hp}. Since $\vec L^2(u)\ge 0$ for real $u$,  its  roots   come in complex conjugate pairs.

\subsection{Reduced solutions}

Let us look for special solutions of  equations of motion \re{Lmotion} such that the Lax vector factorizes into time-dependent and independent parts
\beg\label{LaxReduced}
\begin{split}
{\vec L}^\mathrm{red}(u)= \sum\limits_\bp\frac{{\vec \sigma}_\bp}{u-\eps_\bp}-\frac{(\omega-2\mu)}{g^2}\hat{\bf z}+\\
\frac{2}{g^2}\left[(u-\mu)\hat{\bf z}-\vec{\Delta}\right]=
\left(1+\sum\limits_\p\frac{d_\p}{u-\epsilon_\p}\right){\vec L}_m(u),
\end{split}
\en
 where $\vec \sigma_\p$ (not to be confused with Pauli matrices) denote  spins in this solution that can have arbitrary length to distinguish them from spins $\vec s_\p$ for the quench dynamics that have length 1/2.  Further, $d_\p$ are time-independent constants to be determined later and ${\vec L}_m(u)$ is the Lax vector for an effective $m$-spin system
\beg\label{Lm}
\begin{split}
{\vec L}_m(u)&= \sum\limits_{j=0}^{m-1}\frac{{\vec t}_j}{u-\eta_j}-\frac{(\omega'-2\mu)}{g^2}\hat{\bf z}\\
&+\frac{2}{g^2}\left[(u-\mu)\hat{\bf z}-\vec{\Delta}\right].
\end{split}
\en
Here $\vec t_j$ are new collective spin variables placed at new arbitrary ``energy levels'' $\eta_j$. Note that the bosonic field
$\vec b$ and therefore $\vec \Delta$ are the same in the original and reduced models.

Substituting \eref{LaxReduced} into the equations of motion \re{Lmotion}, we see that $\vec L_m(u)$ satisfies the same equation of motion. This means that variables $\vec t_j$ obey Bloch equations \re{Bloch} with $\eps_\p\to \eta_j$ and $\omega\to\omega'$, and are therefore governed by the same Hamiltonian
\beg\label{Hm}
H_\mathrm{2ch}^\mathrm{red}=\sum\limits_{j=0}^{m-1}2\eta_j t_j^z+\omega'\overline{b}b+g\sum\limits_{j=0}^{m-1}(\overline{b} t_{j}^{-}+bt_{j}^{+}).
\en
We will need at most $m=1$ for analyzing the quench dynamics, so we will be able solve the equations of motions for $\vec t_j$ directly.

Matching the residues at $u=\eps_\p$ on both sides of \eref{LaxReduced}, we express original spins in terms of $\vec t_j$
\beg
 \vec\sigma_\p=d_\p{\vec L}_{m}(\eps_j),
\label{st}
\en
 Constants $d_\p$ are determined from the above equation using ${\vec \sigma\,}^2_\p= \sigma_\p^2$, where $|\sigma_\p|$ is the length of spin 
${\vec \sigma}_\p$. Note that $\sigma_\p$ can be of either sign (for future convenience). We have
\beg
d_\p=-\frac{ \sigma_\p}{\sqrt{ {\vec L}_{m}^2(\eps_\p)}}.
\label{d}
\en
It is important to note that  $\sigma_\p$ are arbitrary constants at this point.  We will determine them later so that the integrals of motion for the reduced solution match those for quench dynamics.

To satisfy \eref{LaxReduced}, we also need to match the residues at $u=\eta_k$ and the $u\to\infty$ asymptotic. This leads to the following
$m+1$ equations:
\beg
\begin{array}{l}
\displaystyle 1+\sum_\p\frac{d_\p}{\eta_k-\eps_\p}=0\quad k=0,\dots,m-1\\
\\
\displaystyle \omega=\omega'-2\sum_\p d_\p.\\
\end{array}
\label{const1}
\en
Equations \re{const1} constrain  the coefficients of the spectral
polynomial 
\beg
Q_{2m+2}(u)=g^4{\vec L}_{m}^2(u)\prod_{k=0}^{m-1}(u-\eta_k)^2,\quad m\ge 0
\label{redsp}
\en
 of the $m$-spin system. Indeed, using Eq.~\re{d},  we  can cast these constraints  into the following form:
\beg
\begin{split}
\sum_\p\frac{\sigma_\p\eps_\p^{r-1} }{\sqrt{ Q_{2m+2}(\eps_\p)}}=-\frac{\delta_{rm}}{g^2},\quad r=1,\dots,m\\
\omega'=\omega+2\sum_{k=0}^{m-1}\eta_k+\sum_\p \frac{2\sigma_\p g^2\eps_\p^m}{\sqrt{ Q_{2m+2}(\eps_\p)} }.
\end{split}
\label{const4}
\en
Here $m\ge 0$. These equations can be viewed as equations for determining the lengths of the collective spins $\vec t_j$.

We thus constructed a class of solutions such that the dynamics reduces to that of a smaller number of spins. These \textit{few spin solutions} however do not match the quench initial conditions, but, as we will see, the long time asymptote of $\Delta(t)$ after the quench coincides with $\Delta(t)$ of an appropriately chosen few spin solution. Specifically, $m=-1,0,$ and 1 are realized depending on the magnitude and the sign of the change in the detuning $\omega$. Let us therefore consider these particular cases.

\subsubsection{m=-1 spin solutions}

$m=-1$ refers to the case when there are no collective spins and  $b=0$, i.e. the oscillator (which can be viewed as an infinite length limit of a spin) is effectively absent as well. In other words, $H_\mathrm{red}=0$ and $\vec L_m(u)=\frac{2u-\omega'}{g^2}\hat {\bf z}$. \eref{st} then implies that all spins in the reduced solution are along the $z$-axis pointing in either positive or negative direction. It is convenient to redefine the sign of $\sigma_\p$ (only for $m=-1$) so that $\vec \sigma_\p=-\sigma_\p\hat{\bf z}$. We see directly from the equations of motion \re{Bloch} that this configuration together with $b=0$ is indeed a solution, a stationary one in the present case.

\subsubsection{m=0 spin solutions}

In this case the reduced problem consists of a free classical oscillator as there are no collective spins, i.e. $H_\mathrm{red}=\omega' \bar b b$. Equations of motion reduce to $\dot b=-i\omega' b$. Therefore 
\beg
\Delta(t)=-gb=c e^{-2i\mu t},
\label{0sdel}
\en
where $c$ is a complex constant and we defined $\mu=\omega'/2$.

Expressions for the original spins follow from the reduced Lax vector
\beg
\vec L_m(u)= -\frac{2}{g^2}\left(\vec \Delta-(u-\mu)\hat z\right),
\en
 \esref{st} and \re{d} imply
 \beg
\vec \sigma_\p= \frac{\sigma_\p }{E(\eps_\p;\Delta, \mu)}\left(\vec \Delta -(\eps_\p-\mu)\hat {\bf z}\right),
\label{0spin}
\en
where $E(\eps_\p;\Delta, \mu)=\sqrt{(\eps_\p-\mu)^2+|\Delta|^2}$. We see that  the ground state \re{gspin} is a one spin solution with $c=\Delta_0$ and $\sigma_\p=1/2$ (to minimize the energy). Excited states are also one spin solutions with different parameters. 
 
There is only one (last) constraint among \esref{const4} for $m=0$, which we recognize as a generalization of the gap equation \re{gapeq}.

\subsubsection{m=1 spin solutions}
\label{1spinsol}

This example is substantially more involved then the previous two. Now there is one collective spin $\vec t$ coupled to an oscillator,
\beg\label{Hm1spin}
H_\mathrm{red}= 2\eta t^z+\omega'\overline{b}b+g (\overline{b} t^{-}+bt^{+}),
\en
making the dynamics rather nontrivial. Our main goal presently is to derive a differential equation for $|\Delta(t)|=g|b(t)|$ and to relate its coefficients to the spectral polynomial $Q_4(u)$ of the reduced $m=1$ problem given in general by  \eref{redsp}.

$H_\mathrm{red}$ conserves $\bar b b+t^z$. It follows that $t^z$ can be expressed through $|b|^2$ as $t^z=c_1 \Omega^2+c_2$, where $c_{1,2}$ are constants and we introduced a notation
\beg
\Delta= \Omega e^{-i\Phi}.
\en
\eref{st} then implies that the $z$-component of the original spins in the reduced solution can be similarly expressed through $|\Delta|$ as
 \beg
 \sigma_\p^z=a_\p\Omega^2+b_\p.
 \label{sz}
 \en
 Note that constants $a_\p$ and $b_\p$ are inversely proportional to $\sqrt{\vec L_m^2(\eps_\p)}$ and therefore to $\sqrt{Q_4(\eps_\p)}$.
 It turns out that an efficient strategy to derive an equation for $\Omega$ and  relate its coefficients to those of $Q_4(u)$ is somewhat indirect. First, we use equations of motion for $\vec \sigma_\p$ together with \eref{sz} to obtain an equation for 
 $\Omega$ and  expressions for $a_\p$ and $b_\p$.  Identifying $\sqrt{Q_4(\eps_\p)}$ in the latter with the help of \eref{d}, we relate the coefficients.

Bloch equations \re{Bloch}  for spins in the reduced solution, $\vec s_p^\mathrm{\, red}\equiv \vec \sigma_\p$, can be written as
\beg
\dot \sigma_\p^z=-i(\sigma_\p^-\bar \Delta-\sigma_\p^+\Delta),\quad \dot \sigma_\p^-=-2i\sigma_\p^z\Delta-2i\eps_\p \sigma_\p^-.
\label{eom}
\en
Substituting $\sigma_\p^z$ from \eref{sz} into the first equation, we obtain
\beg
\sigma_\p^-e^{i\Phi}-\sigma_\p^+e^{-i\Phi}=2ia_\p\dot\Omega.
\label{minus}
\en
 Multiplying the second equation in \re{eom} by $e^{i\Phi}$ and adding the resulting equation to its complex conjugate, we get
\beg
\frac{d\phantom{t}}{dt}\left(\sigma_\p^-e^{i\Phi}+\sigma_\p^+e^{-i\Phi}\right)=4a_\p\eps_\p\dot\Omega-2a_\p\dot\Phi\dot\Omega,
\label{inter}
\en
where we also used \eref{minus}.  Integrating this and adding the result to \eref{minus}, we obtain
\beg
\sigma_\p^-e^{i\Phi}=2a_\p\eps_\p\Omega-a_\p A+ia_\p\dot\Omega,
\label{s-}
\en
where  $A=\int dt \dot\Phi\dot\Omega$.  
\eref{s-} implies
\beg
|\sigma_\p^-|^2=(2a_\p\eps_\p\Omega-a_\p A)^2+a_\p^2\dot\Omega^2.
\label{min}
\en
\esref{min} and \re{sz} combined with the conservation of the length of the spin, $(\sigma_\p^z)^2+|\sigma_\p^-|^2=\sigma_\p^2$,  yield a differential equation for $\Omega$
\beg
(a_\p\Omega^2+b_\p)^2+(2a_\p\eps_\p\Omega-a_\p A+c_\p)^2+a_\p^2\dot\Omega^2=\sigma_\p^2
\en
Dividing the last equation by $a_\p^2$ and rearranging, we obtain
\beg
\begin{split}
\dot\Omega^2+\Omega^4+\Omega^2\left(2\frac{b_\p}{a_\p}+4\eps_\p^2\right) -4\eps_\p A\Omega\\
 +A^2+\frac{b_\p^2-\sigma_\p^2}{a_\p^2}=0\\
 \end{split}
\label{main}
\en
It turns out that $A$ is a certain function of $\Omega$.  To see this, let $x_\p$ be a set of numbers such that $\sum_\p x_\p=0$, multiply \eref{main} by $x_\p$ and sum over $\p$. This yields 
\beg
A=2\mu \Omega+\frac{\kappa}{\Omega},
\label{A}
\en
where $\mu$ and $\kappa$ are arbitrary real constants.   Substituting \eref{A} into \eref{main}, we obtain
\beg
\begin{split}
\dot \Omega^2+\Omega^4+2\Omega^2\left[\frac{b_\p}{a_\p}+2\xi_\p^2\right]+\\ \frac{\kappa^2}{\Omega^2}+
\frac{b_\p^2-\sigma_\p^2}{a_\p^2}-4\kappa\xi_\p=0,\\
\end{split}
\label{omega}
\en
where $\xi_\p=\eps_\p-\mu$. Note that the same equation obtains in the reduced problem with $a_\p\to c_1$, $b_\p\to c_2$ etc. It follows that coefficients
must be $\p$-independent, i.e.
\beg
\frac{b_\p}{a_\p}+2\xi_\p^2=2\rho,\quad \frac{b_\p^2-\sigma_\p^2}{a_\p^2}-4\kappa\xi_\p=4\chi,
\en
where $\rho$ and $\chi$ are $\p$-independent constants. We find
\beg
\begin{split}
 b_\p =-2(\xi_\p^2-\rho)a_\p, \\ 
a_\p=\frac{-\sigma_\p }{2\sqrt{(\xi_\p^2-\rho)^2-\kappa\xi_\p-\chi}}.\\
\end{split}
\label{coef}
\en

As mentioned above $a_\p$ and $b_\p$ are inversely proportional to $\sqrt{Q_4(\eps_\p)}$. \eref{coef} therefore implies
 \beg
Q_4(u)=[(u-\mu)^2-\rho]^2-\kappa(u-\mu)-\chi,
\label{q4}
\en
while the differential \eref{omega} for $\Omega$ reads
\beg
\dot \Omega^2+\Omega^4+4\rho\Omega^2+\frac{\kappa^2}{\Omega^2}+4\chi=0.
\label{last}
\en
This equation can be solved in terms of elliptic function. Let $w=\Omega^2$. We have
\beg
 \dot w^2+4w^3+16\rho w^2+16\chi w+4\kappa^2\equiv \dot w^2+4P_3(w)=0.
 \label{om}
 \en
 Further, let  $P_3(w)=(w-h_1)(w-h_2)(w-h_3)$, where $h_3\ge h_2\ge h_1$, and define
 \beg
 \omega=\Lambda^2+h_1,\quad \Delta_+^2=h_3-h_1,\quad \Delta_-^2=h_2-h_1.
 \label{define}
 \en
 We get
\beg
\dot\Lambda^2=(\Delta_+^2-\Lambda^2)(\Lambda^2-\Delta_-^2)
\en
 with the solution
\beg
\Lambda=\Delta_+ \mbox{dn}\left[ \Delta_+(t-t_0), k'\right],\quad k'=\frac{\Delta_-}{\Delta_+},
\en
 where dn is the Jacobi elliptic function and $t_0$ is an arbitrary integration constant.
 
It also follows from \eref{A} and the definition of $A$ below \eref{s-} that the phase of the order parameter is determined as
 \beg
 \dot\Phi=\frac{dA}{d\Omega}=2\mu-\frac{\kappa}{\Lambda^2+h_1},\quad \Delta=\sqrt{\Lambda^2+h_1}e^{-i\Phi}.
 \label{125}
 \en

\subsection{Matching integrals of motion}

Given the quench initial conditions, we can evaluate all integrals of motion. This is equivalent to evaluating $\vec L^2(u)$ in the initial state as it is conserved and contains all the integrals as residues at $u=\eps_\p$. It turns out that in the thermodynamic limit it is possible to find a reduced (few spin) solution that has the same $\vec L^2(u)$, i.e. exactly the same integrals as the quench dynamics.

In the thermodynamic limit single-particle energies $\eps_\p$  form a continuum on the positive real axis and $\vec L^2(u)$ therefore has a continuum of poles at $u>0$. Additionally,  $\vec L^2(u)$ also has a continuum of roots along the $u>0$ half line as we show in Appendix~\ref{appb}. Thus $\sqrt{\vec L^2(u)}$ has a branch cut along $u>0$ in the continuum limit. There can also be several \textit{isolated} roots whose imaginary parts remain finite in this limit. Isolated roots play an important role in the dynamics; we will determine them below and see that there are at most four such roots (two pairs of complex conjugate roots) for our quench problem.

\eref{LaxReduced} implies
\beg
1+\int d\eps' \frac{d(\eps')\nu(\eps')}{u-\eps'}=-z(u) \sqrt{\frac{ \vec L^2(u)}{ \vec L_m^2(u)}},\quad z(u)=\pm 1,
\label{compare}
\en
where $ \vec L^2(u)$ is  evaluated for the quench initial conditions. Our task is to find the parameters for the reduced problem -- $d(\eps)$ and $\vec L_m^2(u)$ -- so that this equation holds. Then the reduced problem  has the same integrals of motion as the quench dynamics.

Both sides of \eref{compare} have a branch cut along the positive real axis and tend to 1 as $u\to\infty$ for an appropriate choice of the sign $z(\infty)$.  Further, provided the isolated roots of $\vec L^2(u)$ coincide with the roots of $\vec L_m^2(u)$, there are no more branching points and both sides are analytic away from the shared branch cut at $u>0$. If we further ensure that the left and the right hand sides of \eref{compare} have the same jump across the branch cut, then their difference is  an entire function that vanishes at infinity. It is therefore identically zero by Liouville's theorem from complex analysis and \eref{compare} holds. 

To equate  jumps across the branch cut, we take $u\to\eps\pm i0$, apply the well-known formula 
 $1/(x\pm i0)={\cal P}(1 / x) \mp i\pi\delta(x)$,
and subtract one result from another. This fixes $d(\eps)$,
\beg
 d(\eps)=-\frac{z(\eps)}{2i\pi \nu(\eps)} \frac{\sqrt{ {\vec L}^2(\eps_-)}-\sqrt{{\vec L}^2(\eps_+) }}{ \sqrt{ \vec L^2_m(\eps)} },
 \label{jump33}
\en
where $\eps_\pm=\eps\pm i0$. According to  expression \re{d} for $d_\p\equiv d(\eps_\p)$ this is equivalent to fixing the lengths of the spins, $|\sigma_\p|\equiv|\sigma(\eps_\p)|$, in the few spin solution so that
\beg
 \sigma(\eps)=z(\eps)\frac{\sqrt{ {\vec L}^2(\eps_-)}-\sqrt{{\vec L}^2(\eps_+) }}{2i\pi \nu(\eps)  }.
 \label{jump}
\en

Thus, the few spin solution with this $\sigma(\eps)$ and $\vec L_m^2(u)$ whose roots are the same as the isolated roots of $\vec L^2(u)$ has the same integrals of motion as the quench problem.

\subsection{Asymptotic solution for the quench dynamics}
\label{asym-1}

There are altogether $2(N+1)$ initial conditions -- two angles for each classical spin and two initial conditions for the oscillator. So far, we constructed a reduced $m$-spin solution that matches $N+1$ integrals of motion. This satisfies $N+1$ initial conditions. The dynamics of the reduced $m$-spin Hamiltonian contains $2(m+1)$ constants, $m+1$ of which (integrals of motion for $H_\mathrm{red}$) are already fixed since we fixed $\vec L^2_m(u)$. The remaining $m+1$ constants are not sufficient to match the remaining $N\to\infty$ initial conditions for the quench dynamics at finite $m$. This is resolved as follows. We use the known $m$-spin solution to derive  a general asymptotic (i.e. valid at $t\to\infty$) solution of the equations of motion for spins $\vec s_\p$ with the same $\Delta(t)$ and the same  integrals of motion as the $m$-spin solution. Integrals of motion therefore are those for the quench dynamics. In addition, this general solution contains the correct number $N+1$ of independent constants. We therefore  conjecture  that this is the true solution for the quench dynamics at large times after the quench. By construction, to verify this \textit{few spin conjecture}, it is sufficient to show that the true asymptote of $\Delta(t)$  coincides with $\Delta(t)$ in the $m$-spin solution because  given $\Delta(t)$ we obtain  the most general asymptotic solution of equations of motion.

As discussed above \eref{123}, each quench is characterized by three parameters -- the resonance width $\gamma$ and the final $\omega_f$ and initial $\omega_i$ values of the detuning. We determine in the next section that   $\vec L^2(u)$ for the quench dynamics can have 0, 1, or 2 pairs of isolated complex roots for any $\gamma$ depending on $\omega_i$ and $\omega_f$. These by construction must  also be all the roots of $\vec L_m^2(u)$, which has $m+1$ pairs of complex conjugate roots according to \eref{redsp}. Cases relevant for the quench phase diagram are therefore $m=-1, 0,$ and 1.   

It is worthwhile to consider the $m=-1$ case separately in some detail to illustrate this procedure. Suppose $\vec L^2(u)$  evaluated for the quench initial condition has no complex (isolated) roots away from the real axis. Then, there is an $m=-1$ spin solution constructed above that in the $N\to\infty$ limit has the same values of the integrals of motion as the spin dynamics. Spins in this solution are all along the $z$-axis, $\vec \sigma_\p=-\sigma_\p \hat z$, and $\Delta(t)=-gb(t)=0$. It is a particular solution of the equations of motion \re{Bloch} such that $b(t)=0$. 

The general solution of the \textit{spin part} of the equations of motion in \eref{Bloch} with $b(t)=0$ is: spins $\vec s_\p$ precess around the $z$-axis (or equivalently around the reduced spins $\sigma_\p$) with frequencies $2\eps_\p$, i.e.
\beg
s_\p^z=\frac{\sigma_\p^z}{\sigma_\p} \frac{\cos\theta_\p}{2},\quad s_\p^-= 
\frac{\sin\theta_\p}{2} e^{i\alpha_\p(t)},
\label{ssa}
\en
where $\theta_\p$  is the angle $\vec s_\p$ makes with $-\hat{\bf z}$ and $\alpha_\p=-2\eps_\p t+\delta_\p$. Equivalently, this can be expressed as
\beg
\vec s_\p=\frac{\vec \sigma_\p}{\sigma_\p}\frac{\cos\theta_\p}{2}+\vec s_\p^\perp,
\label{ss0}
\en
where $\vec s_\p^\perp$ is the component transverse to $\vec \sigma_\p$, which rotates around $\vec \sigma_\p$ with frequency $2\eps_\p$.
Note that the length of spin $\vec s_\p$ is 1/2 as it should be for the quench initial conditions. 

This spin configuration has $N$ additional constants $\delta_\p$, but it does not satisfy the equation of motion for $b(t)$ in \eref{Bloch} because $b(t)=0$, while $J_-(t)=\sum_\p s_\p^-=\sum_\p f_\p e^{-2\eps_\p t}\ne 0$, where $2f_\p=\sin\theta_\p e^{i\delta_\p}$. However, in the thermodynamic limit $J_-(t)=\int f(\eps) \nu(\eps) e^{-2\eps t}\to 0$ as $t\to\infty$ and this solution becomes self-consistent. 

\begin{figure}[h]
\includegraphics[scale=0.26,angle=0]{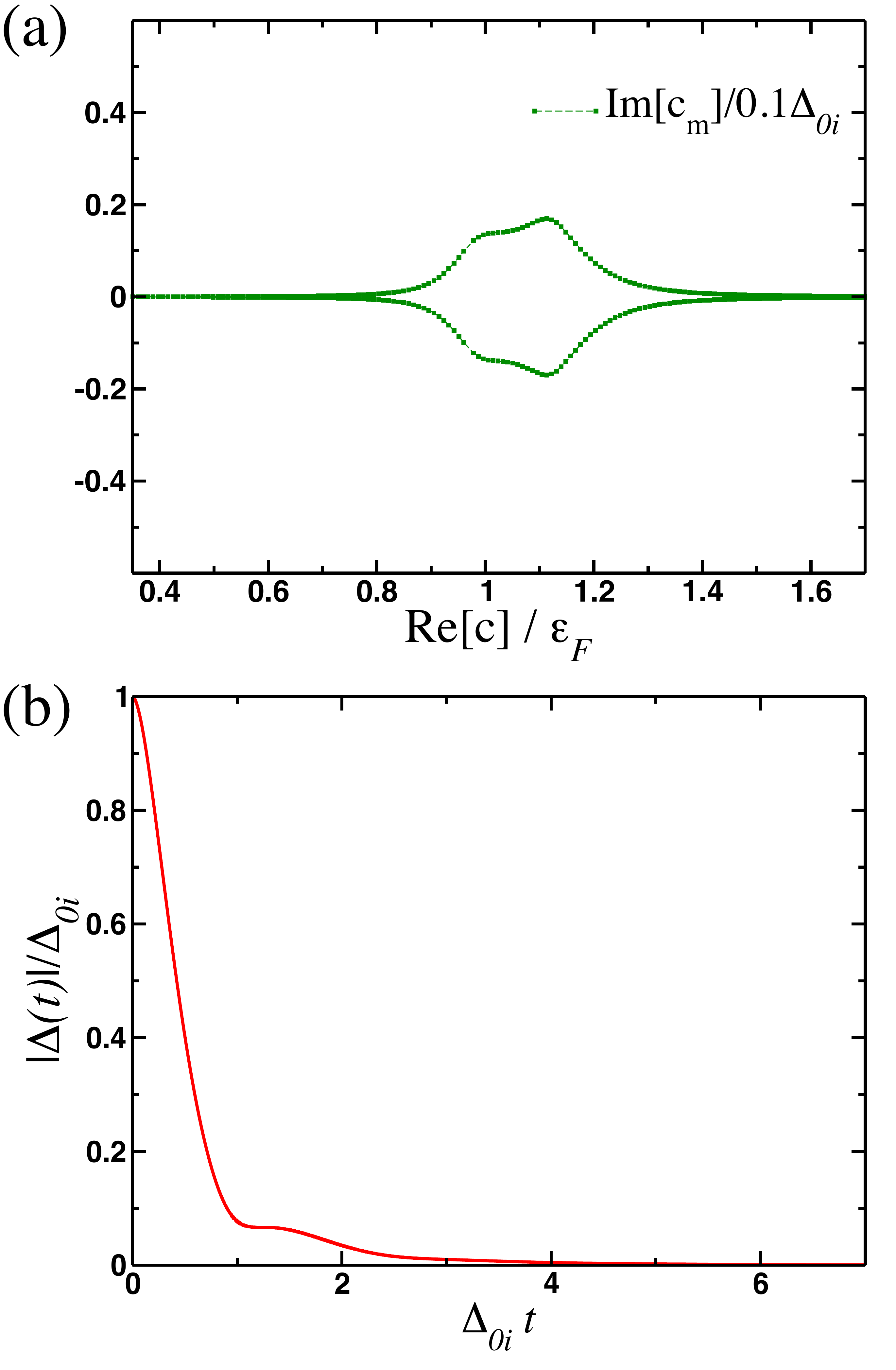}
\caption{(color online) Order parameter $\Delta(t)$ vanishes  whenever   the square of the Lax vector $\vec L^2(u)$ has no isolated roots. Panel (a)   shows real, $\textrm{Re}[c]$, and imaginary,  $\textrm{Im}[c_m]$, parts   of the roots $c_m$ and panel (b) shows the corresponding $|\Delta(t)|$ for a detuning quench in a 3d 2-channel model with $\gamma=0.1$ and
$N=1024$ spins. There are $N+1$ pairs of complex conjugate continual roots whose imaginary parts scale as $1/N$ so that in the $N\to\infty$ limit they form a continuum on the real axis. Here $\Delta_{0i}=0.34\Delta_{\max}, \Delta_{0f}=8.1\times 10^{-3}\Delta_{\max}, \mu_i=0.91\eps_F$, and $\delta\omega=3.45\gamma$.}
\label{0root}
\end{figure}

Next, we set $2\sigma_\p=\cos\theta_\p$ and substitute  $\vec s_\p= \vec \sigma_\p +\vec s_\p^\perp$ into the Lax vector,
\beg
\vec L(u)=\vec L^\mathrm{red}(u)+\sum_\p\frac{\vec s_\p^\perp}{u-\eps_\p}.
\en
The second term vanishes by the Riemann-Lebesgue lemma (dephases) as $t\to\infty$ in the thermodynamic limit for $u$ away from the real axis similarly to $J_-(t)$ and therefore $\vec L(u)\to\vec L^\mathrm{red}(u)$. Constants $\sigma_\p$ are given by \eref{jump} to match the integrals of motion. Then, the solution given by \eref{ssa} with $2\sigma_\p=\cos\theta_\p$  has the same integrals of motion as the quench dynamics and  the right number of additional constants to match the remaining initial conditions.  
As explained above, to verify that this is indeed the true asymptote of the quench dynamics, we only need to show that asymptotic $\Delta(t)$ coincides with $\Delta(t)$ of the $m=-1$ spin solution, i.e. that  $\Delta(t)\to 0$  at large times after the quench whenever $\vec L^2(u)$ has no isolated complex roots (region I in quench phase diagrams above). We confirm this numerically, see e.g.
Figs.~\ref{psd1ch3d}, \ref{0root}, and Refs.~\onlinecite{Yuzbashyan2006,Dzero2006}. There is also a justification of this statement based on the general theory of integrable Hamiltonian dynamics. It works for both $m=-1$ and $m=0$ and we present at the end of the $m=0$ case below \eref{prsebelow}.

  To summarize: if $\vec L^2(u)$ has no isolated complex roots for given (quench) initial conditions, then $\Delta(t)\to 0$ at large times in the thermodynamic limit and the steady state spin configuration   is  
  \beg
s_\p^z= -\frac{\cos\theta_\p}{2},\quad s_\p^-= 
\frac{\sin\theta_\p}{2} e^{-2i\eps_\p t+i\delta_\p},
\label{ssa161}
\en
where
\beg
 \cos\theta(\eps)=z(\eps)\frac{\sqrt{ {\vec L}^2(\eps_-)}-\sqrt{{\vec L}^2(\eps_+) }}{i\pi \nu(\eps)  },
 \label{jump1}
\en
and $\theta_\p\equiv \theta(\eps_\p)$. This expression  evaluates explicitly for quench initial conditions; the answer is given by \eref{distAA}. The sign $z(\eps)=\pm1$ is fixed by requiring that $\cos\theta(\eps)$ be smooth and    spins $\vec s_\p$ point in the negative $z$-direction at $\eps_\p\to \infty$  (so that corresponding single particles states be empty).

The logic for $m\ge 0$ is similar, but the calculation is a bit more involved. To derive the analog of \eref{ssa}, it is convenient to work with the Bogoliubov de Gennes equations \re{BdG}. In addition, there is an equation of motion for $b$ in \eref{Bloch}, which can be viewed as a self-consistency condition.  In terms of $\Delta=-gb$ and Bogoliuobov amplitudes it reads
\beg
\quad \dot{\Delta}=-i\omega  \Delta+ig^2 \sum_\p 2s_\p u_\p\bar v_\p
\label{sc}
\en
The reduced $m$-spin solution is a particular solution $(U_\p,  V_\p)$ of the BdG equations that also satisfies the above self-consistency condition (with $s_\p\to\sigma_\p$). It is straightforward to check that  $(\bar V_\p,  -U_\p)$ is also a solution of the BdG equations with the same $\Delta(t)$. Since for any fixed $\Delta(t)$ these equations are linear in the amplitudes, their most general normalized solution with this $\Delta(t)$ is a linear combination of these two independent solutions
\beg
\pmat
u_\p\\
\\
v_\p\\
\epmat
=
\cos\frac{\theta_\p}{2}
\pmat
U_\p\\
\\
V_\p\\
\epmat
+\sin\frac{\theta_\p}{2}
\pmat
\bar V_\p\\
\\
-\bar U_\p\\
\epmat.
\label{gen}
\en
The coefficients are made real by dropping an unimportant overall time-independent phase and including the relative phase into the common phase of $U_\p$ and  $V_\p$. 
At this point $\theta_\p$ is an arbitrary angle.
This solution does not generally satisfy the self-consistency condition \re{sc} at finite $t$, but, as we will see below, becomes self-consistent as $t\to\infty$.

Let us now determine the spins corresponding to this solution. \eref{gen} implies
\beg
\begin{split}
|v_\p|^2-|u_\p|^2=\left(|V_\p|^2-|U_\p|^2\right)\cos\theta_\p-\\
\sin\theta_\p\left(\bar U_\p \bar V_\p+U_\p V_\p\right),\\
u_\p\bar v_\p=U_\p \bar V_\p \cos\theta_\p+
\frac{\sin\theta_\p}{2}\left( \bar V_\p^2-U_\p^2\right).\\
\end{split}
\label{map}
\en
True spins $\vec s_\p$ are related to $u_\p$, $v_\p$ through \eref{lupvp} with $s_\p=1/2$. Spins $\vec \sigma_\p$ are similarly related to $U_\p, V_\p$. Let
\beg
\begin{split}
U_\p=|U_\p| \exp{\left[i \frac{\alpha_\p-\phi_\p}{2}\right]},\\ 
V_\p=|V_\p| \exp{\left[i \frac{\alpha_\p+\phi_\p}{2}\right]}.\\
\end{split}
\label{amp}
\en
We can express the absolute values of the amplitudes and their relative phase through the spin
components
\beg
|V_\p|^2=\frac{1}{2}+\frac{\sigma_\p^z}{2\sigma_\p}, \, |U_\p|^2=\frac{1}{2}-\frac{\sigma_\p^z}{2\sigma_\p}, \,
e^{-i\phi_\p}=\frac{\sigma_\p^-}{|\sigma_\p^-|},
\label{samp}
\en
while their common phase $\alpha_\p$ needs to be determined separately from the BdG equations.

We obtain in this notation
\beg
\begin{split}
s_\p^z=\frac{\sigma_\p^z}{\sigma_\p} \frac{\cos\theta_\p}{2} -\frac{|\sigma_\p^-|}{\sigma_\p} \frac{\sin\theta_\p}{2}\cos\alpha_\p,\\
s_\p^-=\frac{\sigma_\p^-}{\sigma_\p}\frac{\cos\theta_\p}{2}+
\frac{\sin\theta_\p}{2}e^{-i\phi_\p}
\left(\frac{\sigma_\p^z}{\sigma_\p}\cos\alpha_\p-\right.\\
\biggl. i\sin\alpha_\p\biggr).\\
\end{split}
\label{ss}
\en
Note that $\sigma_\p^z/\sigma_\p$ and $\sigma_\p^-/\sigma_\p$ are components of the unit vector along the spin in the reduced solution $\vec \sigma_\p$. Geometrically \eref{ss} says
that $\vec s_\p$ makes a constant angle $\theta_\p$ (or $\pi-\theta_\p$ for negative $\sigma_\p$) with $\vec \sigma_\p$ and rotates around it with an angular velocity $\dot\alpha_\p$,
\beg
\vec s_\p=\frac{\vec \sigma_\p}{\sigma_\p}\frac{\cos\theta_\p}{2}+\vec s_\p^\perp.
\label{ss1}
\en
To see this, consider a body set of axis for $\vec \sigma_\p$. Take $z'$  along $\vec \sigma_\p$, $x'$-axis along the intersection of the $zz'$ plane with the plane perpendicular to $\vec \sigma_\p$, and $y'$ normal to $x'z'$ to form a right-handed coordinate system as usual. Then $\alpha_\p$ is the angle between $\vec s_\p^\perp$ and the $x'$-axis and \eref{ss} follows.

The contribution of the second terms on the right hand side of \esref{ss} and \re{ss1} (terms containing $\alpha_\p$) to $\vec L(u)$ at $u$ away from the real axis and to $J_-(t)$ vanishes (dephases) at large times at least for $m=0$ and 1  in the thermodynamic limit same as in the $m=-1$ case considered above.   For this to be true it is sufficient that $\alpha_\p$ contain a dispersing linear in $t$ term, i.e. 
\beg
\alpha_\p=-2e_\p t +F_\p(t),
\label{disp}
\en
 where $e_\p$ is a continuous non-constant function of $\eps_\p$ and $F_\p(t)$ is a bounded function of $t$. Note that for $m=-1$, $e_\p=\eps_\p$ and $F_\p(t)=\delta_\p=\mbox{const}$.

To derive the asymptotic state, we follow the same procedure as for $m=-1$ above. We set $2\sigma_\p=\cos\theta_\p$, where $\cos\theta_\p\equiv \cos\theta(\eps_\p)$ is given by \eref{jump1}.  Then, $\vec L(u)\to \vec L^\mathrm{red}(u)$, $\Delta(t)$ is described by this $m$-spin solution at large times and satisfies the self-consistency condition \re{sc}, the asymptotic spin configuration \re{ss} and the  $m$-spin problem have the same integrals of motion as the quench dynamics. The remaining $N+1$ constants required to match the initial conditions are in $\alpha_\p$ (see below) and in the phase of $\Delta(t)$.

To determine $\alpha_\p$, rewrite the BdG eqs as 
\beg
i\partial_t( \ln U_\p)=\eps_\p+\Delta\frac{V_\p}{U_\p},\quad i\partial_t( \ln V_\p)=-\eps_\p+\bar\Delta\frac{U_\p}{V_\p}.
\label{ln}
\en
Adding these equations and using \esref{amp} and \re{samp}, we get after some algebra
\beg
\dot \alpha_\p=-\frac{\sigma_\p(\bar \Delta \sigma_\p^-+\Delta \sigma_\p^+)}{ |\sigma_\p^-|^2}.
\label{alpha}
\en

\subsubsection{m=0}
\label{asym0}

Suppose  $\vec L^2(u)$  has a single pair of isolated complex roots at $u=\mu_\infty\pm i\Delta_\infty$. The 0-spin expression 
\re{0sdel} for $\Delta(t)$ reads
\beg
\Delta(t)=\Delta_\infty e^{-2i\mu_\infty t-2i\varphi}.
\label{0asym}
\en
The notation $\Delta_\infty$ and $\mu_\infty$ anticipates that this is also the long time asymptote for the quench dynamics. \eref{0spin} implies
\beg
\frac{\sigma_\p^-}{\sigma_\p}=\frac{\Delta(t)}{E_\p^\infty },\quad \frac{\sigma_\p^z}{\sigma_\p}=-\frac{\xi_p}{E_\p^\infty },
\label{0saux}
\en
where $E_\p^\infty =E(\eps_\p;\Delta_\infty,\mu_\infty)=\sqrt{(\eps_\p-\mu_\infty)^2+\Delta_\infty^2}$ and $\xi_\p=\eps_\p-\mu_\infty$.

\eref{alpha} obtains $\dot\alpha_\p=-2E_\p^\infty $. We see that $\alpha_\p$ is of the form \re{disp} and therefore the large time asymptote of $\Delta(t)$ according to the few spin conjecture is given by \eref{0asym}. The asymptotic spin configuration is then \eref{ss} with $\cos\theta_\p\equiv\cos\theta(\eps_\p)$ given by \eref{jump1}. Explicitly, using \eref{0saux} and $\alpha_\p=-2E_\p^\infty  t-\delta_\p$, we obtain
\beg
\begin{split}
s_\p^z=-\frac{\xi_\p}{2E_\p^\infty}  \cos\theta_\p  -\frac{ \Delta_\infty}{2E_\p^\infty}  \sin\theta_\p \cos(2E_\p^\infty  t+\delta_\p),\\
s_\p^-e^{2i\mu_\infty t+2i\varphi}= \frac{ \Delta_\infty}{2E_\p^\infty}  \cos\theta_\p 
 -  \frac{\sin\theta_\p}{2} e^{2iE_\p^\infty  t+i\delta_\p}-\\
 \left( \frac{\xi_p}{E_\p^\infty } -1\right) \frac{\sin\theta_\p}{2}\cos(2E_\p^\infty  t+\delta_\p).\\
\end{split}
\label{ss0s}
\en
In a reference frame rotating with frequency $2\mu_\infty$ around  $z$-axis, $\Delta(t)\to\Delta_\infty$ meaning that magnetic field acting on spin $\vec s_\p$ is time-independent. In this frame $\vec s_\p$ rotates around the field or, equivalently, around the reduced spin $\vec \sigma_\p$ with frequency $2E_\p^\infty$ as described by \eref{ss0s}.

We can also determine the Bogoliubov amplitudes corresponding to the 0-spin solution from \esref{amp} and \re{samp}
\beg
\begin{split}
U_\p=\sqrt{\frac{1}{2}+\frac{\xi_\p}{2E_\p^\infty}} e^{-iE_\p^\infty t-i\mu_\infty t-i\varphi},\\
V_\p=\sqrt{\frac{1}{2}-\frac{\xi_\p}{2E_\p^\infty}} e^{-iE_\p^\infty t+i\mu_\infty t+i\varphi}.\\
\end{split}
\label{UpVpm0}
\en
These in turn determine the ``real'' asymptotic amplitudes according to \eref{gen} and therefore the many-body wavefunction 
\re{fullwav}, which allows one to calculate various few-particle Green's functions.

\begin{figure}[h]
\includegraphics[scale=0.26,angle=0]{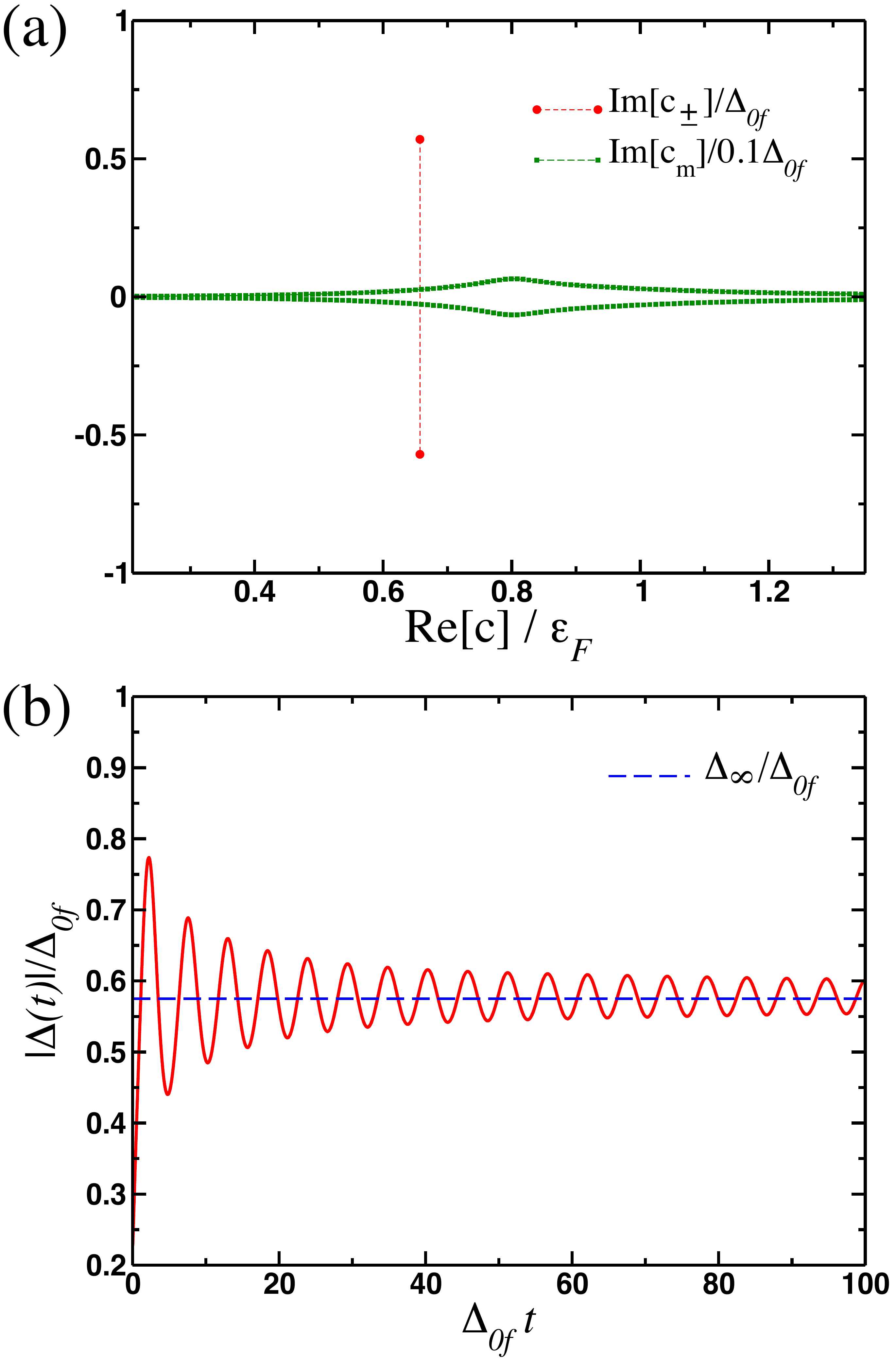}
\caption{(color online) Roots of $\vec L^2(u)$ (top) and    $|\Delta(t)|$ for a detuning quench in a 3d 2-channel model for $N=1024$ spins, $\gamma=1.0$. There is one pair of isolated roots $c_\pm=\mu_\infty\pm i\Delta_\infty$ whose imaginary part remains finite in the  large $N$ limit and $N-1$ continual roots $c_m$ close to the real axis ($\textrm{Im}[c_m]$ is zoomed in by 10).    Observe  $|\Delta(t)|\to\Delta_\infty$ in agreement with the few spin conjecture.  Here $\Delta_{0i}=0.18\Delta_{\max}, \Delta_{0f}= 0.78\Delta_{\max}$, and $\delta\omega=-2.26\gamma$.}
\label{1root}
\end{figure}

As before, to verify the few spin conjecture in the present case it is enough to check that the large time asymptote of $\Delta(t)$ after the quench is given by \eref{0asym} as long as 
$\vec L^2(u)$ has one pair of isolated complex conjugate roots (regions II and II' in quench phase diagrams above).  We do so numerically, see e.g. Figs.~\ref{regIIdeltaplot}, \ref{1root}, and Refs.~\onlinecite{Yuzbashyan2006,Dzero2006}. The large time asymptote of $|\Delta(t)|$ is in excellent agreement with $\Delta_\infty$ derived as the imaginary part of the isolated root, see e.g. Fig.~2 in  Ref.~\onlinecite{Dzero2006}. This is however guaranteed by conservation laws without reliance on the few spin conjecture. Indeed, suppose we find $\Delta(t)\to \tilde\Delta_\infty e^{-2i\tilde\mu_\infty t-2i\tilde\varphi}$. Starting with this, one can retrace the steps that lead to \eref{ss0s} backwards and show that $\vec L^2(u)$ has a single pair of isolated complex conjugate roots at $\tilde\mu_\infty\pm i\tilde\Delta_\infty$. In other words,   $\tilde\mu_\infty=\mu_\infty, \tilde\Delta_\infty=\Delta_\infty$, and the constant $\varphi$ is arbitrary in the 0-spin solution, so we can always set $\tilde\varphi=\varphi$. Let us prove this somewhat differently using Bloch rather than BdG equations.

 Going to a reference frame rotating around  $z$-axis with frequency $2\tilde\mu_\infty$ eliminates  time dependence in the asymptotic $\Delta(t)$. In this frame, the effective magnetic field acting on each spin $\vec s_\p$ in \eref{Bloch} is 
  ${{\vec B}}_\bp=-2\tilde\Delta_\infty\hat{\bf x}+2(\epsilon_\bp-\tilde\mu_\infty)\hat{\bf z}$ and is time-independent. The spin therefore rotates around the field making a constant angle (call it $\pi-\theta_\p$) with it. It is straightforward to determine spin components in this situation. They are given by   \eref{ss0s} with $\mu_\infty\to\tilde\mu_\infty$, $\Delta_\infty\to\tilde\Delta_\infty$, and absent   $e^{2i\mu_\infty t+2i\varphi}$ on the left hand side  in the rotating frame. 
  
  Next, we evaluate  Lax vector \re{Lax} for this spin configuration. For $u$ away from the real axis, summations over $\p$ can be safely replaced with integrations in the continuum limit and contributions from oscillating terms on the right hand side of \esref{ss0s} vanish at $t\to\infty$. The same cancelation occurs in the gap equation of motion \re{123}, so that it becomes \eref{argument} that we will later also need in a different context.  Using this gap equation to simplify the expression for $\vec L(u)$, we obtain
 \beg
\vec L(u) =\left[\tilde\Delta_\infty \hat {\bf x}-(u-\tilde\mu_\infty)\hat {\bf z}\right] L_\infty(u),
\en
where
\beg
L_\infty(u)= \frac{2}{g^2}-\sum_\p\frac{1}{2(u-\eps_\p)E_\p^\infty}.
\label{prsebelow}
\en
We see that $\vec L^2(u)= [\tilde\Delta_\infty^2-(u-\tilde\mu_\infty)^2]L^2_\infty(u)$ has a pair of isolated roots at $u=\tilde\mu_\infty\pm i \tilde\Delta_\infty$, i.e. the parameters of the asymptotic $\Delta(t)$ must coincide with those of an isolated root.

Finally, there is a general argument explaining why the actual quench dynamics at $t\to\infty$ should be described by the above  asymptotic solutions derived from -1 and 0 spin solutions at least when  $\vec L^2(u)$ has none or only one isolated root pair ($m=-1$ and 0). The general motion of a classical Hamiltonian integrable model with $N$ degrees of freedom is quasi-periodic with $N$ independent frequencies, $\vec\omega=(\omega_1,\dots,\omega_N)$, that are determined solely by the values of its integrals of motion\cite{Arnold,Tabor}. There are two types of (quasi)periodic motion: libration and rotation\cite{Goldstein}. Let us explain this terminology with a one-dimensional example.  In libration, the coordinate  returns to its initial value after each period, such as e.g. the coordinate of a harmonic oscillator. In rotation, it increases each time by a fixed amount, such as e.g. the angle of a rotating pendulum. Dynamical variables of libration type can be decomposed in a multi-dimensional Fourier series as follows, 
  \beg
  Q(t)=\sum_{\vec m} c_{\vec m} e^{i\vec \omega\cdot\vec m t},
  \label{quasi1}
  \en
where $\vec m=(m_1,\dots, m_N)$ is a vector with integer components. 
 Dynamical variables of rotation type contain an additional linear term, i.e.
 \beg
  Q(t)=c_0 t+\sum_{\vec m} c_{\vec m} e^{i\vec \omega\cdot\vec m t},
  \label{quasi2}
  \en
see e.g. Ref.~\onlinecite{Goldstein} for further details. In our case, the absolute value of the order parameter, $|\Delta(t)|$ is of libration type, while its phase is of rotation type.
  
    The frequency spectra of  asymptotic solutions constructed above are $\omega(\eps_\p)=2\eps_\p$ for $m=-1$ and $\omega(\eps_\p)=2\sqrt{(\eps_\p-\mu_\infty)^2+\Delta_\infty^2}$ for $m=0$. Important for us is that  the spectra are continuous with no isolated frequencies in the thermodynamic limit. 
  Since setting $2\sigma_\p=\cos\theta_\p$ ensures  the quench dynamics has the same integrals as this solution (lives on the same invariant torus), it also must have an identical frequency spectrum.
  Assuming   $|\Delta(t)|$ is continuously distributed over the spectrum as a collective variable, i.e. the discrete summation in \eref{quasi1} turns into a continuous Fourier transform, it must dephase at large times,   $|\Delta(t)|\to \mbox{const}$. Under the same assumption, the phase of the order parameter according to \eref{quasi2} must tend to a linear in time function as $t\to\infty$. Therefore, $\Delta(t)$ at large times is of the form $\Delta_\infty e^{-2i\mu_\infty t -2i\varphi}$. Since finite $\Delta_\infty$ also implies an isolated root at $\mu_\infty\pm i\Delta_\infty$ while for $m=-1$ there are no isolated roots by definition, we must  have $\Delta_\infty= 0$, i.e.
  $\Delta(t)\to 0$  in this case.
  
 We also prove the few spin conjecture for infinitesimal quenches in Sect.~\ref{validity} independently of above arguments and numerics.
    
\subsubsection{m=1}
\label{asym1}

Suppose we found that  for some initial condition (quench parameters) $\vec L^2(u)$ has two pairs of isolated complex conjugate roots  $c, \bar c, c', \bar c'$. Given $c$ and $c'$, the above method allows us to determine the long time asymptote of $\Delta(t)$, asymptotic spin configuration and time-dependent Bogoliubov amplitudes $u_\p(t), v_\p(t)$ for the dynamics of the two-channel model \re{Model1} starting from this initial condition at $t=0$.

By construction $c, c'$ are also the roots of 
$\vec L_m^2(u)$ furnishing  the spectral polynomial for the reduced problem $Q_4(u)=(u-c)(u-\bar c)(u-c')(u-\bar c')$ and therefore the parameters $\mu, \rho, \kappa, \chi$ through \eref{q4}. We further obtain from \eref{125}
\beg
 \Delta(t)=\sqrt{\Lambda^2+h_1}\exp\left(-2i\mu t-i\int \frac{\kappa dt}{\Lambda^2+h_1}\right), 
 \label{1spindeltafull}
 \en
where $\Lambda$ is the Jacobi elliptic function dn,
\beg
\Lambda=\sqrt{h_3-h_1}\mbox{dn}\left[ \sqrt{h_3-h_1}(t-t_0),  \sqrt{\frac{h_3-h_2}{h_3-h_1}}\right], 
\en
$t_0$ is a constant, and $h_3\ge h_2\ge h_1$ are the roots of the third order polynomial $P_3(w)=w^3+4\rho w^2+4\chi w+\kappa^2$. The amplitude $|\Delta(t)|$ oscillates between a minimum $\Delta_b=\sqrt{h_2}$ and a maximum  $\Delta_a=\sqrt{h_1}$. Plots of $\Delta_a,\Delta_b$, and $h_1$ for various quenches are shown in Figs.~\ref{dinfplot} and \ref{h1111plot}. As we will now see, the parameter $h_1$ also quantifies the deviation from the weak coupling limit where $h_1=0$.

\begin{figure}[h]
\includegraphics[scale=0.26,angle=0]{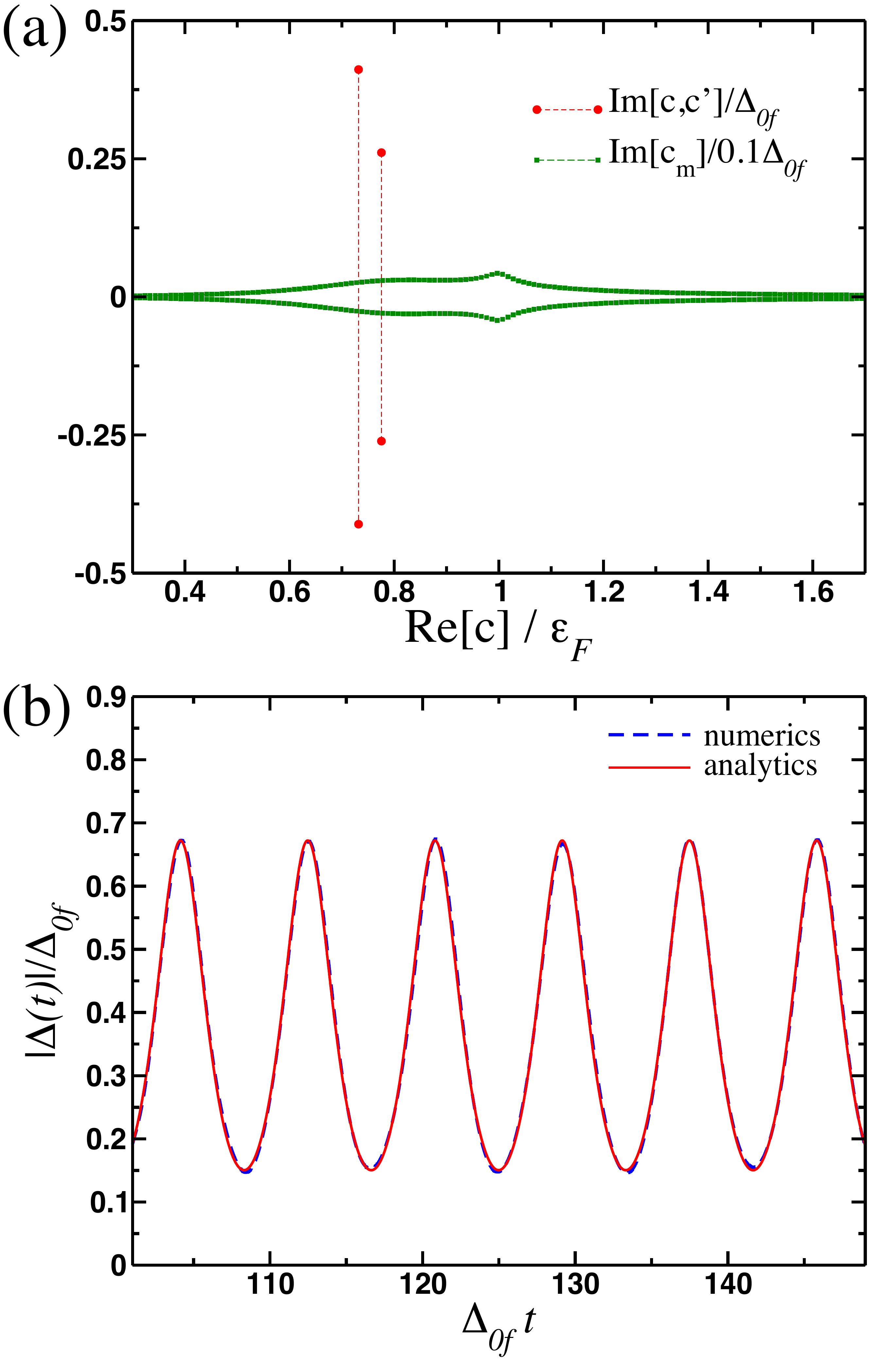}
\caption{(color online) Roots of $\vec L^2(u)$ (top) and   $|\Delta(t)|$ for a detuning quench in a 3d 2-channel model with $N=1024$ spins and $\gamma=1.0$. There are two pairs of  isolated roots $(c, \bar c)$ and $(c', \bar c')$   and $N-2$ continual roots close to the real axis.  The large time asymptote of $|\Delta(t)|$ is described by
\eref{1spindeltafull}, where parameters $h_i$ are extracted from the isolated roots  in agreement with the few spin conjecture. The phase of $\Delta(t)$ is also in excellent agreement, see e.g. Figs.~\ref{regIIIdeltaplot} and \ref{regIIIdeltaplot-a}. Quench parameters are: $\Delta_{0i}=2.68\Delta_{\max}, \Delta_{0f}= 0.76\Delta_{\max}$, and $\delta\omega=-4.13\gamma$.}
\label{2root}
\end{figure}

\begin{figure}[h]
\includegraphics[scale=0.26,angle=0]{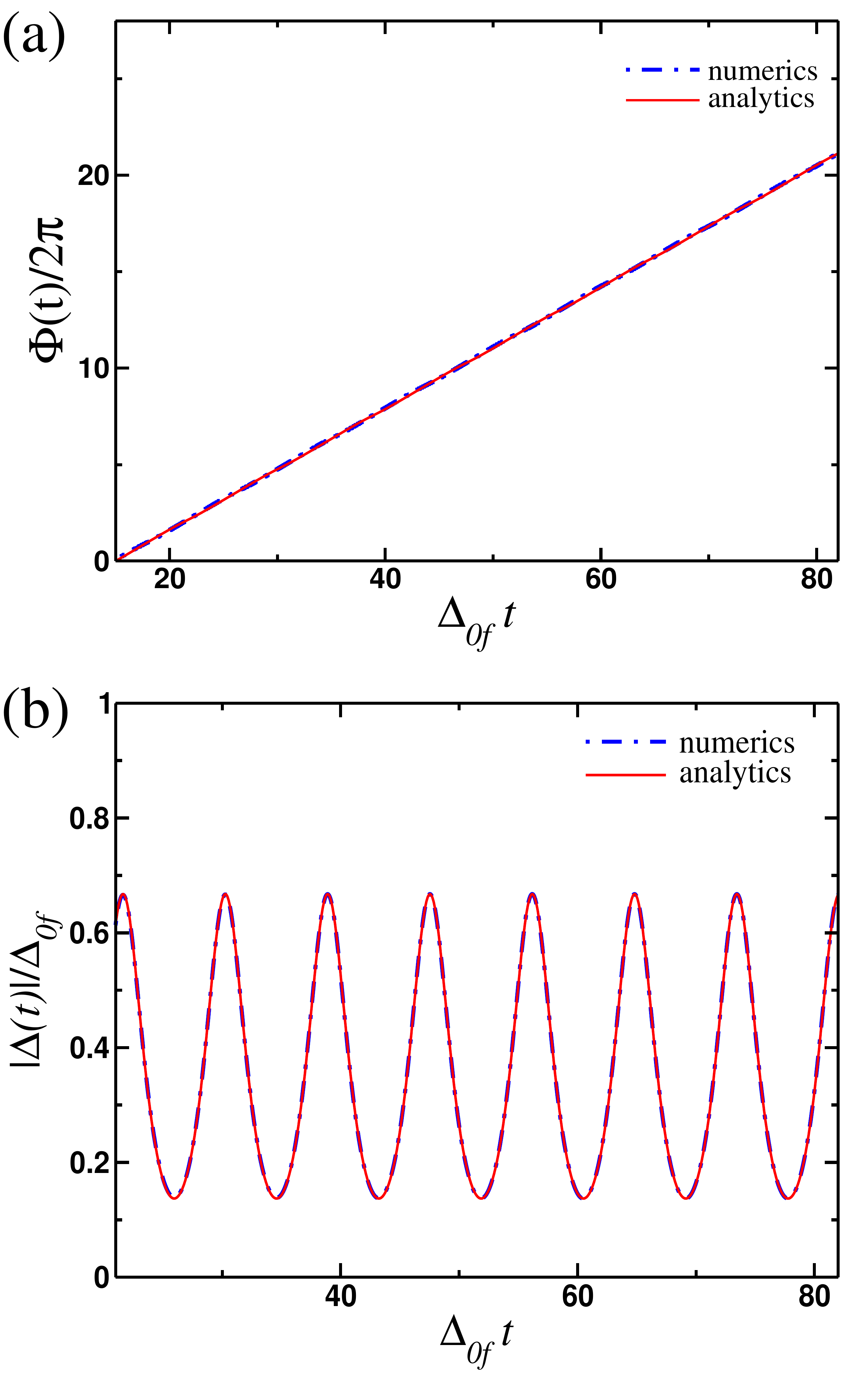}
\caption{(color online) Magnitude   and phase  of  $\Delta(t)$ in region III  (two pairs of isolated roots) after detuning quench from deep BCS to BEC in 3d two-channel model for $\gamma=1$.  Numerical evolution with 5024 spins  against \eref{1spindeltafull}. Parameters $h_1, h_2$, etc. are obtained from isolated roots of $\vec L^2(u)$ as described in the text.    $\Delta_{0i}=2.65\times 10^{-2}\Delta_{\max}, \Delta_{0f}=0.80\Delta_{\max}, \mu_i=1.00\eps_F, \delta\omega=-4.59\gamma$.  }
\label{regIIIdeltaplot-a}
\end{figure}

Of interest is the particular case when the parameter $\kappa=0$. As we will see below, this is realized for quenches  deep within the weak coupling BCS regime in the broad resonance limit   when the two-channel model is equivalent to the BCS Hamiltonian \re{Model2}. $\kappa=0$ implies $h_1=0$, $4\chi=h_2h_3$, $4\rho=-h_2-h_3$, and $Q_4(u)=[(u-\mu)^2-\rho]^2-\chi$. Let $h_3=\Delta_+^2, h_2=\Delta_-^2$ in accordance with the notation of \eref{define}. The roots of $Q_4(u)$ in this case take a simple form with shared real part. Namely, they are
\beg
\mu\pm i\frac{\Delta_+\pm\Delta_-}{2},
\label{spform}
\en
and the expression \re{1spindeltafull} simplifies as well,
\beg
\Delta(t)=\Delta_+ \mbox{dn}[ \Delta_+(t-t_0)]e^{-2i\mu t-2i\varphi}.
\label{spdelta}
\en
This expression for $\Delta(t)$ and the corresponding $m=1$ spin solution were constructed in Ref.~\onlinecite{Spivak2004}.

General expression for the reduced spins obtain from \esref{sz}, \re{s-}, and \re{coef},
 \beg
 \begin{split}
 \frac{\sigma_\p^z}{\sigma_\p}=-\frac{|\Delta|^2-2\xi_\p^2+2\rho}{2\sqrt{Q_4(\eps_\p)}},\\
 \frac{\sigma_\p^-}{\sigma_\p}=-\frac{2\xi_\p\Delta-2\mu\Delta+i\dot\Delta}{2\sqrt{Q_4(\eps_\p)}},\\
 \end{split}
 \label{sigsig}
 \en
where $\xi_\p=\eps_\p-\mu$ and $\Delta$ is given by \eref{1spindeltafull}. Bogoliubov amplitudes  corresponding to the 1-spin solution can now be derived from \eref{ln}. The imaginary and real parts of the right hand sides determine the absolute values of the amplitudes and their phases, respectively,
\beg
\begin{array}{l}
\displaystyle U_\p= \frac{\sqrt{2c^+_\p-|\Delta|^2}}{2Q_4^{1/4}(\eps_\p)} e^{-i\mu t+i\xi_\p t}\exp{\left[i\int \frac{\kappa -4\xi_\p c^+_\p}{2c^+_\p- |\Delta|^2} dt\right]},\\
\\
 \displaystyle V_\p= \frac{\sqrt{2c^-_\p+ |\Delta|^2}}{2Q_4^{1/4}(\eps_\p) } e^{i\mu t-i\xi_\p t}\exp{\left[i\int \frac{\kappa +4\xi_\p c^-_\p}{2c^-_\p+ |\Delta|^2} dt\right],} 
  \end{array}
  \label{UpVpm1}
\en
where $c^\pm_\p=   \sqrt{Q_4(\eps_p)}\pm(\xi_\p^2-\rho)$.

The common phase of the amplitudes $\alpha_\p$ is the sum of their phases in the above equations, i.e.
\beg
\alpha_\p=\int\left[ \frac{\kappa -4\xi_\p c^+_\p}{2c^+_\p- |\Delta|^2}+\frac{\kappa +4\xi_\p c^-_\p}{2c^-_\p+ |\Delta|^2}\right] dt
\label{alf}
\en
The integrand is a periodic function of time. Therefore, $\alpha_\p$ is of the form \re{disp}, which is seen e.g. by expanding the expression under the integral in Fourier series. The linear part $e_\p t$ comes from the zeroth harmonics. We only need to show that
 $e_\p$ is a non-constant (dispersing) function of $\eps_\p$. For this, we expand the integrand for large $\eps_\p$, $e_\p=\eps_\p+O(1)$. Therefore, $e_\p$ is indeed dispersing
 and the contribution of second terms on the right hand sides of \eref{ss} to $\vec L(u)$ and $J_-(t)$ dephases similarly to $m=-1,0$ cases.  By few spin conjecture the asymptotic behavior of $\Delta(t)$  is then given by \esref{1spindeltafull}. The asymptotic spin configuration obtains by substituting \esref{alf} and \re{sigsig} into  \eref{ss}, where  $\cos\theta_\p\equiv\cos(\eps_\p)$ is given by \eref{jump1} and $e^{-i\phi_\p}=\sigma_\p^-/|\sigma_\p^-|$ straightforwardly derives from the second equation in \re{sigsig}.
 
\begin{figure}[h]
\includegraphics[scale=0.26,angle=0]{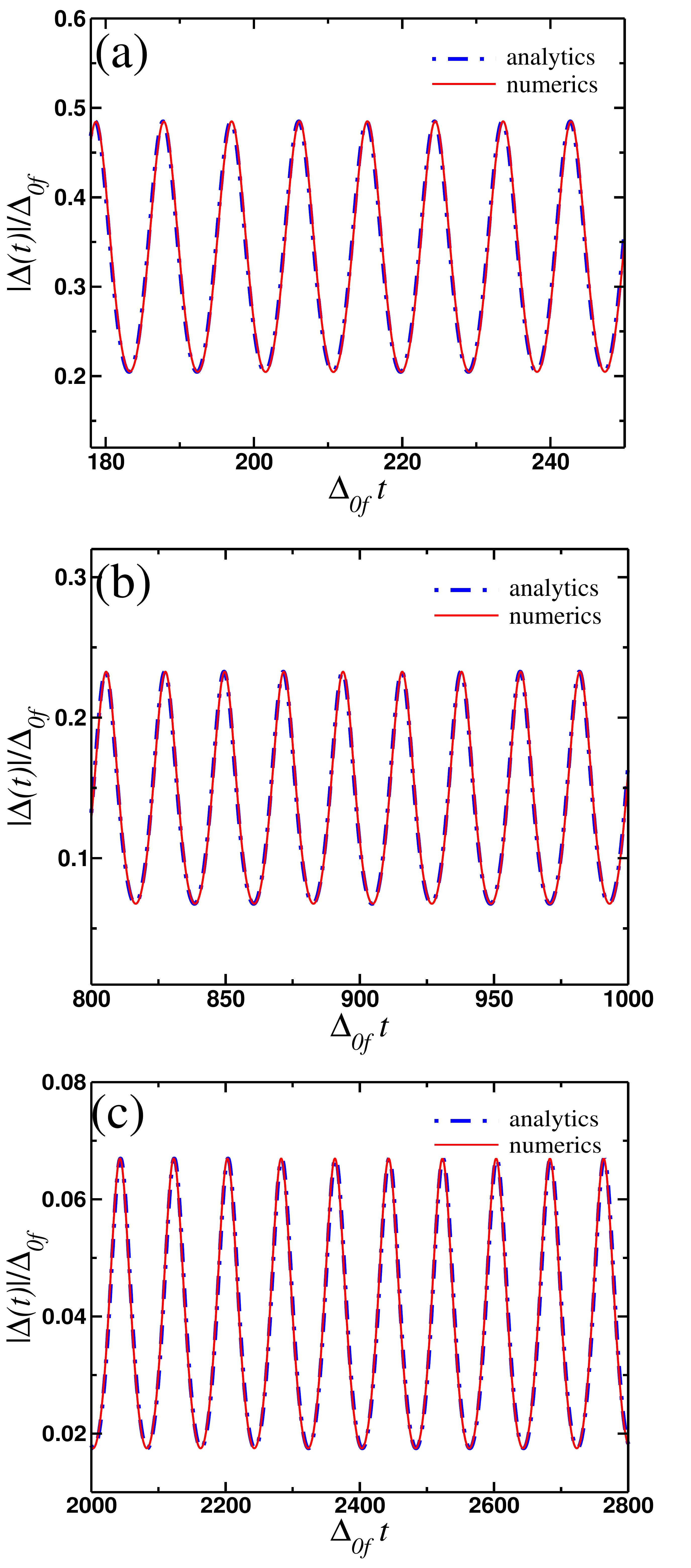}
\caption{(color online) Post-quench $|\Delta(t)|$  for 3d 2-channel model in region III where $\vec L^2(u)$ has two pairs isolated roots. Numerical evolution with 5024 spins  against \eref{1spindeltafull}. $\gamma=0.1$, $\Delta_{0i}=0.035\Delta_{\max}$ in all three panels. $\Delta_{0f}/\Delta_{\max}=0.54, 0.67,$ and $0.85$ in (a), (b), and (c), respectively.}
\label{regIIIdeltaplot-b}
\end{figure}

 As before, to verify the few spin conjecture it is sufficient to check that $\Delta(t)$ at large times after the quench is described  by \eref{1spindeltafull} whenever $\vec L^2(u)$ has two pairs of isolated roots. We do this numerically, see Figs.~\ref{2root}, \ref{regIIIdeltaplot}, \ref{regIIIdeltaplot-a}, and \ref{regIIIdeltaplot-b}. In these plots we compare $\Delta(t)$   from direct numerical evolution of 5024 spins to \eref{1spindeltafull}, where parameters $h_1, h_2, h_3,$ and  $\mu$  obtain from the isolated roots of $\vec L^2(u)$. Note  there are no fitting parameters apart from an overall shift $t_0$ along the time axis.
 
\section{Quench phase diagram and asymptotic spin distribution for the 2-channel model}
\label{qpsd}
 
We established in the previous section that the long time dynamics of the system after a quench are determined by the isolated complex roots of $\vec L^2(u)$. We now proceed to evaluate the roots and thus construct the quench phase diagram -- identify  all possible steady states for quenches throughout the BCS-BEC crossover. We find that depending on the quench parameters $\vec L^2(u)$ has either zero, one, or two pairs of complex conjugate roots and the long time behavior is therefore that described in Sections~\ref{asym-1}, \ref{asym0}, or \ref{asym1}, respectively. Imaginary and real parts of the roots determine the parameters of the asymptotic behavior. For example, in the Volkov and Kogan regime (region II in our quench phase diagrams) where $\Delta(t\to\infty)\to \Delta_\infty e^{-2i\mu_\infty t-2i\varphi}$, the roots are $\mu_\infty\pm i\Delta_\infty$.  We first derive general equations for the roots, lines separating distinct  regimes, and the asymptotic distribution function, and then consider various cases, such as two and three dimensions, wide (one channel) and narrow resonance limits, and deep BCS and BEC regimes.

After the quench the system evolves with the Hamiltonian \re{Model1} where $\omega=\omega_f$ starting from the spin configuration \re{ini} which is the ground state for $\omega=\omega_i$. Since $\vec L^2(u)$ is conserved, we can evaluate it at any $t$. It is convenient to do so at $t=0$. The Lax vector at $t=0$ obtains by plugging the initial condition into the definition \re{Lax}
\beg
\vec L(u)|_{t=0}=\left[\Delta_{0i} \hat {\bf x}-(u-\mu_i)\hat {\bf z}\right]L_0(u)- \frac{\delta\omega}{g^2}\hat {\bf z},
\label{lt0}
\en
where $\delta\omega=\omega_f-\omega_i$ and
\beg
  L_0(u)=-\frac{2}{g^2}+\sum_\p\frac{1}{2(u-\eps_\p)E_i(\eps_\p)},
\en
$E_i(\eps_\p)=E(\eps_\p;\Delta_{0i},\mu_i)=\sqrt{(\eps_\p-\mu_i)^2+\Delta_{0i}^2} $ and we also used the gap equation \re{gapeq}.

Taking the square of the above expression for $\vec L(u)$ and equating it to zero, we obtain an equation for the roots
\beg
\bigl(u-\mu_i\mp i\Delta_{0i}\bigr)\biggl[\frac{2}{g^2}-\sum_\p \frac{1}{2(u-\eps_\p)E_i(\eps_\p)}\biggr]=
\frac{\delta\omega}{g^2}.
\label{rtdis}
\en
Suppose first the single particle levels $\eps_\p$ are discrete and there are  $N\gg 1$ distinct  $\eps_\p$. Then, this is a polynomial equation with $N+1$ pairs of complex conjugate roots. Most of the pairs are close to the real axis, at distances of the order of the spacing between  $\eps_\p$, which is inversely proportional to $N$ (system volume) and goes to zero in the thermodynamic limit.   In thermodynamic limit most of the roots of $\vec L^2(u)$ coalesce to the real axis  merging with its poles   to form a branch cut along the real axis. We fully verify this picture in this section and in Appendix~\ref{appb}. Here we consider the roots whose imaginary part remains finite as $N\to\infty$ and in Appendix~\ref{appb} we   evaluate the roots with vanishing imaginary parts to  order  $1/N$.

\begin{figure}[htb]
\includegraphics[scale=0.26,angle=0]{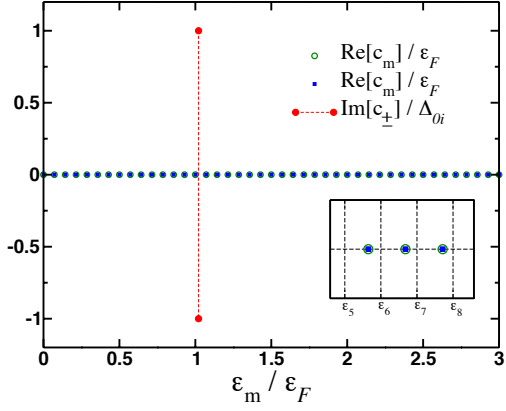}
\caption{(color online) Roots of $\vec L^2(u)$ for the ground state of 3d 2-channel model for $N=54$ spins and $\gamma=1.0$. There are $N$ doubly degenerate real roots $c_m$ (shown as circles and squares), $N-1$ of them located in between discretized  energy levels $\eps_\p\to\eps_m$, and two isolated complex roots $c_\pm=\mu_i\pm\Delta_{0i}$. Here $\Delta_{0i}=0.1\eps_F$. }
\label{groundroots}
\end{figure}

Consider first the ground state. This corresponds to $\delta\omega=0$ in \eref{rtdis} and $\vec L^2(u)=\left[ (u-\mu)^2+\Delta_0^2\right] L^2_0(u)$. There is a pair of complex roots at $c_\pm=\mu\pm i\Delta_0$. The remaining $2N$ roots solve $L_0(u)=0$ and are double degenerate and real, see Fig.~\ref{groundroots}. This is because $L_0(u)$ goes from $+\infty$ to $-\infty$ as $u$ goes from the left vicinity of one pole at $u=\eps_\p$ to the right vicinity of the next one along the real axis always crossing zero in between consecutive $\eps_\p$. In the thermodynamic limit,  spacings between $\eps_\p$ vanish and  real zeroes and poles merge into a continuos line. For $\delta\omega\ne 0$ the real roots acquire imaginary parts each degenerate root splitting into a complex conjugate pair as shown in Figs.~\ref{0root}, \ref{1root}, and \ref{2root}. The imaginary parts  however scale as $1/N$. 

We first take the continuum limit in \eref{rtdis} for $u$ away from the real axis. Then, only isolated complex roots remain and we find that there are only zero, one or two pairs of such roots depending on $\delta\omega$. At $\delta\omega=0$ there are two isolated complex conjugate roots at $u=\mu_i\pm i \Delta_{0i}$. One pair of roots persists for sufficiently small $|\delta\omega|$, but beyond a certain threshold the number of isolated roots changes as we now demonstrate. The continuum limit of \eref{rtdis} reads
\beg
\frac{2}{ u-\mu_i\mp i\Delta_{0i} }\frac{\delta\omega}{\gamma}+\int_0^\infty \frac{f(\eps)d\eps}{(u-\eps)E_i(\eps)}=\frac{4}{\gamma},
\label{isolated}
\en
where as always we measure energies in units of $\eps_F$ and $f(\eps)$ is the dimensionless density of states   defined in \eref{dos}.

As $\delta\omega$ is decreased or increased the single pair of roots can collapse to the real axis or a new pair of isolated roots can emerge from it. The threshold (critical) value of $\delta\omega$ when this occurs is determined by looking for roots of \eref{isolated} with an infinitesimal imaginary part. Replace $u\to u\pm i\delta$ in \eref{isolated} and use $(u-\eps\pm i\delta)^{-1}=P(u-\eps)^{-1}\mp i\pi\delta(u-\eps)$ to separate its real and imaginary parts. The latter yields  critical values of $\delta\omega$ when the number of roots changes
\beg
\frac{|\delta\omega|}{\gamma}= \frac{\pi  f (u)  E_i(u)}{2\Delta_{0i}} ,
\label{beta}
\en
where $u$ is real positive and obtains from the real part of \eref{isolated}
\beg
\dashint_0^\infty \frac{f(\eps)d\eps}{(u-\eps) E_i(\eps)}+\mbox{sgn$(\delta\omega)$}\frac{\pi(u-\mu_i)f(u)}{E_i(u)\Delta_{0i}}=\frac{4}{\gamma}.
\label{realroot}
\en
Dashed integral indicates principal value.

Last two equations determine  critical lines in  quench phase diagrams shown in Figs.~\ref{psd2ch2d}, \ref{psd2ch3d}, \ref{psd2D}, and \ref{psd3D}.  
 We construct  the diagrams in $\bigl(\Delta_{0f}, \Delta_{0i}\bigr)$ plane -- ground state gaps at final and initial detunings $\omega_i$ and $\omega_f$.   The resonance width (dimensionless interaction strength) $\gamma$ is fixed throughout the diagram.  $\Delta_{0i}$, $\Delta_{0f}$, and $\gamma$ uniquely determine $\mu_i$, $\omega_i$, and $\omega_f$ through ground state \esref{mucont} and \re{gapeqcont}.
Each point in this plane represents a particular quench of the detuning $\omega_i\to\omega_f$.   We choose $\Delta_{0i}$ (or equivalently the ratio $\mu_i/\Delta_{0i}$) and the sign of $\delta\omega$ and solve \eref{realroot} for real $u$. \eref{beta} then yields the final detuning $\omega_f$ and therefore $\Delta_{0f}$. We thus obtain a critical line, $\Delta_{0f}$ as a function of $\Delta_{0i}$, in the $\bigl(\Delta_{0f}, \Delta_{0i}\bigr)$ plane. The number of isolated root pairs changes by one as one crosses this line. 

It turns out there is one critical line for either sign of $\delta\omega$. There are therefore three nonequilibrium phases or regimes -- qualitatively different long time behaviors, indicated as regions I, II (including subregion II'), and III in Figs.~\ref{psd2ch2d}, \ref{psd2ch3d}, \ref{psd2D}, and \ref{psd3D}. Region II  contains the $\Delta_{0f}=\Delta_{0i}$ or, equivalently, $\omega_f=\omega_i$ line which corresponds to no quench, i.e. to the system remaining in the ground state at all times. Therefore, in region II \eref{isolated} yields a single pair of isolated complex roots $u=\mu_\infty\pm i\Delta_\infty$. This in turn implies that $\Delta(t)\to \Delta_\infty e^{-2i\mu_\infty t-2i\varphi}$ as $t\to\infty$. For all quenches in region II the system thus goes into the asymptotic state described in Section~\ref{asym0}. 

\begin{figure}[htb]
\includegraphics[scale=0.26,angle=0]{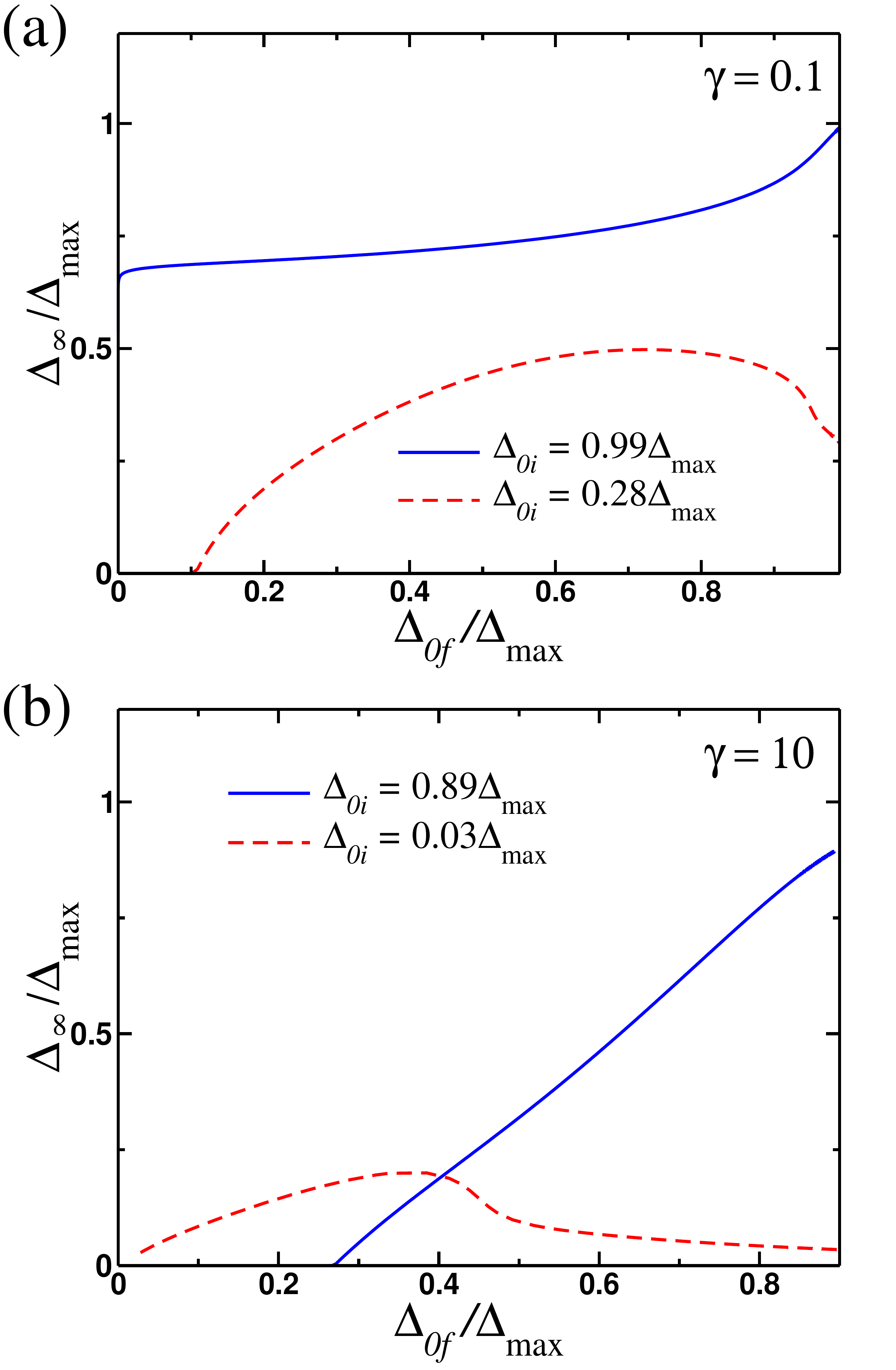}
\caption{(color online)  $\Delta(t)\to \Delta_\infty e^{-2i\mu_\infty t-2i\varphi}$ as $t\to\infty$ after a detuning quench $\omega_i\to\omega_f$ in 3d 2-channel model in region II of the quench phase diagram in Fig.~\ref{psd2ch3d}.    $\Delta_\infty$ extracted from the single isolated root pair of the Lax vector norm is shown as a function  of $\Delta_{0f}$ (ground state gap for  $\omega_f$) at two fixed values of $\Delta_{0i}$ (ground state gap for the initial detuning $\omega_i$). Note   that $\Delta_\infty> \Delta_{0f}$ for BEC to BCS quenches 
$\Delta_{0i}=0.99\Delta_{\max}$ for $\gamma=0.1$.}
\label{delinfpl}
\end{figure}

\begin{figure}[htb]
\includegraphics[scale=0.26,angle=0]{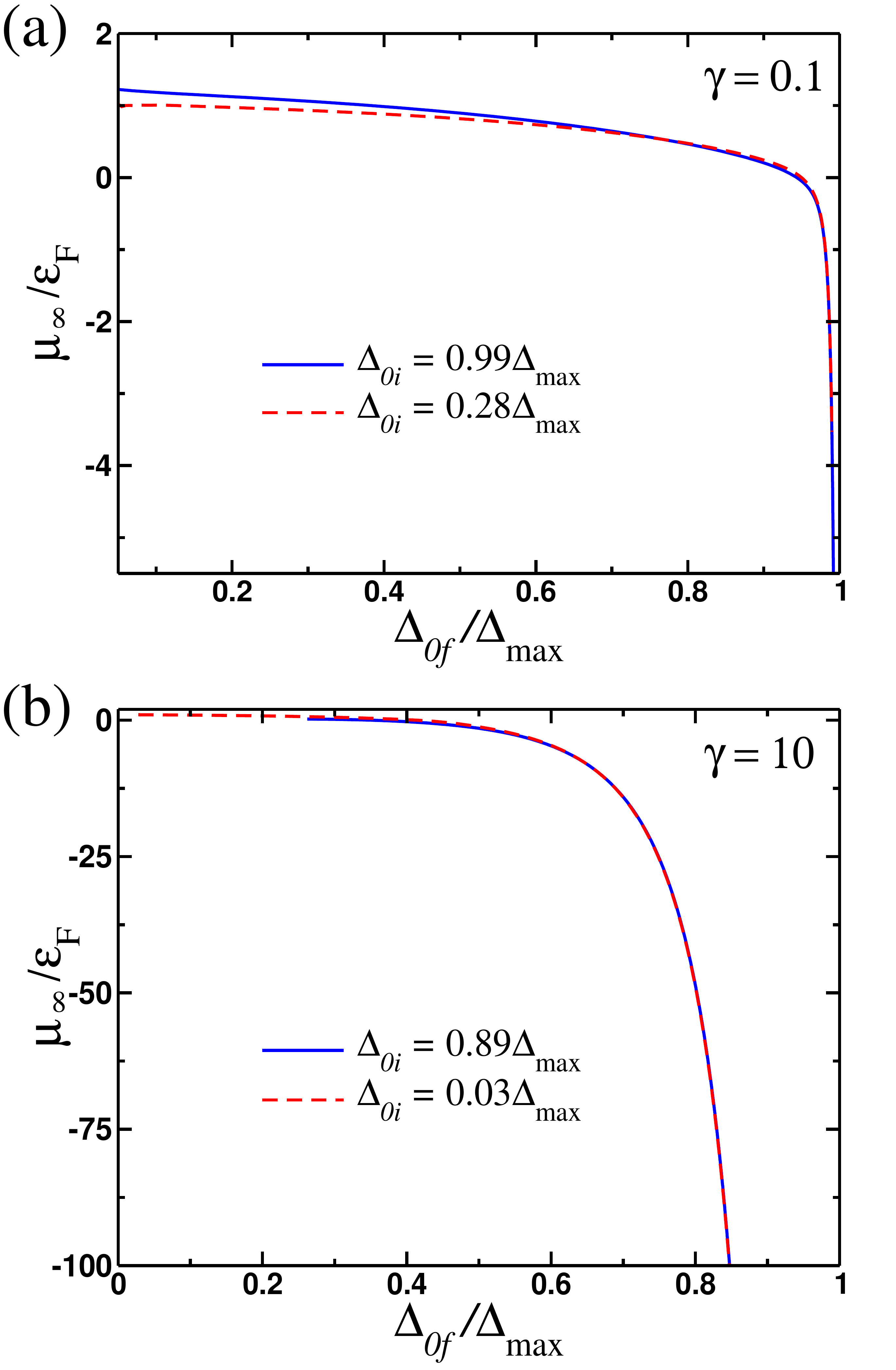}
\caption{(color online)  $\Delta(t)\to \Delta_\infty e^{-2i\mu_\infty t-2i\varphi}$ as $t\to\infty$ after a detuning quench $\omega_i\to\omega_f$ in 3d 2-channel model in region II of the quench phase diagram in Fig.~\ref{psd2ch3d} where $\mu_\infty$ plays the role of the out of equilibrium analog of the chemical potential. Here $\mu_\infty$ is extracted from the single isolated root pair of the Lax vector norm and is shown as a function  of $\Delta_{0f}$ (ground state gap for  $\omega_f$) at two fixed values of $\Delta_{0i}$ (ground state gap for the initial detuning $\omega_i$).  Note that $\mu_\infty$ behaves similarly to the ground state chemical potential in Fig.~\ref{mudelta}.}
\label{muinfpl}
\end{figure}

\begin{figure}[htb!]
\includegraphics[scale=0.26,angle=0]{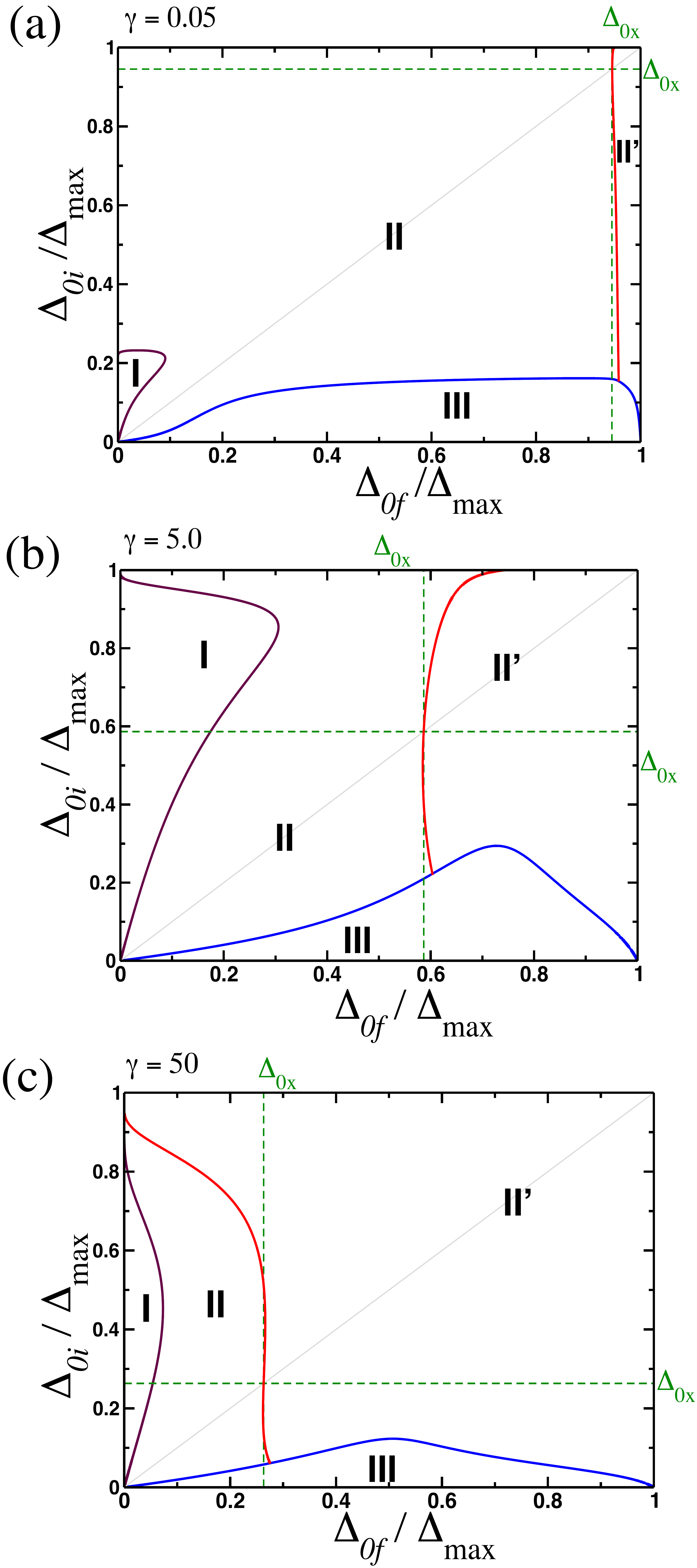}
\caption{(color online) Detuning  quench phase diagrams for 2-channel model   in 2d for  various resonance widths $\gamma$ obtained from \esref{beta2d} and \re{realroot2d}. Each point represents a single quench labeled by $\Delta_{0i}$   and $\Delta_{0f}$  -- pairing gaps the system would have in the ground state for initial and final detunings. At large times the system ends up in one of three steady states shown as regions I, II (including II'), and III.  For quenches in region I the order parameter vanishes. In II
  $\Delta(t)\to\Delta_\infty e^{-2i\mu_\infty t-2i\varphi}$ and in III $|\Delta(t)|$ oscillates persistently. Subregions II and II' differ in the sign of  $\mu_\infty$ (out of equilibrium analog of the chemical potential): $\mu_\infty>0$ in II
 and $\mu_\infty<0$ in II'. The diagonal, $\Delta_{0i}=\Delta_{0f}$, is the no quench line.   $\Delta_{0\times}$ is the ground state gap corresponding to zero chemical potential, i.e. $\Delta_{0\times}$ is given by \eref{mucont} for $\mu=0$.}
\label{psd2D}
\end{figure}

Negative $\delta\omega$ corresponds to $\Delta_{0f}>\Delta_{0i}$. As we cross the critical line going from region II into region III the number of isolated root pairs changes by one. It can be shown both analytically and numerically by analyzing \eref{isolated} that this number increases, i.e. there are two pairs of complex conjugate isolated roots in region III. For quenches in this part of the diagram the large time asymptote of $\Delta(t)$ is given by \eref{1spindeltafull} and the large time state of the system is that obtained in Section~\ref{asym1}.  Plots of $\Delta_\infty$ and $\mu_\infty$  as functions of $\Delta_{0f}$ at two fixed values of $\Delta_{0i}$ are shown in Figs.~\ref{delinfpl} and \ref{muinfpl}.

Similarly, as we enter  region I from region II, $\Delta_\infty\to 0$ and the single pair of isolated roots collapses to the real axis at the critical line. There are hence no isolated roots in region I and therefore $\Delta(t)\to 0$ for quenches in this regime and the system goes into the gapless steady state detailed at the beginning of Section~\ref{asym-1}.

Of interest is the line along which the real part of the root pair $\mu_\infty\pm i\Delta_\infty$ in region II vanishes, i.e. $\mu_\infty=0$ (the line separating subregions II and II' in quench phase diagrams). This can be thought of as a nonequilibrium extension of the BCS-BEC crossover going from a positive to a negative chemical potential. Out of equilibrium, as we will see below, the change of sign of $\mu_\infty$ affects the approach of $\Delta(t)$ to its asymptote. For example, in 3d  the approach changes from $1/t^{1/2}$ in II to $1/t^{3/2}$ in II'. Setting $u=\pm i\Delta_\infty$ in \eref{isolated} and separating the real and imaginary parts, we obtain equations determining this line
\beg
\begin{split}
\frac{\mu_i}{\Delta_{0i}-\Delta_\infty} \mathrm{Im\,}{\cal F}+ \mathrm{Re\,}{\cal F} = \frac{4}{\gamma} \\
\\
\frac{\delta\omega}{\gamma}\frac{2(\Delta_\infty-\Delta_{0i})}{\mu_i^2+(\Delta_\infty-\Delta_{0i})^2}= \mathrm{Im\,}{\cal F},\\
\end{split}
\label{muinf0}
\en
where
\beg
{\cal F}=\int_0^\infty \frac{f(\eps) d\eps}{(i\Delta_\infty-\eps) E_i(\eps)}.
\en
 \eref{muinf0} determines the $\mu_\infty=0$ line via a procedure similar to that for critical lines separating region I from II and II from III. For a given $\Delta_{0i}$, the first equation yields $\Delta_\infty$. We then find $\delta\omega$ and consequently $\omega_f$ and $\Delta_{0f}$ from the second equation. 
 
 Note the intersection of  the $\mu_\infty=0$ line with the $\Delta_{0i}=\Delta_{0f}$ (no quench) line. Along the latter line we also have $\Delta_\infty=\Delta_{0i}$ and therefore at the intersection point $\mu_i=\mu_f=0$ or the first term in the first equation in  \eref{muinf0} would blow up.  In equilibrium $\mu=0$ corresponds to a certain ground state gap $\Delta_0=\Delta_{0\times}$, which obtains from \eref{mucont} and provides a characteristic energy scale for the crossover from the BCS to BEC regime.
 Vanishing
 of $\mu_i$ and $\mu_f$ at the intersection point implies that  straight lines $\Delta_{0i}=\Delta_{0\times}$,  $\Delta_{0f}=\Delta_{0\times}$, and $\Delta_{0i}=\Delta_{0f}$ and the $\mu_\infty=0$ line must cross at the same point, which is indeed seen in all quench phase diagrams in Figs.~\ref{psd2ch2d}, \ref{psd2ch3d}, \ref{psd2D}, and \ref{psd3D}.

Let us  also obtain an explicit expression for the asymptotic spin distribution function \eref{jump1} in all three regimes.
\eref{lt0} implies
\beg
\vec L^2 (u)=\Delta_{0i}^2
L_0^2(u)+\left[ (u-\mu_i)L_0(u)+ \frac{\delta\omega}{g^2}\right]^2.
\en
In the thermodynamic limit
\beg
  L_0(u)=-\frac{2}{g^2}+\int_0^\infty\!\!\frac{f(\eps)d\eps}{2(u-\eps)E_i(\eps)},
\en
We evaluate $L_0(\eps_\pm)$ using $(\eps-\eps'\pm i\delta)^{-1}=P(\eps-\eps')^{-1}\mp i\pi\delta(\eps-\eps')$. This results in
\beg
\begin{split}
\cos\theta(\eps)=\frac{z(\eps)}{i\pi f(\eps)}  \sqrt{A_-^2\Delta_{0i}^2+\left[(\eps-\mu_i)A_-+  \frac{\delta\omega}{\gamma}  \right]^2} \\
 -  \frac{z(\eps)}{i\pi f(\eps)}  \sqrt{A_+^2\Delta_{0i}^2+\left[(\eps-\mu_i)A_+ + \frac{\delta\omega}{\gamma}  \right]^2} ,  \\
\end{split}
\label{distAA}
\en
where
\beg
A_\mp = - \frac{2}{\gamma} \pm\frac{ i\pi  f(\eps)}{2E_i(\eps)}+\dashint_0^\infty\!\!\frac{  f(\eps') d\eps'}{2(\eps-\eps')E_i(\eps')}.
\en 
 The integral here is the same as in \eref{realroot}. We evaluate it in elementary functions in 2d, in weak coupling BCS regime, and in  BEC regime
in Sects.~\ref{phd2d} and \ref{finitegamma3d} below, see also \esref{ge} through \re{gesc} for explicit expressions. Note $\cos\theta(\eps)=1$ for $\delta\omega=0$ (no quench) as it should. Representative plots of  the spin distribution function for two quenches appear in Fig.~\ref{dist_plot}. For future use we also write down the first two terms in large $\eps$ expansion of \eref{distAA}
\beg
\cos\theta(\eps)\approx 1-\left(\frac{\delta\omega}{\gamma}\right)^2\frac{2\Delta_{0i}^2}{E_i^2(\eps)[ H^2(\eps)+\pi^2f^2(\eps)]},
\label{first2}
\en
which are also independently the first two terms in its small $\delta\omega$ expansion. The function $H(\eps)$ is defined in \eref{xkkk}.

Next, we consider two and three dimensions separately as well as various special cases such as wide (single channel limit) and narrow resonance, deep BCS and  BEC regimes.

\subsection{2d}
\label{phd2d}

In 2d the dimensionless density of states $f(\eps)=1$ and all integrals above in this section can be evaluated in terms of elementary functions. It is convenient to introduce a notation
\beg
x=\frac{\mu_i}{\Delta_{0i}},\quad v=\frac{u-\mu_i}{\Delta_{0i}}.
\label{not1}
\en
\eref{isolated} reads
\beg
\begin{split}
\ln\left[-\frac{(v+x)(v+\sqrt{1+v^2})}{\sqrt{1+x^2}\sqrt{1+v^2}-x v+1}\right]=\\
-\frac{2\delta\omega(v\mp i)}{\gamma\sqrt{1+v^2}}+\frac{4\Delta_{0i}}{\gamma}\sqrt{1+v^2}.\\
\end{split}
\label{isolated2d}
\en

The critical lines separating the three asymptotic regimes are determined by \esref{realroot} and \re{beta}, which become 
\beg
\frac{|\delta\omega|}{\gamma}= \frac{\pi}{2} \sqrt{1+v^2},
\label{beta2d}
\en
\beg
\begin{split}
   \ln\left[\frac{(v+x)(v+\sqrt{1+v^2})}{\sqrt{1+x^2}\sqrt{1+v^2}-x v+1}\right]=\\
   -\mbox{sgn$(\delta\omega)$}\pi v+\frac{4\Delta_{0i}}{\gamma}\sqrt{1+v^2},\\
   \end{split}
\label{realroot2d}
\en
where $v$ is real and $v> -x$. It is straightforward to analyze \eref{realroot2d} graphically and to find $v$ and thus the critical lines numerically. 

Positive $\delta\omega$ mean $\Delta_{0f}>\Delta_{0i}$ and the corresponding $v$ determine the critical line separating regions I and II. In this case,  for $\gamma$ above a certain threshold $\gamma_c$ to be determined below, there is  a single root for any $\Delta_{0i}$. This means that a horizontal $\Delta_{0i}=\mbox{const}$ line intersects the  I-II line once for any value of the const and region I therefore extends all the way up to $\Delta_{0i}=\sqrt{\gamma}=\Delta_{\max}$ as seen in Figs.~\ref{psd2ch2d}(c), \ref{psd2D}(b) and \ref{psd2D}(c). When $\gamma<\gamma_c$, the number of roots for positive $\delta\omega$ changes from one to two and then to zero as $\Delta_{0i}$ increases. The I-II  line then displays peculiar reentrant behavior, see the inset in Fig.~\ref{psd2ch2d}(b). 

Negative $\delta\omega$ mean $\Delta_{0f}<\Delta_{0i}$. The roots $v$ in this case yield the II-III critical line. There are two roots for $\Delta_{0i}$ below a certain threshold and no roots above it, implying that a horizontal $\Delta_{0i}=\mbox{const}$ line intersects the II-III critical line twice for a sufficiently small value of the const. 

 The shape of the critical lines as well as the complex roots of \eref{isolated2d} can be determined analytically when the initial and/or final value of the detuning $\omega$ is deep in the BCS or BEC regime.  The BCS limit corresponds to detuning $\omega\to+\infty$.  For the ground state  this implies $\mu\to \eps_F=1$, $\Delta_{0}\to 0$. The gap equation   \re{gapeqcont} then yields
\beg
\ln\frac{4\eps_\Lambda}{\Delta_0^2}=\frac{2\omega-4}{\gamma}.
\label{BCSgapeq}
\en
Deep BEC regime obtains when $\omega\to -\infty$. In this case $\mu\to-\infty$ in the ground state. The gap and chemical potential equations switch roles in the sense that the former determines the chemical potential and the latter the ground state gap.  \eref{gapeqcont} becomes
\beg
\ln\frac{\eps_\Lambda}{|\mu|}=\frac{2\omega+4|\mu|}{\gamma}
\label{BECgapeq}
\en
and \eref{mucont} reads in this limit
\beg
\Delta_0=\left(\frac{1}{\gamma}+\frac{1}{4|\mu|}\right)^{-1/2}.
\label{BECmu}
\en

First, we consider quenches originating deep in the BCS regime, i.e. $\omega_i\to+\infty$ and therefore $\Delta_{0i}\to 0$, $\mu_i\to1$.
Such initial states correspond to $x\to+\infty$.   \eref{realroot2d} becomes
\beg
    \ln\left[\frac{(v+x)(\sqrt{1+v^2}+v)}{x(\sqrt{1+v^2}- v)}\right]=\\
   -\mbox{sgn$(\delta\omega)$}\pi v.\\
   \label{realroot2dBCS}
\en
The roots are: $v\to 0$ for either sign of $\delta\omega$ and $v\to -x+0$ for $\delta\omega <0$. This translates into
\beg
u\approx
\left\{
\begin{array}{l}
\mu_i,\quad \delta\omega>0,\\
\mbox{$\mu_i$ or +0},\quad \delta\omega<0.\\
\end{array}
\right.
\label{bcsroots}
\en
For $v\to 0$ \eref{beta2d} yields $\delta\omega/\gamma=\pm \pi/2$. Therefore, both $\Delta_{0f}$ and $\Delta_{0i}$ are deep in the BCS regime. The gap equation \eref{BCSgapeq} implies $\Delta_0\propto \exp(-\omega/\gamma)$ and hence
\beg
\frac{\Delta_{0i}}{\Delta_{0f}}=e^{\pm \pi/2}.
\label{coup}
\en
This result has been already obtained in Refs.~\onlinecite{Barankov2006,Dzero2006}, which studied quenches within the single channel model in  the weak coupling (BCS) limit.   Weak coupling means small $\Delta_{0i}$ and $\Delta_{0f}$, which corresponds to a vicinity of the origin, $\Delta_{0i}=\Delta_{0f}=0$,  in our phase diagrams. \eref{coup} is the slope of the I-II and II-III critical lines at the origin in Figs.~\ref{psd2ch2d}, \ref{psd2ch3d}, \ref{psd2D}, \ref{psd3D}. 

As we will see below,  \eref{coup} also holds in 3d. This is expected on general grounds because in the BCS limit superconducting correlations  come from a narrow energy window   around the Fermi energy. Main contribution to integrals determining the roots comes from these energies. The density of states is then well approximated by a constant rendering the 2d and 3d cases equivalent.  

The second root at $\delta\omega <0$, $v\to -x+0$, yields $\delta\omega/\gamma\approx -\pi x/2$. This means that the initial state is deep in the BCS regime, while $\omega_f\to-\infty$  and the ground state at $\omega_f$ is in the BEC limit. Further, $\mu_i\to \eps_F=1$, so $x\approx 1/\Delta_{0i}$.
Subtracting  \eref{BCSgapeq} from \eref{BECgapeq}, we obtain
\beg
\ln\frac{\Delta_{0i}^2}{4|\mu_f|}=-\frac{\pi}{\Delta_{0i} } +\frac{4|\mu_f|}{\gamma}+\frac{4}{\gamma}.
\label{fislope}
\en
Here we assume $\gamma$ is finite and treat the single channel limit $\gamma\to\infty$ separately below. Since
$1/\Delta_{0i}$ term diverges much faster then the logarithm in the above equation, we get $4|\mu_f|\approx \pi\gamma/\Delta_{0i}$. \eref{BECmu} now obtains
\beg
\frac{\Delta_{0f}}{\Delta_{\max}}=1-\frac{\Delta_{0i} }{2\pi},
\label{slope}
\en
This equation shows that the II-III critical line terminates at $(\Delta_{0f}, \Delta_{0i})=(\Delta_{\max}, 0)$ linearly with a slope $\Delta_{0i}/(\Delta_{0f}-\Delta_{\max})=-2\pi/\sqrt{\gamma}$.

Simpler  expressions can also be derived for complex roots for quenches within the BCS regime, i.e. in the vicinity of the of the origin in the phase diagrams. By \eref{bcsroots} the real parts of the roots in this regime $\Re[u]\approx\mu_i\approx\eps_F$. Then, $v$ is purely imaginary and also $|v|\ll x$ because $\Im[u]$ is related to the asymptotic value of order parameter amplitude, which is much smaller then $\eps_F$. \eref{isolated2d} becomes
\beg
\ln\left[\frac{v+\sqrt{1+v^2}}{ v-\sqrt{1+v^2}}\right]=-\frac{v\mp i}{\sqrt{1+v^2}}\frac{2\delta\omega}{\gamma}.
\label{smallv}
\en
This equation is symmetric with respect to complex conjugation and with respect to $v\to -v$. The latter symmetry reflects emergence of the particle-hole symmetry in the BCS limit. Note that when there is only one root, these two symmetries together require that it be purely imaginary.

Let $v=-i\cosh\phi$ in \eref{smallv}, where $\phi$ is either purely real or purely imaginary, so that $v$ is purely imaginary. 
\eref{smallv} yields depending on the sign choice on the right hand side
\begin{eqnarray}
\phi=-\frac{\delta\omega}{\gamma}\coth\left(\phi/2\right) \label{coth1},\\
\phi=-\frac{\delta\omega}{\gamma}\tanh\left(\phi/2\right) \label{tanh1}.\\
\nonumber
\end{eqnarray}
Note that in this regime $\delta\omega/\gamma=\ln(\Delta_{0i}/\Delta_{0f})$. It is straightforward to analyze these equations
graphically and to determine when they have solutions. We summarize the results.

\textit{Region I}: $\Delta_{0i}/\Delta_{0f}> e^{\pi/2}$. There are no isolated roots and hence $\Delta(t)\to 0$ at large times.

\textit{Region II}: $e^{-\pi/2}<\Delta_{0i}/\Delta_{0f}<e^{\pi/2}$. There is a single pair of isolated roots at $\mu_\infty\pm i\Delta_\infty$,
\beg
\mu_\infty=\eps_F,\quad \Delta_\infty=\Delta_{0i}\cosh\phi,
\label{qwrty}
\en
where $\phi$ is real for $\delta\omega<0$ and imaginary for $\delta\omega>0$ and is the solution of \eref{coth1}. One can show $\Delta_\infty\le\Delta_{0f}$ for any $\delta\omega$, where the equality is achieved only at $\delta\omega=0$. The long-time dynamics is that described in Section~\ref{asym0}.

It is instructive to evaluate  $\Delta_\infty$, the asymptotic value of the magnitude of the gap, for infinitesimal quenches, when $|\Delta_{0f}-\Delta_{0i}|\ll\Delta_{0i}$. Expanding \esref{coth1} and \re{qwrty} in small $\phi$, we obtain after some calculation
\beg
\Delta_\infty=\Delta_{0f}-\frac{(\Delta_{0f}-\Delta_{0i})^2}{6\Delta_{0f}}.
\label{old1}
\en
Note that within linear analysis $\Delta_\infty=\Delta_{0f}$. As we show in Sect.~\ref{linear}, this is a general feature of    linearized dynamics around the ground state regardless of coupling strength or initial conditions: $|\Delta(t)|$ tends to its ground state value corresponding to the Hamiltonian with which the system evolves at $t>0$.

\textit{Region III}: $\Delta_{0i}/\Delta_{0f}< e^{-\pi/2}$. There are two pairs of complex conjugate roots,
\beg
\eps_F\pm i\Delta_{0i}\cosh\phi_1,\quad \eps_F\pm i\Delta_{0i}\cosh\phi_2,
\label{III}
\en
where $\phi_1$ is the solution of \eref{coth1} and $\phi_2$ is the solution of \eref{tanh1}; $\phi_2$  is real when  $\delta\omega/\gamma=\ln(\Delta_{0i}/\Delta_{0f})\le -2$ and imaginary otherwise. We see that the roots are indeed of the form \eref{spform}. The asymptotic state is that of Section~\ref{asym1}, while $\Delta(t)$ takes the simplified form \eref{spdelta}.

Just as \eref{coup} the above results starting with \eref{smallv} are universal in that they hold for quenches within the BCS regime independent of the dimensionality and also hold for the single channel model.

Next, consider quenches originating deep in the BEC, which corresponds to $\mu_i\to -\infty, \Delta_{0i}\to\sqrt{\gamma}$, and $x\to-\infty$. Since $v>-x$ in \eref{realroot2d}, we also have $v\to\infty$ provided a real root exists. \eref{realroot2d} for $\delta\omega>0$ simplifies to
\beg
\ln\left[\frac{v+x}{|x|}\right]=v\left(\frac{4\Delta_{0i}}{\gamma}-\pi\right)
\label{pigamma}
\en
For $4\Delta_{0i}/\gamma<\pi$, there is a single root at $v\to -x$, which corresponds to $u\approx 0$. Since $\Delta_{0i}\le\sqrt{\gamma}=\Delta_{\max}$, the condition $\Delta_{0i}<\pi\gamma/4$ can be fulfilled only if $\gamma> \gamma_c$, where
\beg
\gamma_c=\frac{16}{\pi^2}.
\en
For $\gamma\ge\gamma_c$ \eref{realroot2d} at $\delta\omega>0$ has a single root for any $\Delta_{0i}$ and, in particular, for $\Delta_{0i}\to\Delta_{\max}$. This means that the I-II critical line extends all the way up to  $\Delta_{0i}=\Delta_{\max}$ terminating at $(\Delta_{0i}, \Delta_{0f})=(\Delta_{\max}, 0)$. 

It is interesting to work out the shape of the I-II critical line near its termination point. First, let $\gamma>\gamma_c$. Since $v\approx -x$, \eref{beta2d} implies $\delta\omega/\gamma\approx \pi|x|/2$. Using \esref{BECgapeq} and \re{BECmu} to determine $\Delta_{0i}$ and $\mu_i$ and \eref{BCSgapeq} for $\Delta_{0f}$, we get
\beg
\begin{split}
\frac{\Delta_{0f}}{\Delta_{\max}}=\frac{1}{\sqrt{2\eps}}\exp\left(-\frac{\alpha}{2\eps}\right),\\ 
\eps=\frac{\Delta_{\max}-\Delta_{0i}}{\Delta_{\max}}, \quad \alpha=\sqrt{\frac{\gamma}{\gamma_c}}-1.\\
\end{split}
\label{shapeabove}
\en
This behavior is seen  in Figs.~\ref{psd2ch2d}(c), \ref{psd2D}(b) and \ref{psd2D}(c). Note the difference between $\gamma=5$ and $\gamma=50$ in Figs.~\ref{psd2D}(b) and \ref{psd2D}(c) that correspond to $\alpha\approx 0.8$ and $\alpha\approx 4.6$, respectively.

Next, let $\gamma<\gamma_c$. In this case, the I-II critical line goes up,   then bends backward reaching a maximum, goes down, and terminates on the $\Delta_{0i}$ axis below $\Delta_{\max}$, see e.g. the inset in Fig.~\ref{psd2ch2d}(b). Near the termination point $\mu_i$ and $\omega_i$ are finite since $\Delta_{0i}< \Delta_{\max}$, while $\omega_f\to\infty$ since $\Delta_{0f}\to 0$. 
\eref{beta2d} implies $v\to\infty$ and $\delta\omega/\gamma\approx \pi v/2$. In this limit \eref{realroot2d} becomes
\beg
\ln\left[\frac{2v}{\sqrt{1+x^2}-x}\right]=v\left(\frac{4\Delta_{0i}}{\gamma}-\pi\right).
\label{neart}
\en
We see that $v$ diverges as $\Delta_{0i}\to \pi\gamma/4=\pi\sqrt{\gamma}\Delta_{\max}/4\equiv \Delta_{\mathrm{th}}$. Therefore, the I-II critical line terminates at $(\Delta_{0i}, \Delta_{0f})=(\Delta_{\mathrm{th}}, 0)$. 
For $\Delta_{0i}$ above $\Delta_{\mathrm{th}}$ and below a certain upper value, which we do not determine explicitly, \eref{realroot2d} has two roots. For $\Delta_{0i}$ below $\Delta_{\mathrm{th}}$ there is one root.

The shape of the I-II critical line as it approaches the termination point for $\gamma<\gamma_c$ obtains from \eref{neart}.
Let $\eps=(\Delta_{0i}- \Delta_{\mathrm{th}})/\Delta_{\mathrm{th}}\ll 1$. \eref{neart} implies 
\beg
v\approx \frac{1}{\pi\eps}\ln\left[\frac{2(\sqrt{1+x^2}+x)}{\pi\eps}\right].
\en
 The gap equation \re{gapeqcont} yields in 2d
\beg
\sqrt{1+x^2}+x=\frac{2(\gamma-\Delta_{0i}^2)}{\gamma\Delta_{0i}}.
\en
Since $\omega_f\to\infty$ corresponds to the BCS limit, we have $\Delta_{0f}\propto e^{-\omega_f/\gamma}\propto e^{-\pi v/2}$. Combining this with the last two equations and using $\Delta_{0i}\approx \Delta_{\mathrm{th}}=\pi\gamma/4$, we get
\beg
\Delta_{0f}=C \exp\left(-\frac{1}{2\eps}\ln \left[ \frac{\gamma_c-\gamma}{\gamma\eps}\right]\right),
\label{shapebelow}
\en
where $C$ is independent of $\eps$.

The I-II critical line for   $\gamma <\gamma_c$ is shown in Figs.~\ref{psd2ch2d}(a), \ref{psd2ch2d}(b), and \ref{psd2D}(a) that correspond to $\Delta_{\mathrm{th}}/\Delta_{\max}\approx 0.25, 0.78,$ and 0.18, respectively.   $\Delta_{\mathrm{th}}$ appears somewhat larger in these  plots since exponentially small, but finite $\Delta_{0f}$ in
\eref{shapebelow} is not noticeable -- the critical line effectively goes down along the $\Delta_{0i}$ axis. In the same way the I-II critical line appears to terminate below $\Delta_{\max}$ in Fig.~\ref{psd2D} for $\gamma=50$ due to exponential smallness of $\Delta_{0f}$ in \eref{shapeabove}.

\begin{figure}[h]
\includegraphics[scale=0.26,angle=0]{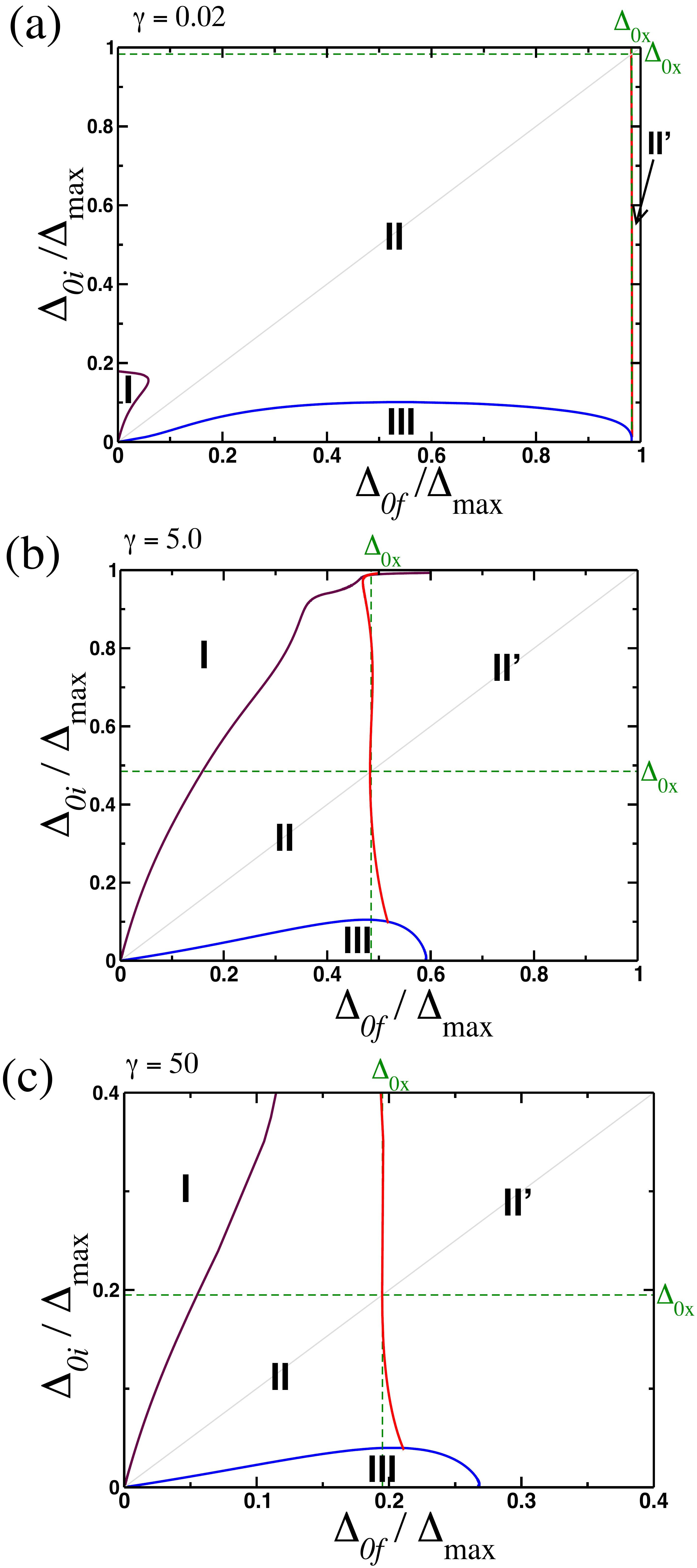}
\caption{(color online) Detuning  quench phase diagrams for 2-channel model   in 3d for  various resonance widths $\gamma$ obtained from \esref{beta3d} and \re{realroot3d}. Otherwise same as Fig.~\ref{psd2D}.}
\label{psd3D}
\end{figure}

\subsection{3d}
\label{finitegamma3d}

3d diagrams for various values of    resonance width $\gamma$ are shown in Figs.~\ref{psd2ch3d} and \ref{psd3D}. Overall they are qualitatively similar to 2d diagrams. A notable difference is that in 3d region III of the oscillating order parameter $\Delta(t)$ for sufficiently large $\gamma$ terminates at $\Delta_{0f}<\Delta_{\max}=\sqrt{2\gamma/3}$.  This means that quenches from infinitesimally weak to sufficiently strong coupling produce no oscillations. Also, in contrast to the 2d case, the
critical line separating the gapless region I in principle always extends all the way up to $\Delta_{0i}=\Delta_{\max}$ and terminates at 
$\Delta_{0f}=\Delta_{0f}^{\mathrm{I-II}}>0$. This is however not noticeable at small $\gamma$ because in this case the value of $\Delta_{0f}^{\mathrm{I-II}}$ is exponentially small.

In 3d the dimensionless density of states $f(\eps)=\sqrt{\eps}$ and \eref{isolated} becomes
\beg
  \int\limits_{-x}^\infty \frac{dy  \sqrt{\Delta_{0i}(x+y)}}{(v-y)\sqrt{y^2+1} }=-\frac{2\delta\omega   }{\gamma(v\pm i)}+\frac{4\Delta_{0i}}{\gamma}, 
\label{v3d}
\en
where $y=\eps/\Delta_{0i}-x$, and $x$ and $v$ are defined in   \eref{not1}. Similarly, \esref{realroot} and \re{beta} determining 
critical lines read
\beg
\frac{|\delta\omega|}{\gamma}= \frac{\pi}{2}  \sqrt{\Delta_{0i}(x+v) (v^2+1)} ,
\label{beta3d}
\en
\beg
   \dashint\limits_{-x}^\infty  \frac{ dy\sqrt{\Delta_{0i}(x+y)} }{(v-y)\sqrt{y^2+1} }+
  \mbox{sgn$(\delta\omega)$}\frac{\pi v  \sqrt{\Delta_{0i}(x+v)}}{\sqrt{v^2+1}} = 
  \frac{4\Delta_{0i} }{\gamma},
\label{realroot3d}
\en
The integral here is a complete elliptic integral. Substitution $y=1/t -x$ reduces it to one of the Carlson elliptic integrals with known asymptotic behaviors in various regimes \cite{Carlson1994,NIST:DLMF}. We however find it more convenient to evaluate the limiting behaviors by a direct analysis of the integral.

First, we consider initial states deep in the BCS regime, i.e. $\omega_i\to+\infty$, which implies $\Delta_{0i}\to 0$, $\mu_i\to 1$, and $x\to 1/\Delta_{0i}\to +\infty$. To evaluate the integral in \esref{v3d} and \re{realroot3d} in this regime, we split the integration range into three intervals: $(-x, -y_\Lambda), (-y_\Lambda, y_\Lambda), (y_\Lambda, \infty)$, where $y_\Lambda$ is such that 
$1\ll y_\Lambda \ll x$. Let the corresponding integrals be $I_1, I_2$ and $I_3$. To the leading order in $1/y_\Lambda$ and $y_\Lambda/x$ we can replace $\sqrt{y^2+1}\to |y|$ in $I_1$, $I_3$   and  $\sqrt{x+y}\to\sqrt{x}$ in $I_2$. The resulting integrals evaluate in terms of elementary functions 
$$
\begin{array}{l}
  I_1+I_3= \dis\frac{2\sqrt{x}}{v}\ln\frac{4x}{y_\Lambda}-\\
 \qquad \qquad \dis\frac{\sqrt{x+v}}{v}\ln \frac{4x(\sqrt{x+v}+\sqrt{x})^2}{y_\Lambda^2-4x(\sqrt{x+v}-\sqrt{x})^2},\\
\\
\dis I_2=\frac{\sqrt{x}}{\sqrt{1+v^2}}
\ln \frac{(\sqrt{1+v^2}\sqrt{1+y_\Lambda^2}+v y_\Lambda)(v+y_\Lambda)}{(\sqrt{1+v^2}\sqrt{1+y_\Lambda^2}-vy_\Lambda)(v-y_\Lambda)},\\
\end{array}
$$
where we used $1\ll y_\Lambda \ll x$ to simplify expressions. The dependence on 
$y_\Lambda$ should of course cancel from $I_1+I_2+I_3$ to the leading order in $1/y_\Lambda$ and $y_\Lambda/x$.

The gap equation \re{gapeqcont} in the BCS regime is handled similarly by splitting the integral into three resulting in
\beg
\frac{\omega}{\gamma}-\frac{2}{\gamma}=\sqrt{\eps_\Lambda}-2+\ln\frac{8}{\Delta_ {0}}.
\label{gap3dbcs}
\en
Suppose the final detuning is also in the BCS regime. The above equation then implies 
\beg
\frac{\delta\omega}{\gamma}=\ln\frac{\Delta_{0i}}{\Delta_{0f}},
\en
same as in 2d. Because $\delta\omega/\gamma$ must remain of order one as $x\to+\infty$, it follows from \eref{beta3d} that $v$ is also of order one for quenches within the BCS regime. Therefore $|v|\ll y_\Lambda$ in the above expressions for $I_1+I_2$ and $I_3$. We obtain $|I_1+I_2|\ll 1$ and
\beg
I_1+I_2+I_3\approx I_3\approx \frac{1}{1+v^2}\ln\left[\frac{v+\sqrt{1+v^2}}{v-\sqrt{1+v^2}}\right].
\en

\eref{v3d} now turns into the 2d \eref{smallv}, \eref{beta3d} yields $|\delta\omega|/\gamma=\pi/2$ and therefore \eref{coup}. Thus quenches within the BCS regime in 3d are identical to those in 2d and all results  from \eref{smallv} to \eref{III} also hold in 3d. As we already commented above, this is expected since in the BCS regime superconductivity comes from the vicinity of the Fermi energy making the dependence of the density of states on the energy and thus the dimensionality inessential. 

\begin{figure}[h]
\includegraphics[scale=0.26,angle=0]{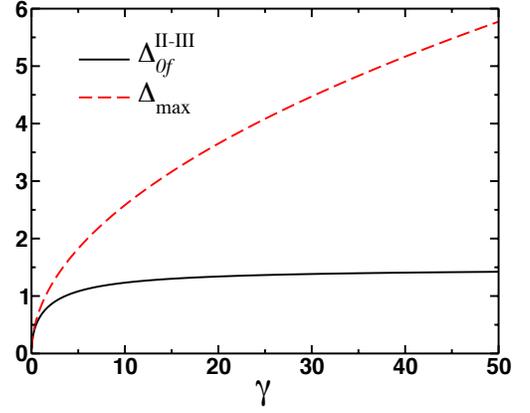}
\caption{(color online) Termination point of the II-III critical line as a function of resonance width $\gamma$ in units of  Fermi energy for 3d 2-channel model. This line encloses region III of persistent oscillations in Figs.~\ref{psd2ch3d} and \ref{psd3D}. It starts at the origin and ends at $\Delta_{0f}^\mathrm{II-III}$ along the $\Delta_{0f}$ axis.  This  reflects an interesting phenomenon: there are no persistent oscillations for quenches to couplings stronger then a certain threshold (i.e. quenches to  detunings $\omega_f$ such that the corresponding ground state gaps  $\Delta_{0f}\ge\Delta_{0f}^\mathrm{II-III}$) no matter how weak the initial coupling is (i.e. for any initial detuning). At $\gamma\to\infty$ (1-channel limit) $\Delta_{0f}^\mathrm{II-III}$ saturates at $1.49\eps_F$ in agreement with \eref{II-III1ch}. }
\label{232ch3D}
\end{figure}

The horizontal $\Delta_{0i}=\mbox{const}$ line for infinitesimal  values of the const intersects
the II-III critical line twice, once near the origin, and the second time near the termination point
of the II-III critical line. The former intersection corresponds to small $v$, as we saw above, and the latter to $v$ of order $x$. To determine the termination point we therefore take $|v|\gg y_\Lambda$ in the above expressions for $I_1+I_3$ and $I_2$. \eref{v3d} becomes
\beg
\begin{split}
\frac{\sqrt{x+v}}{\sqrt{x}}\ln \frac{(\sqrt{x+v}+\sqrt{x})^2}{-(\sqrt{x+v}-\sqrt{x})^2}=\\
-2\left[\frac{\delta\omega}{\gamma}+\ln\frac{8}{\,\,\Delta_{0i}}-\frac{2v\Delta_{0i} }{\gamma}\right]\pm \frac{2i\delta\omega}{v \gamma }.
\end{split}
\label{largev}
\en
The real root of this equation is $v\approx -x\approx -1/\Delta_{0i}$ yielding
\beg
\frac{\delta\omega}{\gamma}=-\ln\frac{8}{\,\,\Delta_{0i}}-\frac{2}{\gamma}.
\en
Combining this with \eref{gap3dbcs}, taking the limit $\Delta_{0i}\to 0$, and plugging into
the gap equation \re{gapeqcont}, we obtain
\beg
4+\frac{4\mu_f}{\gamma}=  \int\limits_0^{\infty}\Biggl[\frac{1}{\eps}-\frac{1}{ \sqrt{(\eps-\mu_f)^2+\Delta_{0f}^2}}\Biggr]\sqrt{\eps} d\eps,
\label{termpt3d}
\en
where we sent the cutoff $\eps_\Lambda$ to infinity. \eref{termpt3d} together with the chemical potential equation \re{mucont}  determine the value of $\Delta_{0f}^\mathrm{II-III}$ where the II-III critical line terminates on the $\Delta_{0f}$ axis. $\Delta_{0f}^\mathrm{II-III}$ is a function of $\gamma$ only, see Fig.~\ref{232ch3D}.

\begin{figure}[h]
\includegraphics[scale=0.26,angle=0]{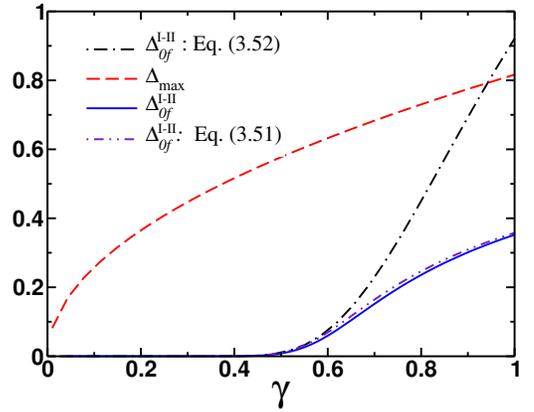}
\caption{(color online) Unlike 2d, in 3d $\Delta_{0f}$ tends to a finite value $\Delta_{0f}^\mathrm{I-II}$ along the I-II critical line as the initial detuning $\omega_i\to-\infty$ ($\Delta_{0i}\to\Delta_{\max}$)  for quenched   2-channel model, see e.g. Fig.~\ref{psd3D}. The gapless regime thus persists even for quenches  from arbitrarily large negative $\omega_i$ to  finite $\omega_f$.  Here we compare $\Delta_{0f}^\mathrm{I-II}$ (in units of the Fermi energy) as a function of the resonance width $\gamma$ extrapolated from actual phase diagrams with that obtained from Eqs.~(3.51) and (3.52). Note that $\Delta_{0f}^\mathrm{I-II}$ is exponentially small at small $\gamma$, so that the I-II critical line \textit{appears} to close earlier at zero $\Delta_{0f}$ in  Fig.~\ref{psd3D}(a). }
\label{termI-II}
\end{figure}

We also note that it follows from the above analysis that, just as in 2d, for initial states deep in the BCS regime there are three roots: $v\to 0$ for either sign of $\delta\omega$ and $v\to -x+0$ for $\delta\omega <0$. Therefore, \eref{bcsroots} holds in 3d as well.

Second, consider quenches from deep  BEC to larger detuning $\omega_f>\omega_i$, i.e. $\omega_i\to-\infty, \delta\omega>0, \mu_i\to -\infty, x\to-\infty, \Delta_{0i}\to\Delta_{\max}$. Since $y\ge |x|\gg 1$ in \eref{realroot3d}, we can replace $\sqrt{y^2+1}\to y$. The principal value integral evaluates to $-\pi\sqrt{|x|}/v$ and \eref{realroot3d} becomes
\beg
-\frac{\pi \sqrt{|x|}}{v}+\pi\sqrt{v-|x|}=\frac{4\sqrt{\Delta_{0i}}}{\gamma},
\en
where we also took into account that we need $v\ge |x|$ so that \eref{beta3d} yields real $\delta\omega$. The solution for large $|x|$ is
\beg
\sqrt{v-|x|}\approx \frac{4\sqrt{\Delta_{0i}}}{\pi\gamma}+\frac{1}{\sqrt{|x|}}.
\en
\eref{beta3d} now yields 
\beg
\frac{\delta\omega}{\gamma}\approx\frac{2|\mu_i|}{\gamma}+\frac{\pi}{2}\sqrt{|\mu_i|}+\frac{32\Delta_{\max}^2}{\pi^2\gamma^3},
\label{deltaobec3d}
\en
where we replaced $\Delta_{0i}^2\to\Delta_{\max}^2=2\gamma/3$ up to terms of order $|\mu_i|^{-1/2}$. The overall correction to this expression is also proportional to $|\mu_i|^{-1/2}$ at large $|\mu_i|$.

Similar simplifications occur in the gap equation \re{gapeqcont}. We replace the square root with $\eps-\mu_i$ to obtain
$$
\frac{\omega_i}{\gamma}\approx\sqrt{\eps_\Lambda}-\frac{2|\mu_i|}{\gamma}-\frac{\pi}{2}\sqrt{|\mu_i|}.
$$
Last two equations determine $\omega_f$ and from the  gap equation \re{gapeqcont} for $\omega=\omega_f$ we obtain
\beg
\frac{128}{3\pi^2\gamma^2}-\frac{4\mu_f}{\gamma}= \int\limits_0^{\infty}\Biggl[\frac{\sqrt{\eps}}{ \sqrt{(\eps-\mu_f)^2+\Delta_{0f}^2}}-\frac{1}{\sqrt{\eps}}\Biggr]  d\eps,
\label{termpt13d}
\en
where we eliminated the cutoff similar to \eref{termpt3d}. This equation combined with \eref{mucont}  determines the termination point $(\Delta_{0i}, \Delta_{0f})=(\Delta_{\max}, \Delta_{0f}^\mathrm{I-II})$ of the I-II critical line. The plot of $\Delta_{0f}^\mathrm{I-II}$ as a function of $\gamma$ is shown in Fig.~\ref{termI-II}.

Note that, in contrast to the 2d case, this critical line formally always extends up to $\Delta_{0i}=\Delta_{\max}$ and $\Delta_{0f}^\mathrm{I-II}$ does not vanish as $\Delta_{0i}\to\Delta_{\max}$. This means that the gapless regime persists even for quenches  to  finite final detunings from initial states lying arbitrarily deep in the BEC regime.  But for small $\gamma$ the value of $\Delta_{0f}^\mathrm{I-II}$ is exponentially small and the critical line \textit{appears} to have closed at smaller $\Delta_{0i}$, see Figs.~\ref{psd2ch3d}(a) and \ref{psd3D}(a). Small $\gamma$ implies large left hand side in \eref{termpt13d} and therefore the final state deep in the BCS regime.  In this regime $\mu_f\to 1$ and
 the  integral in \eref{termpt13d} is  twice the right hand side of \eref{gap3dbcs} without $\sqrt{\eps_\Lambda}$ resulting in
 \beg
 \Delta_{0f}^\mathrm{I-II}=8\exp\left[-\frac{64}{3\pi^2\gamma^2}+\frac{2}{\gamma}-2\right].
 \label{bcsasy}
 \en
We see from Fig.~\ref{termI-II} that $\Delta_{0f}^\mathrm{I-II}$ becomes noticeable for $\gamma\gtrsim 0.45$. For smaller $\gamma$ the gapless region I appears to close at smaller $\Delta_{0i}$ and zero $\Delta_{0f}$.  Fig.~\ref{termI-II} also shows that \eref{bcsasy} provides a reasonable estimate of $\Delta_{0f}^\mathrm{I-II}$ even for large $\gamma$, which will be useful in our analysis of the 1-channel model below.

\section{One channel model}
\label{1channel}

In this section   we collect for reference purposes analogous results for the asymptotic steady state after a quench $\lam_i\to\lam_f$ in the 
one channel model given by \esref{eq:onechannel} and \re{Model}. 

As explained in Sect.~\ref{modelsmf}, the one channel model 
obtains in the broad resonance limit via   replacements
\beg
\frac{\omega}{\gamma}=\frac{\omega}{g^2\nu_F}\to\frac{1}{\lambda},\quad \gamma=g^2\nu_F\to\infty
\en
(in units of $\eps_F$). Our task is to go over  equations of  previous sections performing these replacements. All essential reasoning and methods are
the same.

Chemical potential and gap \esref{mucont} and \re{gapeqcont} now read 
\beg
\frac{4}{ d}= \int_0^{\infty} \biggl(1-\frac{\eps-\mu}{ \sqrt{(\eps-\mu)^2+\Delta_0^2}}\biggr) f(\eps)d\eps,
\label{mucont1ch}
\en
and
\beg
\frac{2}{\lam}= \int_0^{\eps_\Lambda}\frac{f(\eps)d\eps}{ \sqrt{(\eps-\mu)^2+\Delta_0^2}},
\label{gapeqcont1ch}
\en
respectively.

The Lax vector becomes
\beg\label{Lax1ch}
 {\vec L}(u)= \sum\limits_\bp\frac{{\vec s}_\bp}{u-\eps_\bp}-\frac{\hat{\bf z} }{\lam\nu_F}.
 \en
Gaudin algebra, i.e. \esref{fundament} and \re{l2com}, as well as the Lax equation of motion \re{Lmotion} are the same. The numerator of the conserved $\vec L^2(u)$ is now a polynomial of degree $2N$, 
\beg
\vec L^2(u)=\frac{Q_{2N}(u)}{(\lam\nu_F)^2 \prod_\p (u-\eps_\p)^2},
\label{qdef1ch}
\en
where $N$ is the number of nondegenerate  $\eps_\p$. 

Reduced solutions are constructed in the same way with minor modifications. Specifically, the expressions for $\vec L^\mathrm{red}(u)$ in terms of $\vec \sigma_\bp$ and $\vec L_m(u)$ in terms of $\vec t_j$ are replaced in \esref{LaxReduced} and \re{Lm} with the corresponding 1-channel Lax vectors according to \eref{Lax1ch}. The Hamiltonian governing the collective spin variables $\vec t_j$ is
\beg\label{Hm1ch}
H_\mathrm{1ch}^\mathrm{red}=\sum\limits_{j=0}^{m-1}2\eta_j t_j^z -\lam\nu_F\sum\limits_{j,k=0}^{m-1}  t_{j}^{-}t_{k}^{+}.
\en
\esref{st} and \re{d} as well as  constraints \re{const1} are the same, except that the last equation relating $\omega$ and $\omega'$ is absent.
In terms of the $m$-spin spectral polynomial $Q_{2m}(u)$ the constraints become
\beg
\sum_\p\frac{\sigma_\p\eps_\p^{r-1} }{\sqrt{ Q_{2m}(\eps_\p)}}=-\frac{\delta_{rm}}{(\lam\nu_F)^2},\quad r=1,\dots,m.
\en
Further, since the degree of the $m$-spin spectral polynomial is $2m$ rather then $2(m+1)$, an $m$-spin solution of the 2-channel model becomes an 
$(m+1)$-spin solution of the 1-channel model. This name change reflects the fact that the oscillator mode $b$ in the 2-channel model is effectively an additional spin, which was not counted as such.

All remaining equations in Sect.~\ref{method}, i.e. \esref{0sdel} through \re{alf}, are identical for the one channel model, except \eref{Hm1spin} is replaced with \eref{Hm1ch} for $m=2$ and the self-consistency condition \re{sc} is now given by \eref{gap1ch}.

Equations determining  isolated roots, critical lines, and $\mu_\infty=0$ line for the 1-channel model are \eref{isolated}, \esref{realroot} and \re{beta}, and \eref{muinf0}, respectively,  with replacements
\beg
\frac{\delta\omega}{\gamma}\to \frac{1}{\lam_f}-\frac{1}{\lam_i}\equiv \beta,\quad \frac{1}{\gamma}\to 0.
\label{repluu}
\en

Asymptotic spin distribution -- the constant angle the spin $\vec s(\eps)$ makes with the spin $\vec \sigma(\eps)$ in the corresponding $m$-spin solution -- is
\beg
\begin{split}
\cos\theta(\eps)=\frac{z(\eps)}{i\pi f(\eps)}  \sqrt{A_-^2\Delta_{0i}^2+\left[(\eps-\mu_i)A_-+   \delta\beta  \right]^2} \\
 -  \frac{z(\eps)}{i\pi f(\eps)}  \sqrt{A_+^2\Delta_{0i}^2+\left[(\eps-\mu_i)A_+ +  \delta\beta  \right]^2} ,  \\
\end{split}
\label{distAA1ch}
\en
where
\beg
A_\pm =   \pm\frac{ i\pi  f(\eps)}{2E_i(\eps)}+\dashint_0^\infty\!\!\frac{  f(\eps') d\eps'}{2(\eps-\eps')E_i(\eps')}.
\en 
\eref{distAA1ch} is in excellent agreement with the actual spin distribution obtained from direct simulation of spin dynamics\cite{Dzero2006}, see Fig.~3 therein.

\subsection{Quench phase diagram}

Quench phase diagrams for 1-channel model in 2d and 3d are shown in Figs.~\ref{psd1ch2d} and \ref{psd1ch3d}. There is only one diagram in each case extending to positive infinity in both $\Delta_{0i}$ and $\Delta_{0f}$ directions because $\gamma\to\infty$ and therefore $\Delta_{\max}\to\infty$. 

As we commented below \esref{coup}
and \re{III}, the weak coupling part of the diagrams (the region of small $\Delta_{0i}$ and $\Delta_{0f}$ near the origin) is independent of the dimensionality and is exactly the same for 1-channel model. In other words, all results contained in \esref{smallv} through \re{III} and the surrounding text apply to the one channel model in both 2d and 3d; one only needs to replace $\delta\omega/\gamma\to\delta\beta$. 

When either the initial or  final coupling is outside the
deep BCS regime, we need to treat 2d and 3d cases separately.

\subsubsection{2d}

It is straightforward to take the broad resonance limit in \esref{isolated2d} to \re{realroot2dBCS}. In particular, the critical lines are determined by taking this limit in \esref{beta2d} and \re{realroot2d}
\beg
 |\delta\beta|= \frac{\pi}{2} \sqrt{1+v^2},
\label{beta2d1ch}
\en
\beg
\begin{split}
   \ln\left[\frac{(v+x)(v+\sqrt{1+v^2})}{\sqrt{1+x^2}\sqrt{1+v^2}-x v+1}\right]=
   -\mbox{sgn$(\delta\beta)$}\pi v.\\
   \end{split}
\label{realroot2d1ch}
\en
\eref{bcsroots} describing quenches originating in deep BCS remains as is, except the sign of $\delta\omega$ translates into the sign of $\delta\beta$. The two roots $u\approx\mu_i$ for either sign of $\delta\beta$
correspond to quenches also terminating in deep BCS, so they are in the universal regime given by  \esref{smallv} through \re{III}, which is shared  by both models regardless of the dimensionality. 

The analysis for the root $u\approx+0$ at $\delta\beta<0$ leading to \eref{slope} requires some modifications.  $\gamma\to\infty$ limit in \esref{fislope} and \re{BECmu} yields $4|\mu_f|=\Delta_{0i}^2 e^{\pi/\Delta_{0i}}$, $\Delta_{0f}=\sqrt{4|\mu_f|}$ and finally
\beg
\Delta_{0f}=\Delta_{0i} e^{\pi/2\Delta_{0i}},\quad \Delta_{0i}\to0.
\en
This equation gives the asymptotic form of the II-III critical line in the $(\Delta_{0i}, \Delta_{0f})$ plane in Fig.~\ref{psd1ch2d}. We see that this line never terminates in the 2d 1-channel model.

Finally, let us work out the shape of the I-II critical line for large $\Delta_{0i}$, i.e. for quenches originating deep in the BEC regime. \eref{pigamma} becomes
\beg
\ln\left[\frac{v+x}{|x|}\right]=-\pi v.
\en
Now there is always a single root $v\to -x$ ($u\approx 0$). \eref{beta2d1ch} implies 
\beg
\delta\beta=\frac{1}{\lam_f}-\frac{1}{\lam_i}=\frac{\pi|x|}{2}=\frac{\pi|\mu_i|}{2\Delta_{0i}}.
\label{deltabeta}
\en
 We also need the gap equation in BCS and BEC limits and the chemical potential equation in the BEC limit. Sending $\gamma$ to infinity in \esref{BCSgapeq}, \re{BECgapeq} and \re{BECmu}, we obtain 
\beg
\ln\frac{4\eps_\Lambda}{\Delta_{0i}^2}=\frac{2}{\lam_i}, \quad \ln\frac{\eps_\Lambda}{|\mu_f|}=\frac{2 }{\lam_f}, \quad \Delta_{0i}=\sqrt{4|\mu_i|}.
\label{BCSBEC1ch}
\en
Combining these equations with \eref{deltabeta}, we get
\beg
\Delta_{0f}=\Delta_{0i} e^{-\pi\Delta_{0i}/8},\quad \Delta_{0i}\to\infty.
\en
We see that $\Delta_{0f}$ exponentially vanishes along the I-II critical line (gapless regime closes) as $\Delta_{0i}$ increases. The vertical range of Fig.~\ref{psd1ch2d} is not enough to fully display this behavior, though  we see that I-II line does incline towards the $\Delta_{0i}$ axis at large $\Delta_{0i}$.

\subsubsection{3d}

In addition to quenches that fall within the universal weak coupling regime described in   \esref{smallv} to \re{III} and the corresponding text, 
let us derive the termination point of the II-III critical line and analyze the I-II line at large $\Delta_{0i}$.

First, we consider the II-III line. The termination point is given by \eref{termpt3d}. In the $\gamma\to\infty$ limit we have
\beg
4 =  \int\limits_0^{\infty}\Biggl[\frac{1}{\eps}-\frac{1}{ \sqrt{(\eps-\mu_f)^2+\Delta_{0f}^2}}\Biggr]\sqrt{\eps} d\eps.
\label{termpt3d1ch}
\en
Chemical potential equation \re{mucont1ch} provides another relation between $ \mu_f$ and $\Delta_{0f}$.   Numerical solution of these two equations is
\beg
\mu_f^\mathrm{II-III}\approx -1.4602 \eps_F,\quad \Delta_{0f}^\mathrm{II-III}\approx 1.4875\eps_F.
\label{II-III1ch}
\en
 This value of $\Delta_{0f}^\mathrm{II-III}$  agrees with Fig.~\ref{232ch3D}. Unlike 2d, in 3d  region III encloses a finite area, resembling a dome in between the origin and the point $(\Delta_{0i}, \Delta_{0f})=(0, \Delta_{0f}^\mathrm{II-III})$. 

Next, we turn to the critical line separating the gapless region I from region II. For finite $\gamma$ we analyzed the termination point $(\Delta_{0i}, \Delta_{0f})=(\Delta_\mathrm{\max}, \Delta_{0f}^\mathrm{I-II})$  of this line at the end of Sect.~\ref{finitegamma3d}. In the single channel case, $\Delta_{\max} \to\infty$, so the I-II line does not close. As $\Delta_{0i}\to\infty$, the value of $\Delta_{0f}$ for a point on this line tends to $\Delta_{0f}^\mathrm{I-II}$, which is determined by the $\gamma\to\infty$ limit of \eref{termpt13d}, 
\beg
 0= \int\limits_0^{\infty}\Biggl[\frac{1}{ \sqrt{(\eps-\mu_f)^2+\Delta_{0f}^2}}-\frac{1}{\eps}\Biggr] \sqrt{\eps} d\eps,
\label{termpt13d1ch}
\en
together with \eref{mucont1ch}.   The solution of these equations is
\beg
 \mu_f\approx 0.5906 \eps_F ,\quad \Delta_{0f}^\mathrm{I-II}\approx 0.6864 \eps_F.
 \label{bcsasy1ch}
 \en

\section{Transient Dynamics: Linear analysis}
\label{linear}

Here we solve the dynamics for small deviations from the ground state. Linear analysis for the one channel model  in the weak coupling BCS regime was performed  by Volkov and Kogan\cite{Kogan1973}, see also Ref.~\onlinecite{Dzero2006}.  Gurarie\cite{Gurarie2009} extended this study to strongly coupled superconductors.  Both these studies of the linearized dynamics conclude that 
\beg
\Delta(t) \to\Delta_\infty e^{-2i\mu_\infty t-2i\varphi}
\en
 as $t\to\infty$, but the approach to this asymptote is different.
Our analysis adds
several new results to this prior work. We demonstrate that within linear analysis the amplitude of the order parameter  asymptotes to its ground state value for
the Hamiltonian with which the system evolves after  nonequilibrium conditions are created, i.e.
$\Delta_\infty=\Delta_{0f}$, a point that seems to have been  missed by the earlier work. Also, $\mu_\infty=\mu_f$ -- the ground state chemical potential. In other words, $\Delta_\infty-\Delta_0$ and $\mu_\infty-\mu_f$ are second order in the deviation. This is a general result that holds for both one and two channel models and is independent of the type of perturbation that drives the system out of equillibrium. 

Further, we solve linearized equations of motion using the machinery of the exact solution \cite{Altshuler2005,Enolski2005}, which  provides much more detailed information. For example, we also determine the short time behavior, normal modes, full explicit long time form $\Delta(t)$ and individual spins with all  prefactors and phases etc. unavailable to conventional linear analysis.  Note that in quench phase diagrams constructed above small quenches correspond to the vicinity of
the diagonal $\Delta_{0i}=\Delta_{0f}$, see e.g Figs.~\ref{psd2D} and \ref{psd3D}.

\subsection{Asymptotic $\Delta(t)$ and spins}

Consider an infinitesimal quench of the detuning $\delta\omega=\omega_f-\omega_i$. More generally, $\delta\omega$ can be any small parameter that measures the deviation from the ground state in the two or one channel model. We work to linear order in $\delta\omega$.  Suppose  $\Delta(t) \to\Delta_\infty e^{-2i\mu_\infty t-2i\varphi}$. For the detuning or interaction quenches this follows from the few spin conjecture and quench phase diagrams  derived above  and we also  verify it   independently below.  Let us go to a reference frame that rotates with frequency $2\mu_\infty$ around the $z$-axis. In this frame $\Delta(t)=\Delta_\infty$ and the magnetic field $\vec B_\bp=(-2\Delta_\infty, 0, 2\eps_\p-2\mu_\infty)$ acting on spin $\vec s_\bp$ in \eref{Bloch} is time-independent. Note that transformation to the rotating frame results in shifts to $\eps_\p$ and $\omega_f$. Then, the spin rotates around $\vec B_\bp$ making a constant angle $\pi-\theta_\bp$ with it. This is in fact the asymptotic solution described in Sect.~\ref{asym0},
\beg
\vec s_\p(t)=  \frac{\vec n_\p}{2}\cos\theta_\p+\vec s_\p^\perp(t),
\label{ss167}
\en
where $\vec n_\p$ is a unit vector along $-\vec B_\p$,
\beg
 n_p^x=\frac{\Delta_\infty}{E_\p^\infty},\quad n_\p^y=0,\quad n_\p^z=-\frac{\eps_\p-\mu_\infty}{E_\p^\infty}.
 \en
  \eref{Bloch} with $\dot b=0$ further implies $\Delta_\infty=-gb=g^2J_-/(\omega_f-2\mu_\infty)$. The contribution of   $\vec s_\bp^\perp$  to $J_-$ dephases
as $t\to\infty$. The latter is therefore $\sum_\p n_p^x/2$ -- the sum of components of $\vec s_\bp$ along $\vec B_\bp$  projected onto the $xy$-plane,
\beg
\Delta_\infty= \frac{g^2}{\omega_f-2\mu_\infty}\sum_\p\frac{\Delta_\infty\cos\theta_\p}{2\sqrt{(\eps_\p-\mu_\infty)^2+\Delta_\infty^2}}.
\label{argument}
\en
In the ground state $\vec s_\bp$ is aligned with $-\vec B_\bp$, i.e. $\theta_\p=0$. This implies that $\theta_\p$ must be proportional to $\delta\omega$ and therefore corrections to $\cos\theta_\p=1$  are second order in $\delta\omega$. But for $\cos\theta_\p=1$, \eref{argument} is the ground state gap equation \re{gapeq} for $\omega=\omega_f$. 
Moreover, applying the same argument to $J_z$ and \eref{muargument}, we find that $\Delta_\infty$ and $\mu_\infty$ also satisfy the ground state chemical potential equation \re{mu}. It follows that  for small oscillations around the ground state one always has
\beg
\Delta_\infty=\Delta_{0f},  \quad \mu_\infty=\mu_f. 
\label{surprise}
\en
For the same reason  the non-oscillatory part of $\vec s_\p$ (zeroth harmonic) in the steady state is the same as in the ground state at $\omega=\omega_f$, i.e.
is given by \eref{gspin} with $\Delta_0\to\Delta_{0f}$ and $\mu\to\mu_f$.

The same  is true for the one channel model. Note also that infinitesimal quenches in the BCS regime conform to this conclusion, see \eref{old1}. Moreover, this result generalizes to finite spin dynamics, where,
as we show below, zeroth harmonics of $\Delta(t)$ and $\vec s_\bp$ to linear order in $\delta\omega$ coincide with the $\omega=\omega_f$ ground state 
values.

\subsection{Normal modes and finite size dynamics}
\label{normm}

Now we turn to the linear analysis per se.  At this point it is convenient to rewrite summations over $\p$ as  summations over   single particle energies. We adopt the following model of discrete spectrum. Let us discretize the magnitude of the momentum, $p\to p_k$. The corresponding energies are $\eps_k=p_k^2/2m$
 with degeneracy $N_k=N(\eps_k)$ -- the number of states in a momentum shell between $p_k$ and $p_{k+1}$, which is a smooth function of $\eps_k$. 
  The level spacing $\delta_k=\eps_{k+1}-\eps_k$  is also assumed to depend on $\eps_k$ smoothly.  We include  this dependence in $N_k$, so without loss   of generality we take it to be constant, $\delta_k=\delta$.   Our final results depend 
only on the density of states $\nu(\eps_k)=N_k/\delta$ -- the number of states per unit energy.
 Equivalently,  $\eps_i$ can  represent   levels  of some other single particle potential, e.g. a 3d harmonic oscillator potential, see the discussion at the end of Sect.~\ref{modelsmf}.    All quantities and equations, including spins $\vec s_\p$, Hamiltonians, equations of motion and initial conditions, considered in this paper depend on $\p$ only through $\eps_\p$. For any such quantity $A_\p=A(\eps_\p)$
\beg
\sum_\p A_\p=\sum_{k=1}^N N_k A_k \to \int \nu(\eps) d\eps,
\label{sumintrule}
\en
where $A_k=A(\eps_k)$. In particular, the Lax vector \re{Lax} reads
\beg\label{Laxnew}
\begin{split}
{\vec L}(u)= \sum\limits_{k=1}^N\frac{N_k{\vec s}_k}{u-\eps_k}-\frac{(\omega-2\mu)}{g^2}\hat{\bf z}+\\
\frac{2}{g^2}\left[(u-\mu)\hat{\bf z}-\vec{\Delta}\right].
\end{split}
\en

A convenient tool for linear analysis of the dynamics are the separation variables introduced in Refs.~\onlinecite{Enolski2005} and \onlinecite{Altshuler2005} for the one and two channel models, respectively. As we will see, in linearized dynamics these variables are simply the normal modes. 
Separation variables $u_j$ are defined as the solutions of $L_-(u_j)\equiv L_x(u_j)-iL_y(u_j)=0$,   i.e.
\beg
L_-(u)=\frac{2b}{g}+\sum_ {k=1}^N\frac{N_k s_k^-}{u-\eps_k}=0. 
\label{liu}
\en
Because $u=u_j$ are the zeroes of the rational function $L_-(u)$ and $u=\eps_k$ are its poles, we can also write it as
\beg
L_-(u)= \frac{2b}{g} \frac{\prod_j(u-u_j)}{\prod_k(u-\eps_k)}.
\label{liu1}
\en
Matching the residues at $u=\eps_k$ and $u=\infty$ in \esref{liu} and \re{liu1}, we express the spins in terms of $u_j$,
\beg
N_k s_k^-=\frac{2b}{g} \frac{\prod_j(\eps_k-u_j)}{\prod_{m\ne k}(\eps_k-\eps_{m})},
\label{su}
\en
\beg
J_-=\sum_k N_k s_k^-=\frac{2b}{g}\sum_k( \eps_k- u_k).
\label{j-}
\en
Equations of motion   in terms of  new variables are
\beg
\begin{array}{l}
\dis \dot u_k=-\frac{2i \sqrt{Q_{2N+2}(u_k)}}{\prod_{m\ne k}(u_k-u_m)},\\
\\
 \dis  \dot b=-2i b\Biggl(\frac{\omega}{2}+\sum_k( \eps_k- u_k)\Biggr),\\
 \end{array}
\label{dickeev}
\en
see Ref.~\onlinecite{Altshuler2005} for a detailed derivation. Here $Q_{2N+2}(u)$ is the spectral polynomial defined in \eref{qdef}.

Roots of $Q_{2N+2}(u)$ are the same as roots of $\vec L^2(u)$ determined by  \eref{rtdis}. In our new notation
\beg
\bigl(u-\mu \mp i\Delta_{0}\bigr)\biggl[\frac{2}{g^2}-\sum_k \frac{N_k}{2(u-\eps_k)E(\eps_k)}\biggr]=
\frac{\delta\omega}{g^2},
\label{rtdisnew}
\en
where $E(\eps_k)=\sqrt{(\eps_k-\mu)^2+\Delta_0^2}$. Here and everywhere below in this subsection  $\mu$ and $\Delta_0$ without a subscript indicate ground state values $\mu_i$ and $\Delta_{0i}$ for the initial detuning $\omega=\omega_i$. 
 In the ground state $\vec L^2(u)=\left[ (u-\mu_i)^2+\Delta_{0i}^2\right] L^2_0(u)$. There is a pair of complex roots  $c_\pm=\mu_i\pm i\Delta_{0i}$   and $2N$ real double degenerate roots $x_k$ that solve  
\beg
  L_0(x)=-\frac{2}{g^2}+\sum_k\frac{N_k}{2(x-\eps_k)E(\eps_k)}=0,
  \label{l0x}
\en
A plot of $L_0(x)$ reveals that $x_k$ are located between consecutive $\eps_k$, i.e. $\eps_k<x_k<\eps_{k+1}$. 

Since $\vec L^2(x_k)=L_x^2(x_k)+L_y^2(x_k)+L_z^2(x_k)=0$ in the ground state and $x_k$ is real, all components of $\vec L(x_k)$ must vanish, $L_x(x_k)=L_y(x_k)=L_z(x_k)=0$. It follows that $L_-(x_k)=0$ meaning that the separation variables are frozen in the real double roots, $u_k=x_k$. After a quench they start to move from these initial positions, $u_k(t)=x_k+\delta u_k$, where $\delta u_k$ vanishes at $t=0$ and
is proportional to $\delta\omega$ for an infinitesimal quench.
For $\delta\omega\ne 0$  real double roots of $Q_{2N+2}(u)$
split into  pairs of complex conjugate roots $c_k=x_k+\delta c_k$ and $\bar c_k=x_k+\delta \bar c_k$. Therefore, the expression for $Q_{2N+2}(u_k)$,
\beg
\begin{split}
Q_{2N+2}(u_k)=(u_k-c_k)(u_k-\bar c_k) \times\\
(u_k-c_+)(u_k- c_-)
\prod_{m\ne k} (u_k-c_m) (u_k-\bar c_m),\\
\end{split}
\en
to lowest nonzero order in $\delta\omega$ becomes
\beg
\begin{split}
Q_{2N+2}(u_k)=(\delta u_k-\delta c_k) (\delta u_k- \delta \bar c_k) \Omega_k^2\times\\
\prod_{m\ne k} (x_k-x_m)^2,\\
\end{split}
\label{qapp}
\en
$\Omega_k=\sqrt{(x_k-\mu)^2+\Delta_{0}^2}$, not to be confused with function $\Omega(t)$ in Sect.~\ref{1spinsol}. Similarly, the denominator of the equation of motion \re{dickeev} for $u_k$  to the lowest order $\prod_{m\ne k}(u_k-u_m)=\prod_{m\ne k} (x_k- x_m)$, so this equation reads
\beg
\delta \dot u_k=\pm 2i\Omega_k \sqrt{(\delta u_k-\delta c_k) (\delta u_k-\delta \bar c_k)}.
\label{lin}
\en

Corrections to the roots due to the quench obtain by setting $u=x_k+\delta c_k$ in \eref{rtdisnew} and linearizing in $\delta c_k$. Separating  real and imaginary parts, $\delta c_k=a_k+ib_k$,  we have
\beg
 a_k=\frac{\delta\omega (x_k-\mu )}{g^2\Omega_k^2 F_k}, \quad b_k=\frac{\delta\omega  \Delta_0}{g^2\Omega_k^2 F_k},
\label{dck}
\en
where
\beg
F_k=\sum_m \frac{N_k}{2(x_k-\eps_m)^2 E(\eps_m)}.
\label{fk}
\en

Let us also evaluate the correction to the complex root pair $c_\pm=\mu_i\pm i\Delta_{0i}$. Writing the perturbed roots as $\mu'\pm i\Delta'$, we obtain from \eref{rtdisnew} to linear order in $\delta\omega$
\beg
\begin{split}
\mu'-\mu_i=\frac{\delta\omega}{g^2}\frac{\beta_k}{\alpha_k^2+\beta_k^2},\\ 
\Delta'-\Delta_{0i}=-\frac{\delta\omega}{g^2}\frac{\alpha_k}{\alpha_k^2+\beta_k^2},\\
\end{split}
\en
where $\alpha_k$ and $\beta_k$ are defined in \eref{akbk}. Comparing this with first order shifts in the ground state gap and chemical potentials that readily derive from \esref{gapchemnew}, we conclude that
\beg
\mu'=\mu_f, \quad \Delta'=\Delta_{0f},
\label{primef}
\en
as it should be according to Sect.~\ref{asym0}, see the text following \eref{prsebelow} and also below.

\eref{lin} is a harmonic oscillator equation, which yields
\beg
\delta u_k(t)= 
a_k(1-\cos2\Omega_k t)+ i l_k \sin 2\Omega_k t,
\label{14}
\en
where
\beg
l_k=\pm\sqrt{a_k^2+b_k^2}= \frac{\delta\omega}{g^2\Omega_k F_k}.
\label{ambig}
\en
In deriving \eref{14} we took into account the initial condition $\delta u_k(0)=0$ and used expressions \re{dck}.  We set the sign in the last equation in \eref{ambig} to be plus,
which we will justify later in this subsection.

\begin{figure}[h]
\includegraphics[scale=0.26,angle=0]{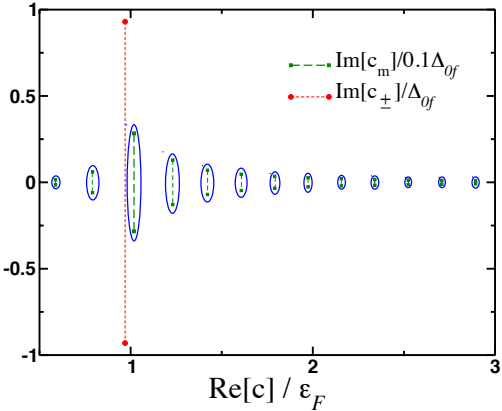}
\caption{(color online) As a result of a  quench doubly degenerate   roots of $\vec L^2(u)$ in Fig.~\ref{groundroots} split into pairs of complex conjugate roots 
$c_m$ (not all $N=54$ pairs of roots are shown).  In linear analysis 
separation variables move periodically on ellipses around the brunch cuts of $(\vec L^2(u))^{-1/2}$  connecting  complex conjugate  $c_m$ without crossing any of the brunch cuts. Each separation variable has its own distinct frequency and corresponds to a normal mode of small oscillations around the ground state. Here $ \Delta_{0f}=0.12\eps_F$, $\delta\omega/\gamma=-0.1$, and other parameters are the same as  in Fig.~\ref{groundroots}. }
\label{ellipse}
\end{figure}

\eref{14} shows that $u_k(t)$ are the normal modes of small oscillations around the ground state and that the normal frequencies are $2\Omega_k=2\sqrt{(x_k-\mu)^2+\Delta_{0}^2}$, where $x_k$ are the roots of \eref{l0x}. \eref{14} also shows that in linear analysis  separation variable $u_k(t)$ moves on an ellipse with semi-axes $a_k$ and $\sqrt{a_k^2+b_k^2}$ around the roots $c_k, \bar c_k$. The latter are the focal points of the ellipse. The function $\sqrt{Q_{2N+2}(u)}$ entering equations of motion for separation variables has  branch cuts connecting pairs of conjugate roots $c_k$ and $\bar c_k$, so one can also say that separation variables move on ellipses around brunch cuts without crossing any of them, see Fig.~\ref{ellipse}.

Next, we determine deviations of the spins $\delta\vec s_k(t)$ and the order parameter $\delta\Delta(t)$ from their  initial ground state configuration \re{grdeltat} and \re{gspin}. We go to a rotating reference frame,
\beg
s_k^-\to s_k^- e^{-2i\mu t},\quad  b\to  b e^{-2i\mu t},
\label{rotate}
\en
to get rid of the time dependence in the unperturbed dynamical variables. This shifts $\omega\to \omega-2\mu$ in the equation of motion \re{dickeev} and now $\dot b=0$ in the ground state \textit{before} the quench, i.e. for $\omega=\omega_i$. Linearizing \eref{su}, we obtain a decomposition of spin deviations in terms of the normal modes
\beg
\frac{\delta s_k^-(t)}{s_k^-(0)}=\frac{\delta\Delta(t)}{\Delta_0}-\sum_j \frac{\delta u_j}{\eps_k-x_j}.
\label{spinnormal}
\en
Similarly, the second equation in \re{dickeev} linearized and integrated in the rotating frame \textit{after} the quench, i.e. with $\omega=\omega_f$,  yields
\beg
\begin{split}
 \frac{\Delta(t)}{\Delta_0}= 1- \sum_k l_k\frac{1-\cos 2\Omega_k t}{\Omega_k}-\\
 i \delta\omega t+2i  t\sum_k a_k-i \sum_k\frac{a_k \sin 2\Omega_k t}{\Omega_k},\\
\end{split}
\label{interm12}
\en
where we took into account $\Delta(t)=-g b(t)$, $\Delta(0)=\Delta_0$ and expressions \re{14}. The $i \delta\omega t$ appears because for unperturbed $u_k$ the   bracketed term in the second equation in \re{dickeev} vanishes for $\omega=\omega_i$, while after the quench $\omega=\omega_f$.

Linearizing spin equations of motion \re{Bloch} directly and plugging expressions \re{spinnormal} and \re{interm12}, one can verify 
that the correct sign in the last equation in \eref{ambig} is indeed plus, even though there is probably a simpler way to show this.

The imaginary part in \eref{interm12} comes from the phase of the order parameter, so we write
\beg
\begin{split}
  \Delta(t) = \left(\Delta_0-\Delta_0 \sum_k l_k\frac{1-\cos 2\Omega_k t}{\Omega_k}\right)\times\\
\exp\left[ -i \delta\omega t+2i  t\sum_k a_k-i \sum_k\frac{a_k \sin 2\Omega_k t}{\Omega_k}\right].\\
\end{split}
\label{deltalinear}
\en
 This coincides with  \eref{interm12} to first order in $\delta\omega$. Moreover, we know from \eref{surprise} that the linear part of the phase (zeroth harmonic in the derivative of the phase) is
$-2\mu_f t$ in the continuum limit, where $\mu_f$ is the ground state chemical potential at detuning $\omega_f$. Similarly, the zeroth harmonic in the amplitude of $\Delta(t)$ is equal to $\Delta_{0f}$. It turns out that this is true even in the discrete case, i.e.
\beg
\begin{split}
\Delta_0 -\Delta_0\sum_k \frac{l_k}{\Omega_k}=\Delta_{0f},\\
2\mu+\delta\omega -2\sum_k a_k= 2\mu_f,\\
\end{split}
\en
where we restored the phase of $\Delta(t)$ to the original reference frame according to \eref{rotate}. Recall that in this subsection  $\mu$ and $\Delta_0$ without a subscript indicate ground state values $\mu_i$ and $\Delta_{0i}$ for the initial detuning $\omega=\omega_i$. With the help of \esref{dck} and
\re{ambig} these relations become
\beg
\begin{split}
\sum_k\frac{x_k-\mu}{\Omega_k^2 F_k}=\frac{g^2}{2}-g^2 \frac{\delta\mu}{\delta\omega},\\
\sum_k \frac{\Delta_0}{\Omega_k^2 F_k}=-g^2\frac{\delta\Delta_0}{\delta\omega},\\
\end{split}
\label{iden11}
\en
where $\delta\mu=\mu_f-\mu$ and $\delta\Delta_0=\Delta_{0f}-\Delta_0$. These are in fact identities, as we prove in Appendix~\ref{appa}. Thus,
\beg
\begin{split}
  \Delta(t) = \left(\Delta_{0f}+\Delta_0 \sum_k l_k\frac{\cos 2\Omega_k t}{\Omega_k}\right)\times\\
\exp\left[  -2i\mu_f t-i \sum_k\frac{a_k \sin 2\Omega_k t}{\Omega_k}\right],\\
\end{split}
\label{deltalin}
\en
in the original reference frame.

An expression for $s_k^-(t)$ obtains similarly from \esref{spinnormal} and \re{interm12} with the help of identity \re{idenuu},
\beg
\begin{split}
s_k^-(t)=s_{kf}^-\Biggl(1+\sum_j  \frac{l_j \cos 2\Omega_j t}{\Omega_j}+\Biggr.\\
\Biggl.\sum_j\frac {a_j\cos 2\Omega_j t}{\eps_k-x_j}-i\sum_j\frac {l_j\sin 2\Omega_j t}{\eps_k-x_j}\Biggr)\times\\
\exp\left[  -2i\mu_f t-i \sum_j\frac{a_j \sin 2\Omega_j t}{\Omega_j}\right],\\
\end{split}
\label{s-lin}
\en
where
\beg
s_{kf}^-\equiv s_f^-(\eps_k)=\frac{\Delta_{0f}}{2\sqrt{(\eps_k-\mu_f)^2+\Delta_{0f}^2}}.
\en
The last term in round brackets in \eref{s-lin} can be as well included into the phase -- to linear 
order the two versions are equivalent. The present form is more convenient for the long time analysis below.
We see that again non-oscillatory parts of the magnitude and phase of $s_k^-(t)$ and magnitude of $s_k^z$ are the same as in the ground state for final detuning $\omega=\omega_f$.

Finally, the expression for $s_k^z(t)$ follows the conservation of $|\vec s_k|=1/2$, $s_k^z=\pm\sqrt{1/4-|s_k^-|^2}$ expanded to the linear order in $\delta\omega$,
\beg
\begin{split}
s_k^z(t)=s_{kf}^z\Biggl(1-\frac{\Delta_{0f}^2}{(\eps_k-\mu_f)^2}\sum_j  \frac{l_j \cos 2\Omega_j t}{\Omega_j}-\Biggl.\\
\Biggl.\frac{\Delta_{0f}^2}{(\eps_k-\mu_f)^2}\sum_j\frac {a_j\cos 2\Omega_j t}{\eps_k-x_j}\Biggr),\\
 \end{split}
\label{szlin}
\en
where
\beg
s_{kf}^z\equiv s_f^z(\eps_k)=-\frac{\eps_k-\mu_f}{2\sqrt{(\eps_k-\mu_f)^2+\Delta_{0f}^2}}.
\en

\subsection{Continuum limit}

\begin{figure}[h]
\includegraphics[scale=0.26,angle=0]{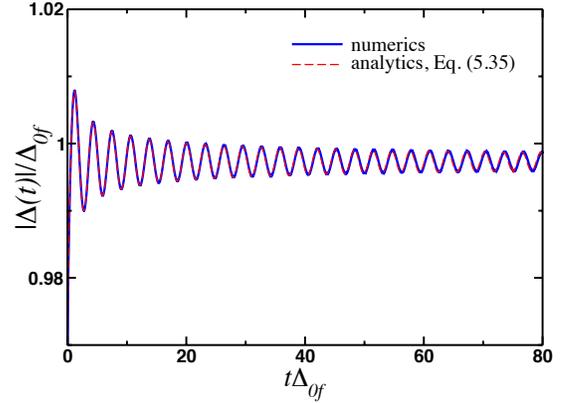}
\caption{(color online)    Comparison of \eref{deltalincont} with $|\Delta(t)|$ computed by numerically evolving $N=5024$ spins in a 3d 2-channel model after a detuning quench.
Here $\gamma=1.0, \Delta_{0i}=0.122\Delta_{\max},  \Delta_{0f}=0.126\Delta_{\max}$, and both \eref{deltalincont} and the spin chain in the numerics are cutoff at $\eps_\Lambda=10\eps_F$. }
\label{sharprise}
\end{figure}

In $N\to\infty$ limit, $x_k\to\eps_k$ and summations in above expressions for $s_k^z(t), s_k^-(t)$, and $\Delta(t)$ turn into integrations. With the help
of \esref{dck}, \re{ambig}, \re{sumintrule}, and \re{feps}, \eref{deltalin} obtains (as always in units of the Fermi energy $\eps_F$)
\beg
  \frac{ \Delta(t)}{\Delta_{0f}} = \Bigl(1+ X_1(t) \Bigr)
\exp\left[  -2i\mu_f t-i X_2(t)  \right],
\label{deltalincont}
\en
where 
\beg
X_1(t)=\frac{\delta\omega}{\gamma}\int\limits_0^\infty \frac{2\cos\left[ 2E(\eps)t\right] f(\eps)d\eps}{E(\eps)\left[\pi^2f^2(\eps)+H^2(\eps)\right]},
\label{x1}
\en
\beg
X_2(t)=\frac{\delta\omega}{\gamma}\int\limits_0^\infty \frac{2(\eps-\mu)\sin[ 2E(\eps)t] f(\eps)d\eps}{E^2(\eps)[\pi^2f^2(\eps)+H^2(\eps)]},
\label{x2}
\en
$E(\eps)=\sqrt{(\eps-\mu)^2+\Delta_0^2}$ and $H(\eps)$ is defined in \eref{xkkk}. In deriving this expression, we also used, $\Omega_k\to E(\eps_k)$, $\nu(\eps)=\nu_F f(\eps), g^2\nu_F=\gamma$, and $\delta=N_k/\nu(\eps_k)$. \eref{deltalincont} is in excellent agreement with numerical results, see e.g. Fig.~\ref{sharprise}.

Expressions \re{szlin} and \re{s-lin} for $s_k^-(t)$ and $s_k^z(t)$ contain two extra summations as compared to $\Delta(t)$. These are handled as in
Appendix~\ref{appa} by splitting each sum into two parts -- over $x_j$ inside and outside a small interval around $\eps_k$. The same method works for summations over $x_j$ because according to \eref{xkkk} $\varrho(\eps)$ is a smooth function and therefore $x_j$ are locally equally spaced with spacing $\delta$ just as $\eps_k$.
The second and the third sums in round brackets in \eref{s-lin} are
 \beg
\begin{split}
Y_1(\eps,t)=\frac{\delta\omega}{\gamma}\dashint\limits_0^\infty\frac{ 2(\eps'-\mu) \cos[2 E(\eps') t]f(\eps')d\eps'}{(\eps-\eps')E(\eps')[\pi^2f^2(\eps')+H^2(\eps')]}\\
-\frac{\delta\omega}{\gamma}\frac{2(\eps-\mu) H(\eps)  \cos[2 E(\eps) t]}{E(\eps) [\pi^2f^2(\eps)+H^2(\eps)]},\\
\end{split}
\label{ups2}
\en
\beg
\begin{split}
Y_2(\eps,t)=\frac{\delta\omega}{\gamma}\dashint\limits_0^\infty\frac{  2\sin[2 E(\eps') t]f(\eps')d\eps'}{(\eps-\eps') [\pi^2f^2(\eps')+H^2(\eps')]}-\\
\frac{\delta\omega}{\gamma}\frac{ 2H(\eps)  \sin[2 E(\eps) t]}{ \pi^2f^2(\eps)+H^2(\eps)},\\
\end{split}
\label{ups1}
\en
respectively. Thus
\beg
\begin{split}
\frac{s^-(\eps,t)}{s_{f}^-(\eps)}=\Bigl(1+X_1(t) + Y_1(\eps,t)-i
Y_2(\eps, t)\Bigr)\times\\
\exp\left[  -2i\mu_f t-i  X_2(t)   \right],\\
\end{split}
\label{s-lincont}
\en
\beg
\frac{s^z(\eps,t)}{s_{f}^z(\eps)}= 1-\frac{  \Delta_{0f}^2}{ (\eps-\mu_f)^2}\Bigl( X_1(t) +
 Y_1(\eps,t)\Bigr).
\label{szlincont}
\en

Functions $X_1$ and $X_2$   are related via differentiation. Define
\beg
\widetilde{X}_1(t)=\int_0^\infty K(\eps) e^{2i\widetilde{E}(\eps) t}d\eps,
\label{x1tilde}
\en
where $\widetilde{E}(\eps)=\sqrt{(\eps-\widetilde{\mu})^2+\Delta_0^2}$ and
\beg
K(\eps)=\frac{\delta\omega}{\gamma} \frac{  2f(\eps)}{E(\eps)\left[\pi^2f^2(\eps)+H^2(\eps)\right]}.
\label{keps}
\en
Then,
\beg
X_1(t)=\left.\mathrm{Re} \widetilde{X}_1(t)\right|_{\widetilde{\mu}=\mu}
\label{x1re}
\en
\beg
X_2(t)=\frac{1}{2t}\Bigl.\mathrm{Re} \frac{\partial \widetilde{X}_1(t)}{\partial \widetilde{\mu}}\Bigr|_{\widetilde{\mu}=\mu}
\label{x2re}
\en
A similar relationship holds for $Y_1$ and $Y_2$.

\subsection{Validity of the few spin conjecture}
\label{validity}

We are now in the position to prove the few spin conjecture for infinitesimal quenches independently of either numerics or arguments of Sect.~\ref{method}. At $t\to\infty$ integrals
in \esref{x1} and \re{x2} vanish by the Riemann-Lebesgue lemma. Therefore,
\beg
\Delta(t)\to\Delta_{0f} e^{-2i\mu_f t}.
\en

According to the few spin conjecture this asymptotic behavior of $\Delta(t)$ occurs when there is a single isolated root pair at $\mu_f\pm i\Delta_{0f}$.   \eref{primef} shows that our $\vec L^2(u)$ does have this pair of roots. Moreover, the remaining $2N$ roots are given by \eref{dck} and we explicitly see from Appendix~\ref{appb} that their imaginary parts scale as $1/N$ at large $N$ and that they merge into a continuum of roots on the real axis in $N\to\infty$ limit. Thus, there is indeed a single isolated root pair at $\mu_f\pm i\Delta_{0f}$ in the thermodynamic limit.

\subsection{Weak coupling limit}
\label{wclimit}

Simpler expressions obtain in the weak coupling (BCS) limit when $\Delta_0$ is much smaller then other energy scales (Fermi energy in gases and Debye energy in metals). This limit describes superconductivity in metals and applies to recent experiments on non-adiabatic BCS dynamics\cite{Shimano2013,Shimano2012}. In our quench phase diagrams (Figs.~ \ref{psd2ch2d} -- \ref{psd1ch2d} etc.) weak coupling regime corresponds to a small neighborhood of the origin.

At weak coupling $\mu\approx \eps_F=1$. Integrals \re{x1} and \re{x2} are dominated by energies close to the Fermi energy, $|\eps-\mu|\sim\Delta_0$,   where $f(\eps)\approx 1$ independent of dimensionality. It is convenient to change the integration variable to $\xi=\eps-\mu$ and extend the integration to the entire real axis. $X_2(t)$ vanishes by particle-hole symmetry (integrand is odd in $\xi$). The error due to these approximations is proportional to $\Delta_0/\eps_F$, which vanishes in weak coupling limit. \eref{deltalincont} implies
\beg
|\Delta(t)|=\Delta_{0f}- 4\delta\Delta_0 \int\limits_{0}^\infty \frac{\cos\left[ 2E(\xi)t \right]  d\xi}{E(\xi) \left[\pi^2 +H^2(\xi)\right]},
\label{deltalinwc}
\en
where $E(\xi)=\sqrt{\xi^2+\Delta_0^2}$, $\delta\Delta_0=\Delta_{0f}-\Delta_{0i}$, and
\beg
 H(\eps)=\ln\left[\frac{E(\xi)-\xi}{ E(\xi) +\xi}\right].
\label{hewc}
\en
In deriving \eref{deltalinwc} we used the weak coupling gap formula $\Delta_0\propto \exp(-\omega/\gamma)$ and \esref{gewc} and \re{xkkk} [Note that
at relevant energies $4E(\eps)/\gamma\propto \Delta_0/\eps_F\to 0$]. We also used the fact that the integrand is even in $\xi$ to convert the integration range from $(-\infty,\infty)$ to $(0,\infty)$.

The phase of the order parameter defined through 
\beg
\Delta(t)=|\Delta(t)|e^{-i\Phi(t)}
\en
 is simply $\Phi(t)=2\eps_F t$. Let us also note that in terms of $\xi=\Delta_0 \sinh(\pi x/2)$ \eref{deltalinwc} reads
\beg
|\Delta(t)|=\Delta_{0f} -2\delta\Delta_0 \int\limits_0^{\infty}
\frac{dx}{\pi} \frac{\cos\left[ 2\tau \cosh(\pi x/2)\right]}{1+x^2}, 
\label{dweak1}
\en 
where $\tau=\Delta_0 t$.

\subsection{Long time behavior of $\Delta(t)$: BCS side}

Integrands in \esref{x1} and \re{x2} are highly oscillatory. The argument of the cosine is stationary at $\eps=\mu$, $E'(\mu)=0$. For $\mu>0$ the stationary point is inside the integration range. For $\mu<0$ there are no stationary points on the integration path. This leads to qualitatively different behavior of $\Delta(t)$ on the BCS ($\mu>0$) and BEC ($\mu<0$) sides. 

Consider first the BCS regime. We evaluate $\widetilde{X}_1(t)$ in \eref{x1tilde} in stationary phase approximation
\beg
\widetilde{X}_1(t)=K(\widetilde{\mu})\sqrt{\frac{\pi\Delta_0}{ t}}e^{2i\Delta_0 t +i\pi/4}+O(1/t),
\en
where we used $\widetilde{E}(\widetilde{\mu})=\Delta_0$, $\widetilde{E}''(\widetilde{\mu})=1/\Delta_0$. With the help of \eref{x1re} we obtain from \eref{deltalincont} for the order parameter amplitude
\beg
|\Delta(t)|=\Delta_{0f}+\sqrt{\pi}K(\mu)\Delta^2_0\frac{\cos(2\Delta_0 t+\pi/4)}{\sqrt{\Delta_0 t}}.
\label{statptex}
\en
The phase of the order parameter   obtains with the help of \eref{x2re}
\beg
\Phi(t)=2\mu_f t+\sqrt{\pi}K'(\mu)\Delta^2_0\frac{\cos(2\Delta_0 t+\pi/4)}{2(\Delta_0 t)^{3/2}}.
\label{phasela}
\en
Coefficients $K(\mu), K'(\mu)$ are given by \esref{keps}, \re{xkkk}, and \re{ge}. Simpler expressions for $G(\eps)$ are available in 2d and in weak coupling  (BCS) limit, see  \esref{ge2d} and \re{gewc}. For example, in the BCS limit ($\Delta_0/\eps_F\to 0$)
\beg
\Delta(t)=\left(\Delta_{0f}-\frac{2\delta\Delta_{0}}{\pi^{3/2}}\frac{\cos(2\Delta_0 t+\pi/4)}{\sqrt{\Delta_0 t}}\right) e^{-2i\mu_f t},
\label{vk11}
\en
where $\delta\Delta_0=\Delta_{0f}-\Delta_{0i}$ and we additionally used $\Delta_0\propto \exp(-\omega/\gamma)$. Note that the second term in \eref{phasela} is proportional to 
$\Delta_0/\eps_F$. This expression for $\Delta(t)$ holds in the BCS limit for both one and two channel models in two and three dimensions. \eref{vk11} for $\mu_f=0$ appeared  in Ref.~\onlinecite{Dzero2006} without derivation.

\begin{figure}[h]
\includegraphics[scale=0.26,angle=0]{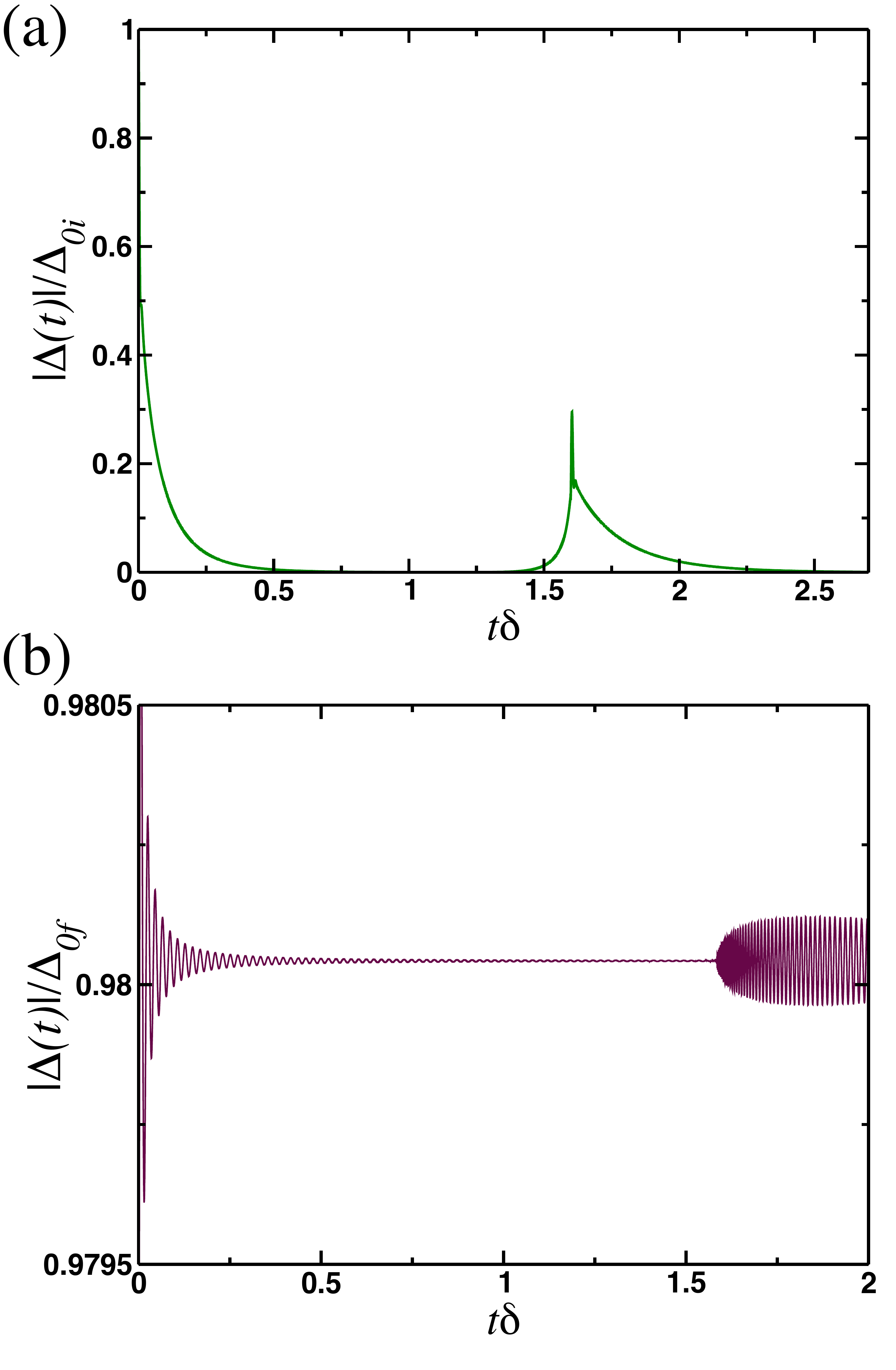}
\caption{(color online)  Finite size effects, such as  partial recurrences in $|\Delta(t)|$, develop at times of order of the inverse level spacing  $\delta\propto 1/N$ between discretized single-particle energy levels $\eps_k$. Long time behaviors derived in our paper apply at times $t\delta\ll 1$. In other words, we take the thermodynamic  limit first and large time limit second. Two detuning quenches in 3d 2-channel model are shown for $N=5024$ and: (a) $\gamma=0.5, \Delta_{0i}=3.0\times 10^{-2}\Delta_{\max}, \Delta_{0f}=2.9\times 10^{-4}\Delta_{\max}, \delta=3.4\times 10^{-3}\Delta_{\max}$, and (b) $\gamma=0.1, \Delta_{0i}=0.97\Delta_{\max}, \Delta_{0f}=0.99\Delta_{\max}, \delta=8.0\times 10^{-3}\Delta_{\max}$.}
\label{blip}
\end{figure}

 Let us also mention that long times for which asymptotes of the order parameter derived in this section apply in practice (e.g. in numerical simulations) mean $t$ such that $1/\Delta_0\ll t\ll 1/\delta$. At times of order of the inverse level spacing $1/\delta$ partial recurrences occur, see Fig.~\ref{blip}. Oscillations with frequency $2\Delta_0$ and $1/\sqrt{t}$ decay in the weak coupling limit of the one channel model were  identified by Volkov and Kogan\cite{Kogan1973}. 
 
\subsection{Long time behavior of $\Delta(t)$: BEC side}
\label{beclong}

In the absence of stationary points integrals of the type of \eref{x1tilde} are dominated by the end point, $\eps=0$ here. Normally, they vanish as $1/t$ at large $t$, but in the present case
$K(0)=0$ in both 2d and 3d, so they vanish faster. Unlike the BCS side, the long time behavior
on the BEC side is not universal in that it depends on the form of $K(\eps)$ at small $\eps$, i.e. 
on the density of states and on the asymptotic spin distribution. As a result, for example, it is different in two and three dimensions. 

We first integrate by parts to obtain
\beg
\widetilde{X}_1(t)=-\frac{1}{2it}\int_0^\infty \left(\frac{K(\eps)}{\widetilde{E}'(\eps)}\right)'
e^{2i\widetilde{E}(\eps) t}d\eps.
\label{byparts}
\en
In 2d the dimensionless density of states $f(\eps)=1$ and it follows from \esref{keps}, \re{xkkk}, and \re{ge2d} that $K(\eps)\propto 1/\ln^2\eps$. We evaluate the large $t$ asymptote of this integral by splitting the integration range into three: $(0, 1/\Lambda t),
(1/\Lambda t, \Lambda/t)$, and $(\Lambda/t, \infty)$, where $\Lambda$ is such that $1\ll \ln\Lambda\ll \ln t$. In the first integral we  expand the integrand in small $\eps$, which leads
to an integral $\int_0^{1/\Lambda t} d(\ln\eps)/\ln^3\eps$ and
\beg
\widetilde{X}_1(t)=\frac{\delta\omega}{\gamma} \frac{ i e^{2i\widetilde{E}(0) t} }{\widetilde{E}'(0)
E(0)} \frac{1}{t\ln^2 t}.
\en
The other two integrals vanish as $1/t\ln^3 t$ and are therefore negligible. \esref{deltalincont}, \re{x1re}, \re{x2re} yield the amplitude and the phase of the order parameter 
\beg
|\Delta(t)|=\Delta_{0f}\left(1-\frac{\delta\omega}{\gamma}\frac{\sin (2E_\mathrm{min} t)}{|\mu| t\ln^2 t}\right),
\label{2dex}
\en
\beg
\Phi(t)=2\mu_f t-\frac{\delta\omega}{\gamma}\frac{\cos (2E_\mathrm{min} t)}{E_\mathrm{min} t\ln^2 t},
\label{phase2dln}
\en
where $E_\mathrm{min}=\sqrt{\mu^2+\Delta_0^2}$.

In 3d $f(\eps)=\sqrt{\eps}$ and $K(\eps)\propto \sqrt{\eps}$ at small $\eps$. This follows from \esref{keps}, \re{xkkk}, and \re{ge} and is, for example, readily verified in the strong coupling limit with the help of the last expression in \eref{gesc}. We split the integration range in 
\eref{byparts}  into two: $(0, 1/\Lambda)$ and $(1/\Lambda, \infty)$, where $t\gg \Lambda\gg1$.  In the first integral we can expand in small $t$, which results in a Gaussian integral that
behaves as $1/\sqrt{t}$ at large $t$. The second integral vanishes faster as $t\to\infty$. We thus determine the following (exact) large time asymptote:
\beg
\widetilde{X}_1(t)=-\frac{\pi^{1/2}}{(2t)^{3/2}}\frac{\delta\omega}{\gamma}\frac{ e^{2i\widetilde{E}(0) t+i\pi/4} }{[-\widetilde{E}'(0)]^{3/2} E(0)H^2(0)}.
\label{strann}
\en
With the help of \esref{x1re} and \re{x2re} we finally derive
\beg
|\Delta(t)|=\Delta_{0f}\left( 1-c\frac{\delta\omega}{\gamma}\frac{\cos(2E_\mathrm{min}t+\pi/4)}{(2|\mu|t)^{3/2}}\right),
\label{ccoeffin}
\en
\beg
\Phi(t)=2\mu_f t-c\frac{|\mu|}{E_\mathrm{min}}\frac{\delta\omega}{\gamma}\frac{\sin(2E_\mathrm{min}t+\pi/4)}{(2|\mu|t)^{3/2}}.
\en
The coefficient $c$ depends on $\mu, \Delta_0$, and $\gamma$. It is known exactly from \eref{strann}, but involves $G(0)$ which in 3d is an elliptic integral according to \eref{ge}. In
the strong coupling BEC limit, $\mu\to-\infty$, $G(\eps)$ is independent of $\eps$ and takes a simple form \re{gesc}. In this case
\beg
c=\sqrt{\frac{\pi|\mu|}{\eps_F}}\left(\frac{4|\mu|}{\gamma\eps_F}+\pi\sqrt{\frac{|\mu|}{\Delta_0}}\right)^{-2},
\en
where we restored the original energy units.

\subsection{Long time behavior of spins}
\label{longspin}

Let us also work out the long time behavior of individual spins given by \esref{s-lincont} and \re{szlincont} and compare it to the asymptotic spin distribution, \esref{ss0s} and \re{distAA}, obtained earlier. The latter result is based on the few spin conjecture, so the agreement with linear analysis provides yet another (though redundant because we already proved the few spin conjecture for infinitesimal quenches in Sect.~\ref{validity}) check.

Functions $X_{1,2}$ vanish as $t\to\infty$, while the large time limit of $Y_{1,2}$   derives from the following identity:
\beg
\lim_{t\to\infty} \dashint\limits_0^\infty \frac{ d\eps' F(\eps') e^{ \pm2i E(\eps') t} }{\eps'-\eps} = \pm i\pi \alpha F(\eps) e^{ \pm 2i E(\eps) t},
\en
where $\alpha$ is the sign of $t dE(\eps')/d\eps'$ at $\eps'=\eps$ and $F(\eps')$ is an arbitrary bounded continuous  function. 

Applying this identity to \esref{ups2} and \re{ups1} and substituting resulting expressions into  \esref{s-lincont} and \re{szlincont}, we obtain
\beg
\begin{split}
\frac{s^-_\infty(\eps,t)e^{ 2i\mu_f t}}{s_{f}^-(\eps)}= 1   -\frac{2\delta\omega}{\gamma}\frac{\exp\left[-2iE(\eps)t-i\phi \right]}{\sqrt{\pi^2f^2(\eps)+H^2(\eps)}}\\
   -\frac{2\delta\omega}{\gamma}\left(\frac{\xi}{E(\eps)}-1\right)\frac{\cos\left[2E(\eps)t+\phi \right]}{\sqrt{\pi^2f^2(\eps)+H^2(\eps)}},\\
\end{split}
\label{s-lincontlarge}
\en
\beg
\frac{s^z_\infty(\eps,t)}{s_{f}^z(\eps)}= 1+\frac{2\delta\omega}{\gamma}\frac{  \Delta_{0f}^2}{ \xi E(\eps)} \frac{\cos\left[2E(\eps)t+\phi \right]}{\sqrt{\pi^2f^2(\eps)+H^2(\eps)}},
\label{szlincontlarge}
\en
where $\xi=\eps-\mu$ and $\phi$ is defined through
\beg
\begin{split}
\cos\phi=\frac{H(\eps)}{\sqrt{\pi^2f^2(\eps)+H^2(\eps)}},\\
 \sin\phi=\frac{\pi f(\eps)\sgn(t\xi)}{\sqrt{\pi^2f^2(\eps)+H^2(\eps)}}.\\
 \end{split}
\en
In our case $t>0$, but we still kept it under the sign function to ensure proper behavior under time reversal, see \eref{timerev}. \esref{s-lincontlarge} and
\re{szlincontlarge} match \eref{ss0s} with
\beg
\theta(\eps)\approx \sin\theta(\eps)=\frac{2\delta\omega}{\gamma}\frac{\Delta_0}{E(\eps)\sqrt{\pi^2f^2(\eps)+H^2(\eps)}}.
\en
(Not that in the present case $\Delta_\infty=\Delta_{0f}$ and $\mu_\infty=\mu_f$.) This indeed agrees with \eref{first2} obtained from the few spin conjecture.

\subsection{Short time behavior}

Here we analyze the short time behavior of $|\Delta(t)|$ for quenches within the universal weak coupling regime.
For  large quenches from  weaker to stronger coupling,  when $\Delta_{0f}/\Delta_{0i}\gg1$, or from the normal state (zero initial coupling) in this regime $|\Delta(t)|$ grows  as $e^{\Delta_{0f}t}$.
This exponential growth reflects the instability of the normal state in the presence of superconducting interactions\cite{Spivak2004,Yuzbashyan2006}.
At the same time, even for small quenches $|\Delta(t)|$ rises or falls sharply at short times, see  Figs.~\ref{regIIdeltaplot} and \ref{sharprise}. Sharp growth is seen in experiment too,  though most of it is probably due to a different mechanism\cite{Shimano2012}.

A direct small $t$ expansion of the cosine in \eref{x1}   diverges at high energies. Cutting off the integral at $\eps_\Lambda$ (Debye energy in the case of metals), one obtains\cite{Leggett2005} $\delta|\Delta(t)|\propto \delta\Delta_0 
(\eps_\Lambda t)^2$. This  is cutoff dependent and applies only to ultra-short times $t\ll 1/\eps_\Lambda$ that vanish as the cutoff is sent to infinity. We are
interested in times $1/\eps_\Lambda\ll t\ll 1/\Delta_0$. 

Consider \eref{dweak1}. The argument of the cosine  is small for $x\ll x_0$, where $x_0$ is determined by
 $e^{\frac{\pi x_0}{2} }=1/\tau$, i.e.  $x_0 = \frac{2}{\pi} \ln (1/\tau)$.
 Let us divide the domain of the integration into three intervals: $[0, x_0-a]$, $[x_0-a, x_0+a]$, and $[x_0+a, \infty)$ and let the corresponding integrals be $I_1$, $I_2$, and $I_3$, respectively.  The auxiliary parameter $a$, $1\ll a\ll x_0$,  is such that $1/a\to0,   a/x_0\to 0$ as $x_0\to\infty$. For example, one can take $a=\sqrt{x_0}$.  Expanding the cosine in small $\tau$ in $I_1$ and integrating, we obtain
 \beg
I_1= \frac{1}{2}-\frac{1}{\pi{x_0}}-\frac{a}{\pi{x_0^2}}+ o(a/x_0^2).
\label{ii1}
\en
In $I_2$ we replace $x^2+1\to x_0^2$ up to terms of order $a/x_0$. After this, a substitution $y=\exp{\pi{x}/2}$ transforms it into the cosine integral
$\int dy\cos y/y$ with known behavior,  leading to
 \beg
I_2=\frac{a}{\pi x_0^2}+ o(a/x_0^2).
\label{j2b}
\en
And integrating by parts in   $I_3$, we see that it is proportional to $e^{-\pi a/2}/x_0^2$, which is negligibly small. Thus,
\begin{equation}
 I_1+I_2+I_3=
\frac{1}{2}-\frac{1}{2|\ln(\tau)|}+o\left(\frac{1}{|\ln(\tau)|}\right) .
\label{Isum}
\end{equation}
Note the cancelation of the auxiliary parameter $a$. Finally, plugging this   into Eq.~\re{dweak1}, we derive  the short time behavior  of the gap function amplitude
\begin{equation}
|\Delta(t)|=\Delta_{0i}+\frac{\Delta_{0f}-\Delta_{0i}}{|\ln (\Delta_0 t) |}.
\label{gapshort}
\end{equation}

\section{Approach to the asymptote in the nonlinear case}
\label{nonlinear}

Here we   discuss the approach of $\Delta(t)$ to its large time asymptote in the nonlinear case. We  will consider regimes I and II -- the gapless phase and the phase where $\Delta(t) \rightarrow \Delta_{\infty} e^{-2 i \mu_{\infty} t-2i\varphi}$. Rather then rigorously deriving the $t\to\infty$ asymptote in its entirety as we did for the linearized dynamics, we present an argument based only on our knowledge of the frequency spectrum that works under certain general assumptions about relevant Fourier amplitudes.

As $t\to\infty$ spins tend to their steady state form, $\vec s(\eps, t)\to \vec s_\infty(\eps,t)$, where $ \vec s_\infty(\eps,t)$ is given by \esref{ssa} and \re{ss0s} in regimes I and II, respectively.
 In   phase II, in a reference frame rotating with frequency $2\mu_\infty$ around $z$-axis,   $\vec s_\infty(\eps)$   rotates with  a constant frequency $2E_\infty(\eps)=2\sqrt{(\eps-\mu_\infty)^2+\Delta_\infty^2}$. As mentioned above, an integrable model with $N$ degrees of freedom is characterized by $N$ incommensurate frequencies \cite{Arnold} that
are determined by the integrals of motion and are fixed throughout its time evolution. The Fourier decomposition of any dynamical quantity can have only these basic frequencies in its spectrum. In particular, 
\beg
|\Delta(t)|=\Delta_\infty+\int\limits_0^\infty F(\eps) \cos[2E_\infty(\eps) t] f(\eps) d\eps
\label{fourgen}
\en
with some unknown function $F(\eps)$.

Terms containing  $\sin[2E_\infty(\eps) t]$ are absent by  time-reversal symmetry [cf. \eref{deltalincont}] of  the equations of motion  \re{Bloch} and \re{123}  [see also \eref{eom}]
\beg
s^z(-t)=s^z(t),\,\,  s^+(-t)=s^-(t),\, \,  \bar\Delta(-t)=\Delta(t),
\label{timerev}
\en
where we suppressed  $\eps$-dependence of spins for compactness. These relations   hold at all times as long as the initial condition at $t=0$ satisfies them, which our initial state \re{ini} does. 

A common practice in previous work is to attempt to determine the approach of $|\Delta(t)|$ to its asymptotic value $\Delta_\infty$ from the steady state spins $\vec s_\infty(\eps,t)$. Consider the one channel case for simplicity. Continuum version of \eref{gap1ch} at $t=\infty$ is
\beg
\Delta_\infty(t)=\lam \int\limits_0^\infty s_\infty^-(\eps,t) f(\eps) d\eps.
\label{deltainfinc}
\en
The constant part of $s_\infty^-(\eps,t)$ yields $\Delta_\infty$, while the contribution of the oscillating part integrated over $\eps$ vanishes (dephases) as $t\to\infty$. One can further determine the large time asymptote of \eref{deltainfinc}  similarly to how we evaluated the large time behavior of  \eref{deltalincont}.
This is however \textit{not} the correct  asymptote of the actual $\Delta(t)$. Not only it does not yield the correct coefficient of the time-dependent part of $\Delta(t)$ [such as the coefficient $c$ in \eref{ccoeffin}], but also the actual time dependence can be different. 

At finite $t$ there is a correction to the steady state value of the spin, $\vec s(\eps,t)=\vec s_\infty(\eps,t)+\delta \vec s(\eps,t)$, so that the actual order parameter is
 \beg
\Delta(t)=\lam \int\limits_0^\infty s_\infty^-(\eps,t) f(\eps) d\eps +\lam \int\limits_0^\infty \delta s^-(\eps,t) f(\eps) d\eps.
\label{deltainfcor}
\en
 Even though $\delta s^-(\eps,t)$ is small as compared to the oscillating part of $s_\infty^-(\eps,t)$ at large times, this is no longer true after integrating these quantities over $\eps$. Consider, for example, \eref{s-lincont}. We showed in Sect.~\ref{longspin} that $s_\infty^-(\eps,t)$ comes from functions $Y_{1,2}(\eps,t)$. But  we see from \eref{deltalincont}  that the integral of these functions over $\eps$ vanishes and as a result  they do not contribute to $\Delta(t)$
 The correction $\delta s^-(\eps,t)$ on the other hand comes from both $X_{1,2}(t)$ and $Y_{1,2}(\eps,t)$. It is this contribution from $X_{1,2}(t)$ to 
 $\delta s^-(\eps,t)$ that actually determines $\Delta(t)$. Thus there is a partial cancelation  between the two integrals in \eref{deltainfcor} and the true large time behavior of $\Delta(t)$ can only be determined by keeping both.
 
 Nevertheless, $\Delta_\infty(t)$ being a legitimate dynamical quantity has the right frequency spectrum and also contains the dimensionless density of 
 states $f(\eps)$. So, it still produces a correct  large time dependence  when, for example, the latter  is set by a stationary point as in
 \eref{statptex} or by the behavior of $f(\eps)$ at small $\eps$ as in \eref{ccoeffin}. The situation on the BEC side in 2d is different. The $\ln^2 t$ dependence in the denominator of \eref{2dex} comes from $K(\eps)\propto 1/\ln^2\eps$ behavior of the Fourier amplitude  at small $\eps$, see \eref{x1tilde} and the text below \eref{byparts}. This is in turn a consequence of $K(\eps)\propto H^{-2}(\eps)$ and $H(\eps)\propto \ln\eps$, which follow
 from \esref{keps}, \re{xkkk}, and \re{ge2d}.  Were we to evaluate the large time asymptote of $|\Delta(t)|$ using \eref{deltainfinc}, we would obtain $1/(t\ln t)$ instead of $1/(t\ln^2 t)$. To see this, note that \eref{s-lincontlarge} implies that the oscillating part of $s_\infty^-(\eps,t)$ is proportional to $H^{-1}(\eps)$, i.e. to $1/\ln\eps$, at small $\eps$ and apply the same steps as in the text below \eref{byparts}. The $1/\ln\eps$ dependence cancels in \eref{deltainfcor} due to the second term on the right hand side.  We note also that   \esref{ss0s} and \re{distAA} imply   
 $s_\infty^-(\eps,t)\propto 1/\ln\eps$   in  all of region II in 2d, not just in the linear approximation.
 
 Similar considerations apply in analyzing \eref{fourgen}. Let us work out the large time behavior  of $|\Delta(t)|$ in steady states I, II, and II' separately.
 
 \subsection{Regime II}
 
 In steady states II and II' $\Delta(t) \rightarrow \Delta_{\infty} e^{-2 i \mu_{\infty} t-2i\varphi}$. For quenches in region II $\mu_\infty>0$, so  it can be viewed as a nonequilibrium extension of the BCS regime. The frequency spectrum $2E_\infty(\eps)$ has a stationary point at $\eps=\mu_\infty$, $E'_\infty(\mu_\infty)=0$, which in regime II lies within the integration range. The large time behavior of \eref{fourgen} obtains with the help of stationary point method [cf. \eref{statptex}]
\beg
|\Delta(t)|=\Delta_{\infty}+\sqrt{\pi}F(\mu_\infty)\Delta^2_\infty\frac{\cos(2\Delta_\infty t+\pi/4)}{\sqrt{\Delta_\infty t}}.
\label{statptexnonlin}
\en
The only assumption about $F(\eps)$ here is that it is smooth.
 This is an extension of  \eref{statptex} to nonlinear regime. In the weak coupling BCS limit this result was published in Ref.~\onlinecite{Yuzbashyan2006}.
 In this limit $\Delta_\infty$ is given by \eref{qwrty} and generally it obtains from \esref{isolated2d} and \re{v3d} in 2d and 3d, respectively, and \eref{not1} as the imaginary part of $u$.  Here we see that expression \re{statptexnonlin} holds throughout the entire region II for both one and two channel models.
 
  \subsection{Regime II'}
  
 Regime II' has the same asymptotic $\Delta(t)$ as II by definition only with $\mu_\infty<0$. There are now no stationary points on the integration path. The approach to the asymptote is therefore determined by the behavior of $F(\eps)f(\eps)$ near the end points,  $\eps=0$ in this case. We  \textit{assume}  this behavior is the same  as in linear analysis, since we expect the time dependence to have the same functional form throughout a given regime.  According to Sect.~\ref{beclong}, this means finite nonzero $F(0)$ in 3d and $F(\eps)\propto 1/\ln^2\eps$ for $\eps\ll 1$ in 2d. 
 
 Expanding \eref{distAA} in small $\eps$ and using \eref{ss0s}, we see that the spin components at $t\to\infty$ do behave the same as in linear analysis, though this in itself does not prove our assumption. Moreover, the asymptotic spin distribution \re{distAA} is continuous across critical lines separating various regimes, so the same small $\eps$ form holds in gapless region I as well.
 
 As long as our assumptions about $F(\eps)$ are correct, the analysis of the integral in \eref{deltainfinc} is the same as that in Sect.~\ref{beclong} resulting in
 \beg
|\Delta(t)|=\Delta_{\infty}\left(1- c_1\frac{\sin (2E_\infty^\mathrm{min} t)}{ t\ln^2 t}\right)\quad\mbox{in 2d,}
\label{2dexnlin}
\en
and
\beg
|\Delta(t)|=\Delta_{\infty}\left( 1-c_2 \frac{\cos(2E_\infty^\mathrm{min}t+\pi/4)}{t^{3/2}}\right)
\quad\mbox{in 3d,}
\label{ccoeffinnlin}
\en
at large times, where $E_\infty^\mathrm{min}=\sqrt{\mu_\infty^2+\Delta_\infty^2}$ and $c_1$ and $c_2$ are real coefficients that depend on $\Delta_{0i}, \Delta_{0f}$, and $\gamma$.

\subsection{Gapless regime}

Finally, we turn to regime I. Now $\Delta(t)\to0 $ at $t\to\infty$. Spins $\vec s_\infty(\eps)$   rotate with frequencies $2\eps$ around $z$-axis, so that the Fourier transform   of the order parameter magnitude is of the form
\beg
|\Delta(t)|= \int\limits_0^\infty F(\eps) \cos(2 \eps t) f(\eps) d\eps,
\label{fourglessgen}
\en
and $\sin(2\eps t)$ term  vanishes by time-reversal symmetry \re{timerev}.

In 3d we similarly assume finite and nonzero $F(0)$ .  Steps outlined below \eref{phase2dln}  
in Sect.~\ref{beclong} now lead to the following large time behavior:
\beg
|\Delta(t)|= \frac{c_3}{t^{3/2}}.
\label{ccoeffinnlingless}
\en

In 2d we speculate that $F(\eps)\propto 1/\ln^r \eps$ at small $\eps$, where $r$ is either 1 or 2. As discussed before in this Section,  $s_\infty^-(\eps,t)\propto 1/\ln\eps$ in 2d, so that $\Delta_\infty(t)\propto 1/(t\ln t)$.  The $1/\ln\eps$ term however cancels from $F(\eps)$ at least in linear analysis and it ends up being proportional to $1/\ln^2\eps$ instead. In the gapless 
case we allow for a possibility that such a cancelation does not occur. The analysis of the integral in \eref{fourglessgen}, analogous to that leading to \eref{ccoeffin}, then yields
\beg
|\Delta(t)|=\frac{c_4}{t\ln^r t}.
\label{cccc2}
\en

 The gapless regime contains the $\Delta_{0i}=\Delta_{0f}=0$ point, the origin of quench phase diagrams. It therefore  includes the weak coupling limit
 $\Delta_{0i}/\eps_F\to 0$ and $\Delta_{0f}/\eps_F\to0$. \eref{fourglessgen} becomes in this limit [see Sect.~\ref{wclimit}]
 \beg
|\Delta(t)|= \int\limits_{-\infty}^\infty F(\xi) \cos(2 \xi t)  d\xi,
\label{fourglessgen1}
\en
 where $F(\xi)$ is even in $\xi$. Now there can be no power law in $t$ contribution at large $t$ coming from integration limits. Instead, $|\Delta(t)|$ vanishes exponentially\cite{Barankov2006,Dzero2006} as $A(t)e^{-2\alpha\Delta_{0i}t}$ independent of dimensionality, where $\alpha\sim1$ and $A(t)$ is a  decreasing power law,  $A(t)\sim \Delta_{0i}$ at  $t\sim 1/\Delta_{0i}$. Recall that throughout this paper we have been using units where $\eps_F=1$.
 To convert to arbitrary units in \esref{ccoeffinnlingless} and \re{cccc2}, one needs to replace $t\to\eps_F t$. Guided by linear analysis we further assume
 that coefficients $c_4$ and $c_5$ are of order $\Delta_{0f}$, which we take to be comparable to $\Delta_{0i}$. It is clear that at any finite
 $\Delta_{0i}/\eps_F\ll1$ power laws in  \esref{ccoeffinnlingless} and \re{cccc2} coming from the lower integration limit will eventually win over the exponential decay. The comparison of $e^{-2\alpha\Delta_{0i}t}$ with $(\eps_F t)^{-1}$ shows that the weak coupling   result is valid at times
 such that $\ln(\eps_F/\Delta_{0i})\gg \Delta_{0i} t\gg1$, while for $\Delta_{0i} t\gg\ln(\eps_F/\Delta_{0i})$ it has to be replaced with \esref{ccoeffinnlingless} and \re{cccc2}.

\section{Experimental signatures}\label{rf}

 Far from equilibrium states of fermionic superfluids described in this paper can be observed in different systems with various experimental techniques. 
 
 Matsunaga et. al.\cite{Shimano2013,Matsunaga2014} directly measured the time-dependent amplitude $|\Delta(t)|$ induced by an ultrafast electromagnetic perturbation in ${\mathrm{Nb}}_{1\mathrm{\text{-}}x}{\mathrm{Ti}}_{x}\mathbf{N}$ films using terahertz pump --  terahertz probe spectroscopy. The underlying system is a BCS superconductor [weak coupling regime of one channel model \re{eq:onechannel}] and for perturbation strength below  certain threshold its non-adiabatic dynamics falls within region II of our quench phase diagrams. Even though we considered BCS interaction quenches in one channel model in this paper, it is clear from our arguments that our results apply more generally to any kind of non-adiabatic global perturbation. Therefore, we expect $|\Delta(t)|$ to be described by \eref{statptexnonlin} derived originally  in nonlinear regime by Yuzbashyan et.al.  \cite{Yuzbashyan2006}. These experiments indeed measure damped oscillations with frequency $2\Delta_\infty$, where $\Delta_\infty$ is the  asymptotic value of $|\Delta(t)|$ even when the system is deep in the nonlinear regime and $\Delta_\infty$ is much different from the ground state gap. The power law approach however appears to be faster than $1/t^{1/2}$.
 
 In this paper we primarily focused on detuning or interaction quenches  in cold fermions. Experiments addressing superfluidity in these systems include measurements of the molecular condensate fraction\cite{Jin2004,Ketterle2004}, radio frequency  absorption spectra\cite{Chin2004}, and observation of vortices\cite{Ketterle2005}. Signatures of ``far from equilibrium phases" I, II, and III -- gapless, gapped (Volkov-Kogan), and oscillatory -- in these experiments can be derived from the many-body wavefunction $\Psi(t)$
 determined above. 
 
   The pseudospin (fermionic) part of $\Psi(t)$ is a direct product of spin-1/2 wavefunctions $\prod_\p (\bar u_\p |\downarrow\rangle + \bar v_\p |\uparrow\rangle)$ found in Sect.~\ref{asym-1}. In
 the gapless steady state
 \beg\label{gaplessuv}
\pmat
u_\p\\
\\
v_\p\\
\epmat
=
 \cos\frac{\theta_\p}{2}\!\!
\pmat
 1\\
\\
 0\\
\epmat
e^{-i\eps_\p t}
+ \sin\frac{\theta_\p}{2}\!\!
\pmat
 0\\
\\
 1\\
\epmat 
 e^{i\eps_\p t-i\delta_\p},
\en
where $\cos\theta_\p\equiv\cos\theta(\eps_\p)$ is given by \eref{distAA}    in all three phases. The second term represents an occupied  pair of  states  $\pm \p$ (pseudospin up), the first -- empty (pseudospin down).  $\Psi(t)$ in the gapless phase is a coherent superposition of eigenstates of a free Fermi gas with different energies reflecting the fact that $\Delta(t)\to 0$ implies vanishing of interactions between fermions on the mean-field level.  Effectively the system is  governed by a non-interacting Hamiltonian   at $t\to\infty$.  It nevertheless retains superconducting correlations. For example, in the weak coupling  regime its superfluid density   is half  that in the ground state and in phase II\cite{Dzero2006}. Phase I   is therefore a nonequilibrium gapless superfluid.

  In  the gapped steady state  \esref{gen} and \re{UpVpm0}  imply
 \beg\label{geneg}
 \begin{split}
\pmat
u_\p e^{i\mu_\infty t}\\
\\
v_\p e^{-i\mu_\infty t}\\
\epmat
=
\cos\frac{\theta_\p}{2}
\overbrace{\pmat
|U_\p|\\
\\
|V_\p|\\
\epmat e^{-iE_\p^\infty t}}^{\textrm{ground state pair}} 
+\\
\sin\frac{\theta_\p}{2}
\overbrace{\pmat
|V_\p|\\
\\
-|U_\p|\\
\epmat e^{iE_\p^\infty t }}^{\textrm{excited pair}},
\end{split}
\en
where  
\beg
|U_\p|=\sqrt{\frac{1}{2}+\frac{\xi_\p}{2E_\p^\infty}} ,\quad |V_\p|=\sqrt{\frac{1}{2}-\frac{\xi_\p}{2E_\p^\infty}} ,\\
\label{absUpVp}
\en
$\xi_\p=\eps_\p-\mu_\infty$, and we dropped the nonessential constant phase $\varphi$. Bogoliubov amplitudes $|U_\p|$ and $|V_\p|$ are the same as in the BCS ground state\cite{Schrieffer} with gap $\Delta_\infty$ and chemical 
potential $\mu_\infty$. 
The two wavefunctions on the right hand side of \eref{geneg} are the two orthonormal eigenstates of the BdG Hamiltonian
\beg
H_{\mathrm{BdG}}=
\left(\begin{matrix} 
 \xi_\p & \Delta_\infty \\ 
 \Delta_\infty & -\xi_\p \end{matrix}\right).
 \en
 The first one is a Cooper pair wavefunction in the BCS ground state  and corresponds to an alignment of the pseudospin $\vec s_\p$ antiparallel to the effective magnetic field. The second one is an excited state of the Cooper pair
($\vec s_\p$ parallel to the effective magnetic field) termed an \textit{excited pair} in the original BCS work\cite{BCS}. It is interesting to note that these excitations of the condensate in superconducting metals carry no charge and spin, so non-adiabatic dynamics considered here provides a unique venue for creating and measuring them\cite{Coleman2007}. The steady state in phase II therefore is a coherent mixture of ground state and excited pairs -- a superposition of eigenstates of the BCS Hamiltonian with gap $\Delta_\infty$ and chemical 
potential $\mu_\infty$.

A similar    interpretation of the oscillatory state obtains by Fourier transforming the amplitudes \re{UpVpm1}
\beg\label{genegosc}
 \begin{split}
\pmat
u_\p e^{i\tilde\mu t}\\
\\
v_\p e^{-i\tilde\mu t}\\
\epmat
=
\sum\limits_{n=-\infty}^\infty \left\{ \cos\frac{\theta_\p}{2} 
 \pmat
a_{\p n}\\
\\
b_{\p n}\\
\epmat e^{-i(e_\p - n\omega_\Delta) t} 
+\right.\\
\left. \sin\frac{\theta_\p}{2} 
 \pmat
\bar b_{\p n}\\
\\
-\bar a_{\p n}\\
\epmat e^{i(e_\p - n\omega_\Delta) t }\right\},
\end{split}
\en
where $\omega_\Delta$ is the oscillation frequency of $|\Delta(t)|$, $\tilde\mu$ and $-2e_\p$ are the zeroth harmonics of the phase of $\Delta(t)$ and the common phase of the amplitudes, see \esref{amp} and \re{disp}, and we again dropped the constant phase $\varphi$. This expression derives by first going to a frame rotating with frequency $2\tilde\mu$ to get rid of the linear term in the phase of $\Delta(t)$. This makes $e^{-i\phi_\p}$, the  term involving the relative phase,  periodic according to \eref{samp} and it does  not contribute to the momentum dependent phases on the right hand side. Phase III therefore can be understood as a superposition of generalized excited and ground state pairs with dispersions $\pm e_\p$ and     quanta of the amplitude (Higgs) mode $|\Delta(t)|$. As noted in Sect.~\ref{asym1}, $e_\p\to\eps_\p$ at large $\eps_\p$.

The knowledge of the steady state allows us to compute  far from equilibrium correlation and Green's functions in all three phases. For example\cite{note177},
\beg
\begin{array}{r}
i{\cal G}_{\p,>}(t,t')=\langle \hat a^{\phantom{\dagger}}_{\p\up}(t) \, \hat a^\dagger_{\p\up}(t')\rangle= \bar u_{\p}(t) u_{\p}(t'),\\
-i{\cal G}_{\p,<}(t,t')=\langle \hat a^\dagger_{\p\up}(t')  \, \hat a^{\phantom{\dagger}}_{\p\up}(t)\rangle= \bar v_{\p}(t') v_{\p}(t),\\
{\cal G}_{\p}^+(t,t')=\langle \hat a^{ \dagger}_{-\p\dn}(t) \, \hat a^\dagger_{\p\up}(t')\rangle=  v_{\p}(t) \bar u_{\p}(t').\\
\end{array}
\en
 With these we can  evaluate various observables such as the superfluid density mentioned earlier in this section.  
 Note also that the steady state momentum distribution $n^\infty_p(t) dp$ is simply related to the $z$-component of the pseudospin according to \eref{pseudodef}.  Taking into account that $\p$ and $-\p$ are both included in $s_\p^z$ and integrating over the angles, we have
\beg
n^\infty_p(t)=2p^2(2 s_\p^z+1).
\en
Expressions for $s_\p^z$ in phases I, II, and II appear in \eref{ssa161}, \eref{ss0s}, and \esref{ss} and \re{sigsig}, respectively.

Finally, let us discuss the signatures of nonequilibrium phases in radio frequency (RF) spectroscopy\cite{Torma2000,Regal2003,rf1,Kinnunen2004,rf2,rf3,rf4}. Recall that in an atomic Fermi gas the pairing occurs between atoms in two different hyperfine states,  $|\up\rangle\equiv|1\rangle$ and $|\dn\rangle\equiv|2\rangle$.  The RF photon transfers atoms from one of these states, say $|2\rangle$, to the third hyperfine state 
$|3\rangle$ that does not interact with $|1\rangle$ and $|2\rangle$. In an unpaired Fermi gas where atoms $|2\rangle$ are free, the RF absorption spectrum has a peak at the atomic transition energy $\omega={\cal  E}_{23}$. In the paired ground state, the peak shifts to $\omega>{\cal  E}_{23}$ by an amount equal to the minimum binding energy  of Cooper pairs\cite{Torma2000}. 

The RF response of steady states I, II, and III was calculated in Ref.~\onlinecite{Coleman2007} for quenches within the BCS regime and in Ref.~\onlinecite{Yuzbashyan2013} for quenched $p$-wave superfluids. The calculation in the present case is identical\cite{note476}, so we will not reproduce  it here. The RF spectrum of phase I is similar to that of the normal state -- a peak at $\omega={\cal  E}_{23}$. In phase II there are two peaks -- at $\omega>{\cal  E}_{23}$ and $\omega<{\cal  E}_{23}$, which come from the ground state and excited pairs, respectively, see \eref{geneg}. The first peak corresponds to a process in which an RF photon breaks a ground state pair, the second -- excited pair. The RF response of phase III similarly reflects the structure of the corresponding steady state wavefunction \re{genegosc}. There are two series of peaks spaced by $\omega_\Delta$, the frequency of oscillations of $|\Delta(t)|$, coming from processes where an RF photon breaks a ground state (excited) pair and absorbs or emits several quanta of the amplitude (Higgs) mode $|\Delta(t)|$.
 
\section{Conclusion}

In this paper we studied the coherent dynamics of an isolated BCS-BEC condensate   in  two and one channel (BCS) models in two and three spatial dimensions.  Our main focus was on
detuning quenches $\omega_i\to\omega_f$ (interaction quenches $\lam_i\to\lam_f$ in the one channel model). We constructed  exact quench phase diagrams   and predicted the order parameter dynamics $\Delta(t)$ and the full  time-dependent wavefunction $\Psi(t)$ of the system at large times for any pair of values~$(\omega_i, \omega_f)$. In contrast to most previous work, we considered   quenches beyond the weak coupling limit of BCS to BCS  quenches. We add to this BCS to BEC and BEC to BCS quenches across the Feshbach resonance as well as quenches on the BEC side. We showed that the weak coupling limit is universal in that it is model and dimension independent. Outside of this limit, there are several qualitatively new features, the two channel model having richer quench phase diagram  as it contains an extra parameter -- dimensionless resonance width~$\gamma$. All results for the one channel model obtain from the two channel ones by taking the broad resonance, $\gamma\to\infty$, limit.

We find the same three main nonequilibrium phases (asymptotic states) as in the weak coupling regime. Interestingly, this seems to be a  universal, model independent feature of  quench dynamics of fermionic condensates, at least when there is a global complex order parameter, so that the  Cooper pairs interact only through this collective mode. The same three phases occur, for example, in $p$-wave superconductors\cite{Yuzbashyan2013,2014PwavePRL}, spin-orbit coupled superfluids\cite{Pu2014}, and $s$-wave superconductors with energy-dependent  interaction\cite{Levitov2007}. One can speculate that similar universality according to the order parameter type exists among quench phase diagrams of multicomponent superfluids, such as  three fermion species with pairing interactions or multi-band  superconductors.

The above three main phases are: phase I where $\Delta(t)$ vanishes, phase II where $\Delta(t)\to\Delta_\infty e^{-2i\mu_\infty t}$ up to a constant phase factor, and phase III where $|\Delta(t)|$ oscillates persistently. It turns out $\mu_\infty$ plays the role of a nonequilibrium analog of the chemical potential. For quenches within the weak coupling regime $\mu_\infty\approx \eps_F$, while for quenches to deep BEC $\mu_\infty\to -\infty$. Some of the new effects as one moves beyond the weak coupling regime are as follows. The oscillatory approach of 
$|\Delta(t)|$ to a constant (Volkov-Kogan behavior) changes from $1/\sqrt{t}$ for $\mu_\infty>0$ to $1/t^{3/2}$ in three dimensions and $1/(t\ln^2 t)$ in two dimensions for $\mu_\infty<0$, and the oscillation frequency changes from $2\Delta_\infty$ to $2\sqrt{\mu_\infty^2+\Delta_\infty^2}$. For resonance width below a certain threshold, the asymptotic gap amplitude $\Delta_\infty$ can be much larger than $\Delta_{0f}$ -- the ground state gap at final detuning $\omega_f$. Similarly, exponential vanishing of $|\Delta(t)|$ in phase I gives way to a power law behavior. Persistent oscillations in phase III are first  suppressed for stronger quenches and then disappear altogether.  For example,   in three dimensional  one channel model  there is a critical coupling $\lam_c$, such that even  quenches from an infinitesimally small $\lam_i$ to $\lam_f>\lam_c$ produce no such oscillations. As $\lam_f$ approaches $\lam_c$ from below, the oscillation amplitude first increases, then decreases and finally vanishes at $\lam_f=\lam_c$.

The post-quench asymptotic state of the condensate is a coherent superposition of ground state and excited pairs at each momentum (multiple bands of such pairs shifted by the oscillation frequency of $|\Delta(t)|$ in phase III).  These are two orthogonal eigenstates of a Cooper pair in the self-consistent field, and, for instance, the BCS ground state is a direct product of ground state pair wavefunctions.  Our steady state in phases I and II is a direct product of such time-dependent superpositions. In the Anderson pseudospin language, ground state (excited) pairs correspond to the alignment of pseudospin antiparallel (parallel) to the magnetic field. Even though we refer to these states as ground state or excited pairs, we should stress that they are not the same as similar states of Cooper pairs in the ground or excited states of the BCS Hamiltonian since the self-consistent field is different. Excited pairs are elusive excitations in superconductors -- it is difficult to couple to them as they carry no charge or spin. Non-adiabatic dynamics of the BCS-BEC condensate provides an opportunity to access them, e.g. in the RF absorption spectrum. 

Our treatment of the dynamics of the BCS-BEC condensate neglects the coupling  to  the non-condensed modes (mean field approximation) -- molecules with nonzero momenta $\q$ in the two channel model. We check the validity of this approximation for the two channel model by estimating the rates of the decoherence processes due to these terms for post-quench steady states in phase II and comparing them to the typical timescale on which the quench dynamics occurs. Our  preliminary results indicate that the mean field approach is justified for quenches sufficiently far from the $\mu_\infty=0$ line in the quench phase diagrams, e.g. quenches  within deep BEC, deep BCS, or across the resonance from deep BCS to deep BEC and vice versa. A more thorough study of these effects is necessary to fully clarify the situation. 
 
 In mean field   various pairing Hamiltonians, e.g. one and two channel models considered here,  chiral $p$-wave BCS,  a certain class of $d$-wave BCS models\cite{Richardson2002}, are equivalent to integrable classical spin (or spin-oscillator) chains with long-range interactions. The most remarkable general feature of their dynamics is a reduction in the number of effective degrees of freedom  as $t\to\infty$. Consider e.g. the one channel   model. As explained above,  its  dynamics in the thermodynamic limit at long times after the quench can be described in terms of just a few -- zero (phase I), one (phase II), or two (phase III) -- new collective classical spin variables. In other words, the number of spin at long times reduces from infinity to zero, one, or two. Moreover, the new spins time-evolve with the same Hamiltonian only with ``renormalized'' parameters. For example, in phase I the effective Hamiltonian at large times is simply $H=0$, and in phase II it is $H=2\mu_\infty S_z - g S_-S_+$, where $\vec S$ is the collective spin  of length $|\vec S|=\Delta_\infty/g$,  and $g$ is the original BCS coupling constant. The order parameter $\Delta(t)$ coincides with that of the few spin problem, while the original spins  relate to the collective ones in a more involved fashion. 
 
 It is this  feature of the dynamics together with the integrability of the underlying model that allowed us to explicitly  determine the exact post-quench asymptotic state of the system.  In this paper we presented for the first time a comprehensive, consistent overview of a general method to explicitly evaluate the large time asymptotic solution in classical integrable systems that support this kind of reduction.  We are not aware of any similar method for other integrable nonlinear models, the rather different soliton resolution conjecture\cite{solres} being the closest analog we were able to identify.
 
 An interesting open question is whether a similar reduction in the number of degrees of freedom in the course of time evolution occurs also in non-integrable pairing models. This can explain the aforementioned universality of the quench phase diagrams among  systems characterized by a global complex order parameter. It   seems non-accidental indeed that the non-integrable  spin-orbit coupled superfluid\cite{Pu2014} has the same three main post-quench phases and that, moreover, $\Delta(t)$ in phase III is given by an elliptic function dn. Presumably a generalization of this method to non-integrable models would   rely on more general considerations without recourse to integrability-specific techniques and thus would clarify the underlying physical mechanism. It would also  make a number  of interesting problems, such as e.g. the  competition between chiral and anti-chiral components in $p$-wave superconductors upon switching on superconducting interactions and more generally the dynamical interplay among various components in a multicomponent superfluid, potentially amenable to in depth analysis.

\begin{acknowledgments}

This work was supported in part  by the David 
and Lucile Packard Foundation (M.S.F. and E.A.Y.), by the Welch Foundation under Grant No.~C-1809 and by
an Alfred P. Sloan Research Fellowship (BR2014-035) (M.S.F.).

\end{acknowledgments}

\appendix
\section{Pair-breaking rates}
\label{pairbreak}

In this appendix, we  perform a preliminary analysis of the validity of neglecting $\q\ne 0$ terms far from equilibrium in  the Hamiltonian \re{eq:twochannel}. So far we have studied the quench dynamics of the condensate decoupled from these non-condensed modes. 
There are two kinds of relevant processes due to the $\q\ne0$ terms: (i) excitation of molecules out of  the condensate and (ii) excitation of fermionic quasiparticles through two-particle collisions. We estimate characteristic timescales of both processes in the post-quench steady state. We find that sufficiently far from the $\mu_\infty=0$ line in our quench phase diagrams (see  Figs.~\ref{psd2ch3d} and \ref{psd3D})  these timescales are much larger than the characteristic time of the quench dynamics. This means that dropping $\q\ne 0$ terms is indeed justified at times it takes for the quench dynamics to develop and reach the steady state. At much later times, after the quench dynamics plays out, these terms set in, presumably leading to decoherence and eventual thermalization of our (isolated) system.  We note also that the $\mu_\infty=0$ line can be  very roughly interpreted  as a far from equilibrium generalization of the unitarity point. Quenches away from this line  are from  BCS or BEC initial detuning to far BCS and BEC side   including quenches across the  resonance.  

  In what follows we consider a three-dimensional condensate and, for simplicity, we content ourselves with steady states in phase II (including II') where pairing amplitude asymptotes to a constant, $|\Delta(t\to\infty)|=\Delta_{\infty}$. 
  
  \subsection{Steady state molecular production}
  
 Here we compute the rate at which molecules with non-zero momentum are produced in steady state II  where initially all molecules have zero momentum. To the lowest order in the interaction, the   corresponding scattering amplitudes  are\cite{Phillips}
 \beg\label{AmpMol}
\begin{split}
{\cal A}_b(\p_1,\p_2)&\delta(E_\mathrm{fin}-E_\mathrm{in})=\int\limits_{-\infty}^\infty\langle \Psi_\mathrm{fin}| \hat V(t)|\Psi_\mathrm{in}\rangle dt,
\end{split}
\en 
where $|\Psi_\mathrm{in}\rangle$ and $E_\mathrm{in}$ are the steady state wavefunction and energy.  $|\Psi_\mathrm{fin}\rangle$ obtains from $|\Psi_\mathrm{in}\rangle$ 
by  destroying two pairs and creating a molecule with momentum $\bq=\p_1+\p_2$ and  two unpaired atoms with momenta $\p_1$ and $\p_2$.   The energy
of the final state is 
\beg
E_\mathrm{fin}=E_\mathrm{in}+\zeta_\bq\pm E_{\p_1}^\infty\pm E_{\p_2}^\infty,
\en
where plus  (minus)  corresponds to a ground (excited) pair  and 
\beg
\zeta_\bq=\frac{q^2}{4m}+\omega_f-2\mu_\infty,
\en
 is the energy of the molecule.  The interaction $\hat V(t)$  is described by the last term in \eref{eq:twochannel} 
\beg\label{Vt}
\begin{split}
\hat V(t)=g \sum_{\p_1, \p_2}\!\! &\left[ \b_{\p_1+\p_2}^\dagger(t) \a_{ \p_1 \uparrow}(t) \a_{ \p_2\downarrow}(t) +\right.\\&\left. \b_{\p_1+\p_2}(t) \ad_{ \p_2\downarrow}(t) \ad_{ \p_1\uparrow}(t) \right].
\end{split}
\en

Since our initial state does not contain molecules with non-zero momentum, only the first term in \eref{Vt}
contributes to the matrix element \re{AmpMol}. One also needs to keep in mind that  our steady state   contains
 superpositions of a ground state pair with   energy $-E_{\bp}^\infty$ and an excited   pair with 
  energy $+E_\bp^\infty$  for each $\bp$. \esref{geneg} and \re{AmpMol}  then yield the following four   scattering amplitudes\cite{note177}:  
\beg\label{MatElement}
\begin{split}
{\cal A}_b^{(--)}(\p_1,\p_2)&=g\cos\frac{\theta_{\p_2}}{2}\cos\frac{\theta_{\p_1}}{2}|V_{\p_2}||V_{\p_1}|, \\
{\cal A}_b^{(+-)}( \p_1,\p_2)&=g\sin\frac{\theta_{\p_2}}{2}\cos\frac{\theta_{\p_1}}{2}|U_{\p_2}||V_{\p_1}|, \\
{\cal A}_b^{(++)}( \p_1,\p_2)&=g\sin\frac{\theta_{\p_2}}{2}\sin\frac{\theta_{\p_1}}{2}|U_{\p_2}||U_{\p_1}|, \\
\end{split}
\en
where $-$ $(+)$ describes breaking a ground state (excited) pair,   and ${\cal A}_b^{(-+)}( \p_1,\p_2)={\cal A}_b^{(+-)}( \p_2,\p_1)$.    

 Molecular production rate per atom at zero temperature  obtains from these  amplitudes  and Fermi's Golden Rule\cite{Phillips}  
 
\beg\label{RateB}
\begin{split}
\tau_\textrm{mol}^{-1}=\frac{2\pi}{N_f}&\!\sum\limits_{\p_1\p_2\alpha\beta}|{\cal A}_b^{(\alpha\beta)}(\p_1,\p_2) |^2\times\\&  \delta\left(\zeta_{\p_1+\p_2}-\alpha E_{\p_2}^\infty-\beta E_{\p_1}^\infty\right).
\end{split}
\en
In this expression $N_f$ is the total number of fermions in the absence of molecules and  we took into account that there are no molecules with non-zero momentum in our steady state.  

Let us specialize to quenches into either deep BCS ($\omega_f\to+\infty$) or deep BEC ($\omega_f\to-\infty$).  We expect a much higher rate in the latter case, because in the BCS regime $\zeta_\q\to+\infty$ requiring excited pairs of extremely high energy to create a molecule. For quenches to far BEC side $\mu_\infty\to-\infty$, while $\Delta_\infty$ remains finite regardless of the initial detuning, see e.g. Figs.~\ref{delinfpl} and \ref{muinfpl}.  It follows that $E_\bp^\infty\approx\xi_\bp= |\mu_\infty|+p^2/2m$ and \eref{argument} implies $\omega_f\approx 2\mu_\infty$. For $\alpha=\beta=-1$ the argument of the delta-function in \eref{RateB} is always positive, i.e. energy conservation cannot be satisfied meaning that 
the ground state pairs do not contribute to the rate. Similarly, if $\alpha=\beta=1$ (two excited pairs)
\beg
\begin{split}
&\zeta_{\p_1+\p_2}-E_{\p_1}^\infty-E_{\p_2}^\infty\approx\frac{(\p_1+\p_2)^2}{4m}+\omega_f-\\&\frac{p_1^2+ p_2^2}{2m}=
\omega_f-\frac{(\p_1-\p_2)^2}{4m}<0.
\end{split}
\en

  Therefore, only scattering processes   involving one fermion from an excited and  another from a ground state pair  contribute.   Expression \re{RateB}
for the rate in this case is
\beg\label{RateBMix}
\begin{split}
\tau_{\mathrm{mol}}^{-1} \approx\frac{4\pi g^2 }{N_f} &\sum\limits_{\p_1,\p_2}
\sin^2\frac{\theta_{\p_2}}{2} \cos^2 \frac{\theta_{\p_1}}{2}|U_{\p_2}|^2|V_{\p_1}|^2 \times
\\&\delta\left( \frac{3p_1^2+2\p_1\cdot\p_2-p_2^2}{4m}\right),
\end{split}
\en
   Next, we go from summations to integrations, integrate over the angle between $\p_1$ and $\p_2$, and change integration variables from momenta to energies, which results in
\beg
\begin{split}
\tau_{\textrm{mol}}^{-1} \approx \frac{3\gamma}{2\eps_F}\int\limits_0^\infty d\eps_2 \sin^2\frac{\theta(\eps_2)}{2}|U(\eps_2)|^2\times\\
\int\limits_{\eps_2/9}^{\eps_2} d\eps_1 \cos^2\frac{\theta(\eps_1)}{2}|V(\eps_1)|^2.
\end{split}
\en
 We replace the cosine with one, use $ |V(\eps_1)|^2\approx \Delta_\infty^2/4(\eps_1+|\mu_\infty|)^2$, which follows from \eref{absUpVp} together
with $ |U(\eps_1)|^2\approx 1$, and integrate over $\eps_1$. According to \eref{first2}, the probability of finding an excited pair
\beg
 \sin^2\frac{\theta(\eps_2)}{2}\to \frac{\Delta_{0i}^2(\delta\omega)^2}{16 E_i^4(\eps_2)}\quad\mbox{as $\eps_2\to\infty$}.
 \label{sinext}
 \en
A larger rate obtains for finite $\omega_i$ than for $\omega_i$ close to $\omega_f$. In this case $\delta\omega\approx 2\mu_\infty$ and  $\sin^2[\theta(\eps_2)/2]$ appreciably differs from zero at energies about $\sqrt{\Delta_{0i}|\mu_\infty|}$. We obtain
\beg
\tau_{\textrm{mol}}^{-1}\sim\frac{\gamma \Delta_\infty^2\Delta_{0i}}{\eps_F |\mu_\infty|}\to 0.
\en
In deriving \eref{sinext} we assumed finite resonance width $\gamma$. A separate estimate for the broad resonance limit  for quenches to deep BEC finds a rate that also vanishes, but as $\gamma^{-1/3}|\mu_\infty|^{-1/2}$.
 
This result for the molecular production rate should be compared with the typical time scale $\tau_{\mathrm{dyn}}$ of the quench dynamics for quenches to the far BEC side.  \esref{ccoeffin} and \re{ccoeffinnlin} imply
\beg\label{taudyn}
\tau_{\textrm{dyn}}^{-1}\sim|\mu_\infty|.
\en
We see that  indeed $\tau_{\textrm{dyn}}\ll \tau_{\textrm{mol}}$.

\subsection{Two-particle collisions}

 Next, we estimate the relaxation rate due to two-particle collisions. In contrast   to the molecular production,  we find that here the    contribution  coming from just the ground state pairs is  of the same order of magnitude or larger than  that from collisions that involve excited pairs.  We therefore consider   ground state pairs only  and take the probability of finding such a   pair at  a given momentum $\p$ to be $\cos^2(\theta_\p/2)\approx 1$.  
Let us analyze quenches to the far BCS side of the Feshbach resonance from any initial detuning.  In this case, $\omega_f\gg\mu_\infty\approx\epsilon_F$, see e.g. Fig~\ref{muinfpl}.  The total scattering amplitude for this case has been  studied in Ref.~\onlinecite{Galaiko1972} [see Eq.~(71) therein], which also estimates the corresponding rate as
\beg\label{eq:rateBCS}
{\tau}_{\textrm{in}}^{-1}\sim\left(\frac{g^2\nu_F}{\omega_f}\right)^2\frac{\Delta_\infty^2}{\epsilon_F}=\gamma^2\epsilon_F\left(\frac{\Delta_\infty}{\omega_f}\right)^2.
\en
In fact, this  is the well-know  Fermi liquid result for the quasiparticle lifetime. Indeed, $\lam=g^2\nu_F/\omega_f$ is the strength of the effective interaction between fermions
[see  \eref{eq:parmap}] and $\Delta_\infty$ is the typical excitation energy -- the energy scale at which spins deviate appreciably from their ground state positions.

\eref{eq:rateBCS} has to be compared with the characteristic timescale of the dynamics for quenches to the far BCS side. According to \eref{statptexnonlin} this timescale is
\beg\label{taudynBCS}
\tau_{\textrm{dyn}} \sim\frac{1}{\Delta_\infty}.
\en
We see that $\tau_{\textrm{dyn}}\ll \tau_{\textrm{in}}$ for any finite resonance width $\gamma$ since $\omega_f\to\infty$ in deep BCS. In the broad resonance limit too
$\tau_{\textrm{dyn}}/\tau_{\textrm{in}}=\lam^2\Delta_\infty/\eps_F \ll 1$. This is because at large $\gamma$ quenches to far BCS in phase II are only possible from initial detunings also on the far BCS side, see e.g. Figs.~\ref{psd2ch3d}(c) and \ref{psd1ch3d}. It then follows from \eref{qwrty} that $\Delta_\infty \le \Delta_{0f}\ll \eps_F$.

 A preliminary analysis for quenches to the far BEC side shows that, at least for a finite resonance width $\gamma$ and sufficiently large $|\omega_f|$, one still has $\tau_{\textrm{dyn}}\ll\tau_{\textrm{in}}$. Thus, neglecting two-particle collisions is justified at times it takes  the quench dynamics to fully develop and reach its asymptote.

 \section{Finite size corrections to the roots}
\label{appb}

As mentioned in Sect.~\ref{qpsd}, in the thermodynamic limit   $\vec L^2(u)$ for quench initial conditions  has a continuum of roots along the positive real axis. Here we verify this and determine  finite size corrections to these roots.

Roots of $\vec L^2(u)$ are determined by \eref{rtdis} or, equivalently, by \eref{rtdisnew} in notation explained in the beginning Sect.~\ref{normm}, which we employ here as well. The level spacing $\delta$ is of order $1/N$. Thermodynamic limit means $N\to\infty$, so $\eps_k$ become continuous with density $\nu(\eps)$.

Let us look for a pair of complex conjugate roots close to $\eps_m$ writing it as $c_m=\eps_m +\varsigma_m \delta$. 
We take $\varsigma_m\equiv\varsigma(\eps_m)$ to be of order 1, to be confirmed below. Note that $\varsigma_m$ is generally complex. Our goal is to evaluate $c_m$ to first order in $1/N$. We split the summation  in \eref{rtdisnew} into two parts -- over $\eps_k$ in a small interval $(\eps_m-\Delta\eps, \eps_m+\Delta\eps)$ and over remaining $\eps_k$. The interval is however sufficiently large so that it contains many $\eps_k$.  Specifically, $\Delta\eps\to 0$, but $\Delta\eps/\delta\to \infty$ in thermodynamic limit. For example,  
$\Delta\eps=\delta\sqrt{N}$  fulfills these conditions. The latter summation becomes a principal value integral in the $N\to\infty$ limit, while the former one
to leading order in $1/N$ reads
\beg
\begin{split}
\frac{N(\eps_m)}{ 2E(\eps_m) \delta}\sum_{p=0}^\infty \left[\frac{1}{p+\varsigma_m}-\frac{1}{p+1-\varsigma_m} \right]=\\
\frac{\pi \nu(\eps_m) }{ 2 E(\eps_m) }\cot \pi\varsigma_m.\\
\end{split}
\label{cot}
\en
The first sum is from $\eps_k<\eps_m$, the second -- from $\eps_k>\eps_m$. Here it is important that  the degeneracy $N_k\equiv N(\eps_k)$ and the spacing between $\eps_k$ vary smoothly with $\eps_k$. As long as this is the case, we can include any variation of the spacing into $N_k$.

Thus, \eref{rtdisnew} to leading order in $1/N$ becomes
\beg
\begin{split}
\frac{2}{g^2}-\dashint_0^\infty \frac{\nu(\eps')d\eps'}{2(\eps_m-\eps')E(\eps')}-\frac{\nu(\eps_m) }{ 2E(\eps_m) }\cot \pi\varsigma_m=\\
\frac{\delta\omega}{g^2}\frac{\eps_m-\mu\pm i\Delta_0}{E^2(\eps_m)}.\\
\end{split}
\en
Recalling that $\nu(\eps)=\nu_F f(\eps)$ and $g^2\nu_F=\gamma$ in units of Fermi energy, we obtain
\beg
\begin{split}
\pi\cot \pi\varsigma(\eps)=\frac{4E(\eps)}{\gamma f(\eps)}-\frac{G(\eps)}{f(\eps)} -
\frac{2\delta\omega}{\gamma}\frac{\eps-\mu\pm i\Delta_0}{E(\eps)f(\eps)},\\
\end{split}
\label{conroots}
\en
where
\beg
G(\eps)= E(\eps) \dashint_0^\infty \frac{f(\eps')d\eps'}{(\eps-\eps')E(\eps')}.
\label{ge}
\en
This principal value integral  is the same as in \eref{realroot}. We evaluated it in elementary functions for various cases
in Sects.~\ref{phd2d} and \ref{finitegamma3d}. Specifically, in 2d 
\beg
G_\mathrm{2d}(\eps)=\ln\left[\frac{\eps\left(\eps-\mu+E(\eps)\right)}{ E(\eps)\sqrt{\mu^2+\Delta_0^2}+\mu^2+\Delta_0^2-\mu\eps}\right],
\label{ge2d}
\en
in weak coupling (BCS) limit, $\mu\approx \eps_F\gg\Delta_0$, for energies not too far from the Fermi energy, in both 2d and 3d
\beg
G_\mathrm{wc}(\eps)=\ln\left[\frac{E(\eps)+\eps-\mu}{ E(\eps) -\eps+\mu}\right],
\label{gewc}
\en
in  strong coupling (BEC) limit in 2d and 3d,
\beg
G_\mathrm{sc}^\mathrm{2d}(\eps)=\ln\frac{\eps}{|\mu|},\quad 
G_\mathrm{sc}^\mathrm{3d}(\eps)=
-\pi\sqrt{\frac{|\mu|}{\Delta_0}}.
\label{gesc}
\en

Ground state continual roots $x_k=\eps_k+\varrho_k \delta$ obtain by setting $\delta\omega=0$ in \eref{conroots}, i.e.
\beg
 \pi\cot \pi\varrho(\eps)=\frac{4E(\eps)}{\gamma f(\eps)}-\frac{G(\eps)}{f(\eps)} \equiv 
 \frac{H(\eps)}{f(\eps)}.
\label{xkkk}
\en
The quantity $F_k\equiv F(\eps_k)$ defined in \eref{fk} evaluates similarly to \eref{cot},
\beg
\begin{split}
F(\eps)=-\frac{N(\eps)}{ 2E(\eps) \delta^2}\frac{\partial}{\partial\varrho}\sum_{p=0}^\infty \left[\frac{1}{p+1-\varrho}- \frac{1}{p+\varrho} \right]=\\
\frac{\nu(\eps)}{ 2E(\eps) \delta} \frac{\pi^2f^2(\eps)+H^2(\eps)}{f^2(\eps)}.\\
\end{split}
\label{feps}
\en

\section{Identities}
\label{appa}

In this appendix we prove \eref{iden11}. To this end, consider a function
\beg
R(u)=L_0(u)\left[ (u-\mu)^2+\Delta_0^2\right],
\label{ru1}
\en
where $L_0(u)$ is given by \eref{l0x}. Since zeroes of $L_0(u)$ are $x_k$ and its poles are $\eps_k$ it alternatively can be written
\beg
R(u)= -\frac{2}{g^2}\frac{\prod_{k=1}^N (u-x_k)}{\prod_{k=1}^N(u-\eps_k)}\left[ (u-\mu)^2+\Delta_0^2\right],
\label{ru2}
\en
\eref{iden11} follows by matching two leading terms in $1/u$ expansions of function $1/R(u)$ obtained with the help of these two alternative forms.

Because $1/R(u)$ is a rational function with poles at $u=x_k$ and $\mu\pm i\Delta_0$, we have
\beg
\begin{split}
\frac{1}{R(u)}=\sum_k \frac{1}{(u-x_k)L'_0(x_k) \Omega_k^2}+\\
\frac{1}{2i\Delta_0 (u- c_+)L_0(c_+)}-\frac{1}{2i\Delta_0 (u- c_-)L_0(c_-)},\\
\end{split}
\label{1ru}
\en
where $c_\pm=\mu\pm i\Delta_0$ and we took into account that the square bracket in \eref{ru1} evaluated at $u=x_k$ is equal to $\Omega_k^2$.
Note also that $L'_0(x_k)=-F_k$, see \eref{fk}.  
\eref{l0x} implies
\beg
L_0(c_\pm)=-\beta_k\mp i\alpha_k,
\en
where
\beg
\begin{split}
\alpha_k=\sum_k\frac{N_k \Delta_0}{2 \left[E(\eps_k)\right]^{3/2}},\\
\beta_k=\frac{2}{g^2}+\sum_k\frac{N_k (\eps_k-\mu)}{2 \left[E(\eps_k)\right]^{3/2}}.\\
\end{split}
\label{akbk}
\en

The leading term in $1/u$ expansion of $1/R(u)$ according to \eref{ru2} is $-2/(g^2 u^2)$. Therefore, the coefficient at $1/u$ in \eref{1ru} vanishes and
that at $1/u^2$ is $-2/g^2$. This yields 
\beg
\begin{split}
\sum_k\frac{\Delta_0}{F_k\Omega_k^2}=\frac{\alpha_k}{\alpha_k^2+\beta_k^2},\\
\sum_k\frac{x_k-\mu}{F_k\Omega_k^2}=\frac{g^2}{2}-\frac{\beta_k}{\alpha_k^2+\beta_k^2}.\\
\end{split}
\label{iden33}
\en

Gap and chemical potential equations \re{gapeq} and \re{mu} in the notation of Sect.~\ref{normm} read
\beg
\begin{split}
\frac{\omega-2\mu}{g^2}=\sum_k\frac{N_k}{2E(\eps_k)},\\
 2n=\frac{2\Delta_0^2}{g^2}+\sum_k N_k\left(1-\frac{\eps_k-\mu}{E(\eps_k)}\right).
 \end{split}
 \label{gapchemnew}
\en
Differentiation of these equations with respect to $\omega$  obtains $\delta\mu/\delta\omega$ and $\delta\Delta_0/\delta\omega$ and comparison of the resulting quantities with the right hand side of \eref{iden33} proves \eref{iden11}.

Another identity used in Sect.~\ref{normm} derives by noting that according to \eref{ru2} $1/R(\eps_k)=0$. Setting $u=\eps_k$ in \eref{1ru}, we obtain after some algebra
\beg
\sum_j \frac{1}{(\eps_k-x_m) F_m \Omega_m^2}=\frac{\alpha_k(\eps_k-\mu)-\Delta_0\beta_k}{2\Delta_0(\alpha_k^2+\beta_k^2)\left[ E(\eps_k)\right]^2}.
\label{idenuu}
\en

\bibliography{paper}

\begin{thebibliography}{83}
\bibitem[{not({\natexlab{a}})}]{notelin}
\bibinfo{howpublished}{\textrm{The integrand in \eref{phiintro} generally has a
  nonvanishing zeroth Fourier mode, i.e. it contributes to both linear and
  periodic parts.}}
\end{thebibliography}

\end{document}